\documentclass[11pt]{article}
\pdfoutput=1
\usepackage{jheppub}
\usepackage{upgreek}
\usepackage{float, extarrows, tikz-cd}
\usepackage{graphicx}
\usepackage{slashed}
\usepackage{tabularx,ragged2e}
\usepackage{amssymb}
\usepackage{subcaption}
\usepackage{amsmath,amssymb,amscd}
\usepackage{slashed}
\usepackage{caption}
\usepackage{xcolor}
\usepackage{dsfont}
\usepackage{verbatim}
\usepackage{mathtools, xcolor,ytableau, amsfonts,tikz}
\usepackage{graphicx}
\usepackage{physics}
\usepackage{simpler-wick}
\usepackage{microtype}
\usepackage{lineno}
\usepackage[nobottomtitles*]{titlesec}
\newcolumntype{C}{>{\Centering\arraybackslash}X}

\numberwithin{equation}{section}

\def\le{\left}
\def\ri{\right}
\newcommand\ov{\over}

\newcommand{\es}[2] {\begin{equation} \label{#1} \begin{split} #2 \end{split} \end{equation}}

\def\<{\langle}
\def\>{\rangle}

\newcommand\Om{\Omega}

\author[a]{Hanzhi Jiang}        
\author[b]{and M\'ark Mezei}

                \affiliation[a]{Rudolf Peierls Centre for Theoretical Physics, University of Oxford, Oxford OX1 3PU, U.K.}                                                              
               \affiliation[b]{Mathematical Institute, University of Oxford, Woodstock Road, Oxford, OX2 6GG, U.K.}  

\emailAdd{hanzhi.jiang@physics.ox.ac.uk}     
\emailAdd{mezei@maths.ox.ac.uk}

\begin{document}

\title{New horizons for inhomogeneous quenches and Floquet CFT}

\abstract{A fruitful avenue in investigating out-of-equilibrium quantum many-body systems is to abruptly change their Hamiltonian and study the subsequent evolution of their quantum state. If this is done once, the setup is called a quench, while if it is done periodically, it is called Floquet driving. We consider the solvable setup of a two-dimensional CFT driven by Hamiltonians built out of conformal symmetry generators: in this case, the quantum dynamics can be understood using two-dimensional geometry. We investigate how the dynamics is reflected in the  holographic dual three-dimensional spacetime and find new horizons. We argue that bulk operators behind the new horizons are reconstructable by virtue of modular flow.}

\maketitle
\flushbottom

\section{Introduction}

Understanding the out-of-equilibrium dynamics in interacting many-body systems is challenging: there are no conventional small parameters to expand in and numerical simulations are hampered by the exponentially large Hilbert space that the system explores. There are a handful of special setups, where analytical progress can be made: in random, dual unitary, and stabiliser quantum circuits~\cite{Fisher:2022qey}; in models of Sachdev-Ye-Kitaev type~\cite{Rosenhaus:2018dtp,Chowdhury:2021qpy}; in systems with a holographic gravity dual~\cite{Son:2002sd,Balasubramanian:2011ur,Shenker:2013pqa}; and in conformal field theories (CFTs) driven by unconventional Hamiltonians. This paper examines the interplay between the latter two, both of which describe quantum dynamics in terms of geometry: in terms of conformal transformations of the complex plane and in terms of gravity in three-dimensional asymptotically AdS spacetimes. 

1+1 dimensional CFTs with inhomogeneous deformations have received considerable attention recently \cite{Wen:2018vux,Wen:2018agb,Fan:2019upv,Wen:2020wee,Han:2020kwp,Fan:2020orx,Wen:2022pyj}, as they provide exactly solvable models that describe quantum critical systems out of equilibrium.  These deformations include the so-called sine square deformation (SSD) \cite{Wen:2018vux}, which is a special case of $SL(2)$ deformations of 2d CFTs. In SSD, one changes the CFT Hamiltonian $H_0=\int_0^l dx\,T_{00}(x)$ on a circle of circumference $l$ to a new Hamiltonian with a sine-square enveloping function
\begin{align}
    H_1=2\int_0^l dx\ \sin ^2\Big(\frac{\pi x}{l}\Big)T_{00}(x)
\end{align}
SSD was first introduced in quantum many-body systems to eliminate boundary effects \cite{Gendiar:2008udd,Hikihara:2011mtb,gendiar2011suppression}. Later on, 2d CFT with SSD Hamiltonian and its generalization, the Mobius deformed Hamiltonian were studied in \cite{Katsura:2011ss,Ishibashi:2015jba,Ishibashi:2016bey,Okunishi:2016zat,Wen:2016inm,Tamura:2017vbx}, where it was shown \cite{Ishibashi:2015jba,Ishibashi:2016bey} that the spectrum of SSD CFT is continuous. In \cite{Wen:2018vux} 
explicit expressions for entanglement entropy (EE) after a quantum quench were found. It was also pointed out in \cite{Wen:2018vux} that SSD CFT can be interpreted as a CFT on curved spacetime. 

Floquet CFT \cite{Wen:2018agb,Fan:2019upv,Wen:2020wee,Han:2020kwp,Fan:2020orx,Wen:2022pyj}\footnote{see also \cite{Andersen:2020xvu,Choo:2022lgm,Lapierre:2024lga}.}  is constructed by $\emph{periodically}$ driving the CFT with ordinary and SSD Hamiltonians, $H_0$ and $H_1$, see Figure \ref{fig:FloquetFig}. The physical motivation is to provide an exactly solvable toy model that captures some of the features of the more realisable periodically driven systems including Floquet topological insulators \cite{oka2009photovoltaic,kitagawa2010topological,lindner2011floquet,Rechtsman:2012qj,cayssol2013floquet,rudner2013anomalous,titum2015disorder,titum2016anomalous,klinovaja2016topological,Thakurathi:2016ypz}, Floquet symmetry-protected phases \cite{Iadecola:2015tra,von2016phase,else2016classification,potter2016classification,po2017radical}, as well as Floquet time crystals \cite{else2016floquet,khemani2016phase,von2016absolute,else2017prethermal,zhang2017observation,yao2017discrete}. Floquet CFT describes driven quantum many-body systems at criticality. Floquet CFT was found to have heating and non-heating phases \cite{Wen:2018agb,Fan:2019upv}, corresponding to different types of $SL(2)$ transformations. In the heating phase, the energy is localised in peaks that also dominate entanglement entropy. If initially prepared in a thermal state, the system outside these peaks cools down and behaves like the vacuum state, leading to the name "Floquet's refrigerators" \cite{Wen:2022pyj}. In the non-heating phases, the energy and entanglement show oscillatory behavior without a fixed period.

Given these recent advances in inhomogeneously deformed CFT$_2$, it is then a natural question to ask what the holographic duals of these deformed CFT$_2$ are.\footnote{We consider CFTs with large central charges $c$ and large gaps, i.e.~holographic CFTs.} The AdS/CFT correspondence \cite{Maldacena:1997re,Gubser:1998bc,Witten:1998qj} 
serves as a powerful tool to study the time evolution of strongly coupled conformal field theories in $d$ spacetime dimensions. For example, the computation of entanglement entropy in the dual $(d+1)$-dimensional Lorentzian spacetime boils down to the areas of codimension-2 extremal surfaces 
\cite{Ryu:2006bv,Ryu:2006ef,Hubeny:2007xt}. Moreover, as we will see shortly, in the holographic dual of SSD CFT, there are new $\emph{horizons}$ emerging in the bulk (see Figure \ref{fig:AdSHor} and \ref{fig:BTZHor}), and bulk excitations falling across these new horizons remain reconstructable.  Another motivation for exploring the holographic dual of SSD and Floquet CFT comes from holographic complexity. In particular, these holographic constructions may serve as testing grounds for various complexity measures and their possible relations, including the  Bogoliubov-Kubo-Mori information geometry studied in \cite{deBoer:2023lrd} in the Floquet CFT context.  

The aim of this paper is to study the gravity duals of SSD and Floquet CFT. There are already holographic studies of SSD and Floquet CFTs in the literature. In \cite{MacCormack:2018rwq}, a $(2+1)$-dimensional bulk metric with SSD boundary is constructed. Holographic dynamics of SSD quench with thermal initial states was investigated in \cite{Goto:2021sqx}, and the time evolution of mutual information under this framework was further explored in \cite{Goto:2023wai}. There are also recent works  \cite{Bernamonti:2024fgx,Kudler-Flam:2023ahk} exploring holographic aspects of SSD or Mobius quench in boundary conformal field theory, whose gravity dual involves end-of-the-world branes. We will construct the gravity dual spacetime for Floquet CFT at general times, whereas prior work primarily focused on stroboscopic, i.e.~discrete times \cite{deBoer:2023lrd,Das:2022pez}. There are also previous works whose setups are related to but different from this paper's: AdS/CFT with periodic time-dependence was studied in \cite{Auzzi:2013pca,Auzzi:2012ca}. Holographic aspects of 2d CFT with quantum KdV deformed Hamiltonian $H=\sum_k \mu_{2k+1}\hat{Q}_{2k+1}$ (where $\hat{Q}_{1}$ is the holomorphic part of the CFT Hamiltonian $H_0$) were addressed in \cite{Dymarsky:2020tjh}.

\subsection{Outline and summary}

In Section \ref{specquench} we analyse a special quantum quench from $H_0$ to the SSD Hamiltonian $H_1$. In Subsection \ref{SSD: Boundary Horizons} we point out that SSD Hamiltonian leads to a different $\emph{foliation}$ of the 2-dimensional spacetime as is shown in Figure \ref{fig:Triangle}. Furthermore, there is a $\emph{triangle}$-shaped Killing horizon \eqref{triangle} with $\emph{zero}$ temperature. 

Afterwards in Section \ref{Gravity dual of SSD}, we investigate holographic aspects of SSD CFT. In Subsection \ref{4.1}, we show that the gravitational duals of SSD CFT prepared in the ground and thermal states are dual to pure AdS$_3$ and BTZ black hole \cite{Banados:1992gq,Banados:1992wn} with non-trivial $\emph{foliations}$, respectively. Then Subsection \ref{New Horizon on AdS_3} and \ref{New Horizon on BTZ} demonstrate the existence of new $\emph{horizons}$ in both pure AdS$_3$ \eqref{AdSKilling} and BTZ \eqref{BTZNullGeod2}, which are plotted in Figure \ref{fig:AdSHor} and \ref{fig:BTZHor}, respectively. In Subsection \ref{New Horizon on AdS_3} we show that the AdS$_3$ new horizon, which is a Killing horizon, also has $\emph{zero}$ temperature. In Subsection \ref{New Horizon on BTZ} we also compare our new horizon in BTZ geometry with the BTZ black hole horizon and find that the new horizon is $\emph{larger}$ (see Figure \ref{fig:HorCompare}). 

In Subsection \ref{bulk reconstructions}, we comment on the reconstructions of bulk operators in the presence of the aforementioned new horizons and conclude that bulk operators remain $\emph{reconstructable}$ even after falling into these horizons (as long as they remain outside the BTZ black hole horizon).

In Section \ref{SSDq} we generalise the SSD quench results to quenches governed by other $SL(2)$ Virasoro subalgebras. In Section \ref{Floquetsetup} we turn our attention to Floquet CFT whose physics has mostly been analysed at stroboscopic times; here we spell out what happens for continuous time between the discrete stroboscopic time steps; see Figure \ref{fig:FloquetGrid} for the relevant plot.   Subsequently, in Section \ref{Gravity dual of Floquet CFT 1} we study the holographic dual of Floquet CFT, and argue that there are no new horizons in the bulk anymore. We also find that in the thermal state cases, the BTZ black hole horizon asymptotically approaches the boundary at two points, which is plotted in Figure \ref{fig:FloquetHor}. This is in agreement with the results in \cite{deBoer:2023lrd} obtained in the slow-driving limit. 

Finally, Section \ref{Holographic Entanglement Entropy} is devoted to the study of holographic entanglement entropy \cite{Ryu:2006bv,Ryu:2006ef,Hubeny:2007xt} in the gravity dual of SSD and Floquet CFT prepared both in the ground and thermal state. We find that due to the non-trivial foliations of the gravity dual of SSD CFT in the bulk, the holographic entanglement entropy is computed by spacelike geodesics anchoring on the boundary that is, in general, at non-equal times, see e.g.~Figure \ref{fig:EEGroundState} for an example. The time evolution of large-radius cutoff surfaces (plotted in Figure \ref{fig:CutoffSurface3d}) plays a crucial role. In cases where CFTs are initially prepared in thermal states, we discover that holographic entanglement entropy shows the entanglement plateau phenomenon \cite{Hubeny:2013gta}. We find precise agreements between our holographic results and field-theoretic calculations in all comparable cases and reproduce holographically the cooling and heating effects \cite{Wen:2022pyj,Goto:2021sqx} discovered earlier in the field theory context.  

Section \ref{conclusion} concludes this work and points out possible future directions. To ease the flow of the main text, we relegated most technical details to often very detailed appendices.

\section{A special quantum quench}\label{specquench}
\subsection{Setup}\label{Set up}

We consider CFT$_2$ on a $\emph{cylinder}$ $\mathbb{R}\times S^1$ of circumference $l=2\pi$ with $\emph{periodic}$ boundary conditions,\footnote{The references \cite{Wen:2018vux,Wen:2018agb,Fan:2019upv,Wen:2020wee,Han:2020kwp,Fan:2020orx,Wen:2022pyj} sometimes adopt $\emph{open}$ boundary condition. We will not use this boundary condition in our work but will point out its appearance when referring to their works.} in either the $\emph{ground}$ state $|0\rangle$ or the $\emph{thermal}$ state $\rho=e^{-\beta H}$ at temperature $\beta^{-1}$. We are interested in the $\emph{Lorentzian}$ time evolution of driven CFTs, although it is technically easier to work in $\emph{Euclidean}$ and then analytically continue to Lorentzian signature. The coordinate on the Euclidean cylinder is 
\begin{equation}
    \begin{aligned}
        w&=\tau+i x\\
        \overline{w}&=\tau-ix
    \end{aligned}
    \label{Eucledian_w}
\end{equation}
Here, we use $\tau$ and $t$ to denote imaginary and real time, respectively. The spatial variable $x$ is periodic with periodicity $2\pi$: $x\sim x+2\pi$. The analytic continuation is given by $\tau\to it$, which makes $w$ and $\overline{w}$ (real) light cone coordinates.

We will mostly work in the $\emph{Heisenberg\ picture}$: we take a primary operator $O$ of weights $(h, \overline{h})$ on the $t=0$ time slice with $w=i x$, $\overline{w}=-ix$ and time evolve it using the operator $e^{-\tau H}$ with $H$ made out of the Virasoro generators. Such a time evolution is $\emph{geometric}$, the operator moves $O$ to $\emph{new}$ positions $(w_{{\rm new}}(\tau,x),\overline{w}_{{\rm new}}(\tau,x))$,\footnote{Here $w=i x$, $\overline{w}=-ix$. In later sections we will use the notations $\frac{\partial w_{{\rm new}}}{\partial w}$ and $\frac{\partial w_{{\rm new}}}{\partial (ix)}$, $\frac{\partial \overline{w}_{{\rm new}}}{\partial \overline{w}}$ and $\frac{\partial \overline{w}_{{\rm new}}}{\partial (-ix)}$ interchangeably. }
\begin{align}
    O^{(H)}(\tau,x)\equiv e^{\tau H }O(ix,-ix)e^{-\tau H }=\Big(\frac{\partial w_{{\rm new}}}{\partial w}\Big)^h\Big(\frac{\partial\overline{w}_{{\rm new}}}{\partial\overline{w}}\Big)^{\overline{h}}O(w_{{\rm new}},\overline{w}_{{\rm new}})\,, \label{Heisenberg}
\end{align}
whose explicit expression depends on the Hamiltonian $H$ that drives the system. This is a powerful result that applies irrespective of the state of the system or the boundary conditions (provided they preserve $H$).

 For example, if $H$ is the CFT Hamiltonian 
\begin{align}
    H_0=\int_0^{2\pi} dx\ T_{00}(x)=L_0+\overline{L}_0-\frac{ c}{12}\label{H0}
\end{align}
with the last term being the Casimir energy, then after (imaginary) time $\tau$, $(w_{{\rm new}},\overline{w}_{{\rm new}})$ are simply $(w,\overline{w})$ in \eqref{Eucledian_w}.

\subsection{The sine-squared deformed time evolution} \label{SSDsetup}
The fundamental ingredient of the deformation we would like to study is the sine-squared deformation (SSD) \cite{Wen:2018vux}, which deforms the CFT Hamiltonian \eqref{H0} with the enveloping function
\begin{align}
    v_1(x)=2 \sin^2\Big(\frac{x}{2}\Big)\label{v(x)}
\end{align}
The resulting deformed Hamiltonian is 
\es{1}{
    H_1&=2\int_0^{2\pi} dx\ \sin ^2\Big(\frac{x}{2}\Big)\,T_{00}(x)\\
    &=L_0-\frac{L_{-1}+L_1}{2}+\overline{L}_0-\frac{\overline{L}_{-1}+\overline{L}_1}{2}-\frac{c}{12}
}
Note that the SSD enveloping function $v_1(x)$ is $\emph{inhomogeneous}$ in space, in particular, it vanishes at $x=0$ (or $2\pi$). From \eqref{1} one observes $[H_0,H_1]\neq 0$, because $[L_0, L_{\pm 1}]=\mp L_{\pm 1}$; however, $H_0$ and $H_1$ share the same ground state $|0\rangle$, as $L_n|0 \rangle =0$  for $n\ge-1$. 

The use of $H_1$ as a Hamiltonian is unconventional. However, $H_1$ can be conformally mapped to a dilation \cite{Okunishi:2016zat,Wen:2018vux,Fan:2019upv}; mapping back to the cylinder, we obtain $(w_{{\rm new}},\overline{w}_{{\rm new}})$ given by the following logarithms of Mobius transformations \cite{Wen:2018vux, Goto:2021sqx} (see Appendix \ref{SSDreview} for details)
\begin{align}
    w_{{\rm new}}&=\log \Big[\frac{\big(1+\frac{i t}{2}\big)e^{ i x}-\frac{i t}{2}}{\frac{i t}{2}e^{i x}+\big(1-\frac{i t}{2}\big)}\Big]\,,\label{wnew}\\
    \overline{w}_{{\rm new}}&=\log \Big[\frac{\big(1+\frac{i t}{2}\big)e^{-i x}-\frac{i t}{2}}{\frac{i t}{2}e^{-i x}+\big(1-\frac{i t}{2}\big)}\Big]\,,\label{wnewbar}
\end{align}
where we have analytically continued to real-time $\tau\to it$. The logarithm is a multi-valued function; we fix the appropriate sheet by demanding continuity in $t$. The terms inside the logarithm are $SL(2)$ transformations of the variables $z=e^{i x}$ and $\overline{z}=e^{-i x}$ on the unit circle.\footnote{$z$ and $\overline{z}$ are coordinates on the $\emph{complex\ plane}$, related to the cylinder via the exponential maps $z=e^{w}$ and $\overline{z}=e^{\overline{w}}$.} The $SL(2)$ transformation has a $\emph{fixed\ point}$, i.e.~a point that stays invariant under the $SL(2)$ transformation at $x=0$ (identified with $x=2\pi$). In analogy with \eqref{Eucledian_w}, we can define \cite{Goto:2021sqx}
\begin{equation}
\label{tnewxnew}
    \begin{aligned}
        w_{{\rm new}}&= i(t_{{\rm new}}+x_{{\rm new}})\,,\\
        \overline{w}_{{\rm new}}&= i(t_{{\rm new}}-x_{{\rm new}})\,.
    \end{aligned}
\end{equation}
$t_{{\rm new}}$ and $x_{{\rm new}}$ are $\emph{real}$  functions of $t$ and $x$.\footnote{As $SL(2)$ transformations preserve unit circles, $w_{{\rm new}}$ and $\overline{w}_{{\rm new}}$ are completely determined by the argument of the logarithm function and are, therefore, purely imaginary. } We will see that the $SL(2)$ transformation \eqref{wnew} and \eqref{wnewbar} determines (almost) all the physical phenomena of interest: in the $(w_{{\rm new}},\overline{w}_{{\rm new}})$ coordinate, the CFT is in equilibrium, and the time-dependent conformal transformations \eqref{wnew} and \eqref{wnewbar} allow us to deduce non-equilibrium phenomena from known equilibrium physics \cite{Wen:2018vux,moosavi2021inhomogeneous}. 

There is an alternative way of thinking about the SSD CFT: the driving by the Hamiltonian \eqref{1} is equivalent to the natural time evolution of the CFT on the curved spacetime \cite{Wen:2018vux,MacCormack:2018rwq,deBoer:2023lrd}:
\es{SSDMetric3}{
ds^2=-v_1(x)^2dt^2+dx^2\,.
}
For the review of the argument, see Appendix \ref{SSDcurvedreview}. Thus we have three equivalent ways of thinking about correlation functions
\es{3desc}{
\le\langle O^{(H_1)}(\tau,x) \dots \ri\rangle&=\Big(\frac{\partial w_{{\rm new}}}{\partial (i x)}\Big)^h\Big(\frac{\partial\overline{w}_{{\rm new}}}{\partial(-ix))}\Big)^{\overline{h}}\dots \le\langle O(w_{{\rm new}},\overline{w}_{{\rm new}})\dots\ri\rangle_{ds^2=dw_{{\rm new}} d\overline{w}_{{\rm new}}}\\[4pt]
&=\le\langle O(\tau,x) \dots \ri\rangle_{ds'^2=v_1(x)^2d\tau^2+dx^2}\,.
}

Quantum quenches with SSD CFT at $t>0$ have been studied for systems prepared in ground \cite{Wen:2018vux}  and thermal \cite{Goto:2021sqx} states.\footnote{Note that in \cite{Wen:2018vux}, $\emph{open}$ boundary conditions has been imposed to set up a quench problem, which makes the ground state of $H_0$ into an excited state of $H_1$. 
 In this work, we  study the non-trivial time evolution of $H_1$ with periodic boundary conditions. Although this setup cannot strictly be regarded as a quench if the initial state is the ground state, the expression for $w_{{\rm new}}$ \eqref{wnew} and $\overline{w}_{{\rm new}}$ \eqref{wnewbar} remain the same.} Some of the most interesting phenomena that have been found are as follows: In \cite{Goto:2021sqx} it was found that the CFT possesses two energy peaks   that move towards the fixed point $x=0$; the entanglement entropy decreases and approaches the $\emph{vacuum}$ entanglement entropy if the subregion $A$ does not include the fixed point $x=0$ (identified with $2\pi$). In the following, we will provide a geometric perspective on these, predominantly in the context of holography.

\subsection{Boundary triangle} \label{SSD: Boundary Horizons}
\begin{figure}[htbp]
\centering
\includegraphics[width=.68\textwidth]{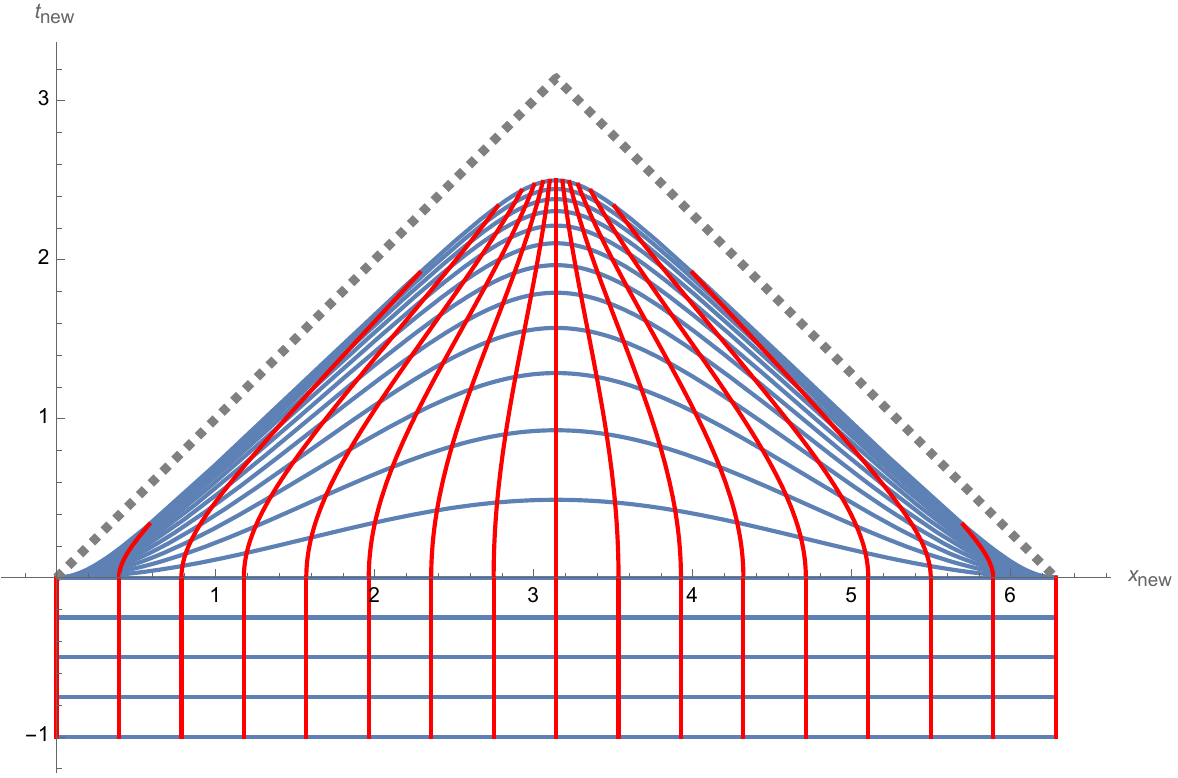}
\qquad
\caption{Constant $t$ lines (blue) and constant-$x$ lines (red) in the $(t_{{\rm new}}, x_{{\rm new}})$-plane. At $t<0$ when the driving Hamiltonian is $H_0$, constant-$t$ lines and constant-$x$ lines are uniformly distributed horizontal and vertical lines, respectively. At $t>0$ when the driving Hamiltonian is suddenly switched to $H_1$, constant-$t$ lines asymptote a triangle-shaped curve \eqref{triangle}, which is plotted as a dashed gray line; constant-$x$ lines converge to the top tip of the triangle, $(\pi,\pi)$. \label{fig:Triangle}} 
\end{figure}

We have seen in Section \ref{SSDsetup} that the non-trivial time evolution of SSD CFTs is completely determined by the coordinate transformations \eqref{wnew} and \eqref{wnewbar}. The spacetime coordinates $t_{{\rm new}}(t,x)$ and $x_{{\rm new}}(t,x)$ then follow from \eqref{tnewxnew}. In Figure \ref{fig:Triangle}, we plot the constant-$t$ lines and constant-$x$ lines on the $(t_{{\rm new}}, x_{{\rm new}})$ plane.\footnote{It is more convenient to plot with the trigonometric form of $t_{{\rm new}}$ \eqref{tnewtrig} and $x_{{\rm new}}$ \eqref{xnewtrig}.}

Figure \ref{fig:Triangle} shows that the action of $H_1$ as Hamiltonian changes the $\emph{foliation}$ of the $(t_{{\rm new}},x_{{\rm new}})$ spacetime in a non-uniform manner.\footnote{This reflects the $\emph{inhomogeneous}$ nature of $H_1$; in particular, the fixed point $x=0$ (which maps to $(t_{{\rm new}},x_{{\rm new}})=(0,0)$) does not time evolve at all. } As time evolves, constant-$t$ lines approach a $\emph{triangle}$-shaped curve 
\begin{equation}
\label{triangle}
    t_{{\rm new}}=
    \left\{
    \begin{aligned}
        &x_{{\rm new}},\ \ 0\leq x_{{\rm new}} < \pi\\
        &2\pi-x_{{\rm new}},\ \ \pi\leq x_{{\rm new}} < 2\pi
    \end{aligned}
    \right.
\end{equation}
instead of moving towards $t_{{\rm new}}\to\infty$ homogeneously as is in the $H_0$ evolution case; constant-$x$ lines converge to the top tip of the triangle, $(\pi,\pi)$. Indeed, from the expression of $t_{{\rm new}}$ \eqref{tnewtrig} and $x_{{\rm new}}$ \eqref{xnewtrig}, one can verify that
\begin{align}
    \lim_{t\to\infty}t_{{\rm new}} = \pi && \lim_{t\to\infty}x_{{\rm new}} = \pi\label{tnewxnewLimit}
\end{align}
regardless of $x$ ($x\neq0,2\pi$). As the triangle \eqref{triangle} separates points that are connected to infinity (outside the triangle) from those that are not (inside the triangle), it can be regarded as a new $\emph{horizon}$ in the $(t_{{\rm new}},x_{{\rm new}})$ plane. Note that the new horizon is ultimately the consequence of the fixed points $x=0,2\pi$ on the circle. If there were no such fixed points, the entire constant $t$ line in $(t_{{\rm new}},x_{{\rm new}})$ would evolve as $t$ grows, and there would be no new horizons.

The SSD Hamiltonian  \eqref{1} evolves the CFT along a conformal Killing vector field, hence the horizon \eqref{triangle} is a $\emph{Killing}$ horizon associated with this vector field. From the expression~\eqref{1}, the conformal Killing vector $k_1$ corresponding to $H_1$ can be written as 
\begin{align}
    k_1&=-z_{{\rm new}}\partial_{z_{{\rm new}}}-\overline{z}_{{\rm new}}\partial_{\overline{z}_{{\rm new}}}+\frac{z_{{\rm new}}^2+1}{2}\partial_{z_{{\rm new}}}+\frac{\overline{z}_{{\rm new}}^2+1}{2}\partial_{\overline{z}_{{\rm new}}}\\
    &= (\cosh  \tau_{{\rm new}}\ \cos x_{{\rm new}}-1)\partial_{\tau_{{\rm new}}}+\sinh \tau_{{\rm new}}\ \sin x_{{\rm new}}\partial_{x_{{\rm new}}}\label{ConfKillingVec}
\end{align}
where $z_{{\rm new}}=e^{\tau_{{\rm new}}+ix_{{\rm new}}}$ and $\overline{z}_{{\rm new}}=e^{\tau_{{\rm new}}-ix_{{\rm new}}}$.\footnote{The conformal Killing vector~\eqref{ConfKillingVec} is a time translation along the $\tau$ direction in the $(\tau_{{\rm new}},x_{{\rm new}})$ coordinate, as can be seen from~\eqref{Heisenberg}. We will only  rarely use $\tau_{{\rm new}}$ as a time coordinate.} One can easily check that $k_1$ obeys the conformal Killing equation
\begin{align}
    \partial_{(\mu}(k_1)_{{\nu})}=\delta_{\mu\nu}\partial_{\rho}(k_1)^{\rho}\,.
\end{align} 
The analytic continuation of the conformal Killing vector \eqref{ConfKillingVec} to Lorentzian signature is
\begin{align}
    k_1=(1-\cos t_{{\rm new}}\ \cos x_{{\rm new}})\partial_{t_{{\rm new}}}+\sin t_{{\rm new}}\ \sin x_{{\rm new}}\partial_{x_{{\rm new}}}\,,\label{ConfKillingVecLor}
\end{align}
where we have removed the overall $i$ so that the vector field implements Lorentzian time evolution. The conformal Killing horizon is located where the conformal Killing vector becomes null, so we compute its norm squared:
\begin{align}
   |k_1|^2=(\cos t_{{\rm new}}-\cos x_{{\rm new}})^2=0\label{3.10} && \quad \implies \quad  t_{{\rm new}}=\pm x_{{\rm new}}+2n\pi\ (n\in\mathbb{Z}) 
\end{align}
As $t_{{\rm new}}\in[0, \pi]$, $t_{{\rm new}}=x_{{\rm new}}$ or $t_{{\rm new}}=-x_{{\rm new}}+2\pi$, which is exactly the expression of the triangle \eqref{triangle}. 
Since \eqref{3.10} has a double zero,
this is an $\emph{extremal}$ conformal Killing horizon, hence it has zero temperature.\footnote{A simple calculation on \eqref{ConfKillingVec} (after analytic continuation) confirms this reasoning, 
\begin{align}
    T=\frac{1}{2\pi}\sqrt{\nabla_{\mu} |k_1|\nabla^{\mu} |k_1|}\Big|_{{\rm hor.}}=\sqrt{-\sin ^2t_{{\rm new}}+\sin ^2x_{{\rm new}}}=0\label{TemperatureBdy}
\end{align}
where $|k_1|$ is given by \eqref{3.10}. } The physical reason for this is that $H_0$ and $H_1$ share the \emph{same} ground state.\footnote{This is in contrast to the Unruh effect, where the Minkowski vacuum is an excited state of and Rindler observer and vice versa.} 

We close this section by comparing the $H_1$ time evolution to those under CFT modular Hamiltonians in the literature, see the comprehensive recent references~\cite{Cardy:2016fqc,deBoer:2021zlm,Das:2024vqe}. The operators $K_\text{full}=s_0L_0+s_1L_1+s_{-1}L_{-1}$ are full/complete modular Hamiltonians associated to an interval $A$ for the vacuum of a 2d CFT on a circle, $K_\text{full}=K_A-K_{\bar A}$. Just like $H_1$, $K_\text{full}$ also annihilates the vacuum.  If we consider the reduced density matrix of the interval, it is a thermal state under modular evolution $\rho\propto e^{-K_A}$. In our study, we do not consider subregions and focus on $H_1$ defined on the entire $S^1$.

\section{Gravity dual for the special quantum quench}\label{Gravity dual of SSD}

\subsection{Bulk geometry}\label{4.1}

The gravity dual of a generic (global) quench involves the changing of boundary conditions for the gravity fields, corresponding to uniform injection of energy into the bulk from the near boundary region. In the simplest scenario, the spacetime reaches equilibrium by settling to a black hole, the dual of thermalisation in the boundary QFT.  

In the special quench we analyse, the Heisenberg evolution on the cylinder can be described by the $t$-depending coordinate transformation $(t_{{\rm new}},x_{{\rm new}})$ \eqref{tnewxnew} expressing non-equilibrium physics from equilibrium results. Correspondingly, the matter stress tensor in the bulk remains zero and the spacetime does not change. What changes is the cutoff surface that the field theory lives on. This among other things leads to the change in energy in the boundary theory.

To construct the holographic dual of SSD CFT, we introduce the emergent bulk radial coordinate $r_{{\rm new}}\in[0,\infty)$. Holographic CFTs in their ground and thermal states are dual to pure AdS$_3$ and BTZ black holes \cite{Banados:1992gq,Banados:1992wn}, respectively. As the CFT in $(t_{{\rm new}},x_{{\rm new}})$ are in equilibrium, the dual bulk metrics in the $(t_{{\rm new}},r_{{\rm new}},x_{{\rm new}})$ coordinate are given by those of the pure AdS$_3$ and BTZ black hole metrics, 
\begin{align}
    &|0\rangle:  && ds^2=-(r_{{\rm new}}^2+1)dt_{{\rm new}}^2+\frac{dr_{{\rm new}}^2}{r_{{\rm new}}^2+1}+r_{{\rm new}}^2dx_{{\rm new}}^2\label{AdSMetric}\\
    &\rho=e^{-\beta H_0}: && ds^2=-(r_{{\rm new}}^2-r_+^2)dt_{{\rm new}}^2+\frac{dr_{{\rm new}}^2}{r_{{\rm new}}^2-r_+^2}+r_{{\rm new}}^2dx_{{\rm new}}^2\label{BTZMetric}
\end{align}
where $r_+$ is the radius of the BTZ black hole, related to the inverse temperature $\beta$ via $r_+=\frac{2\pi}{\beta}$. 

\begin{figure}[htbp]
\centering
\includegraphics[width=.48\textwidth]{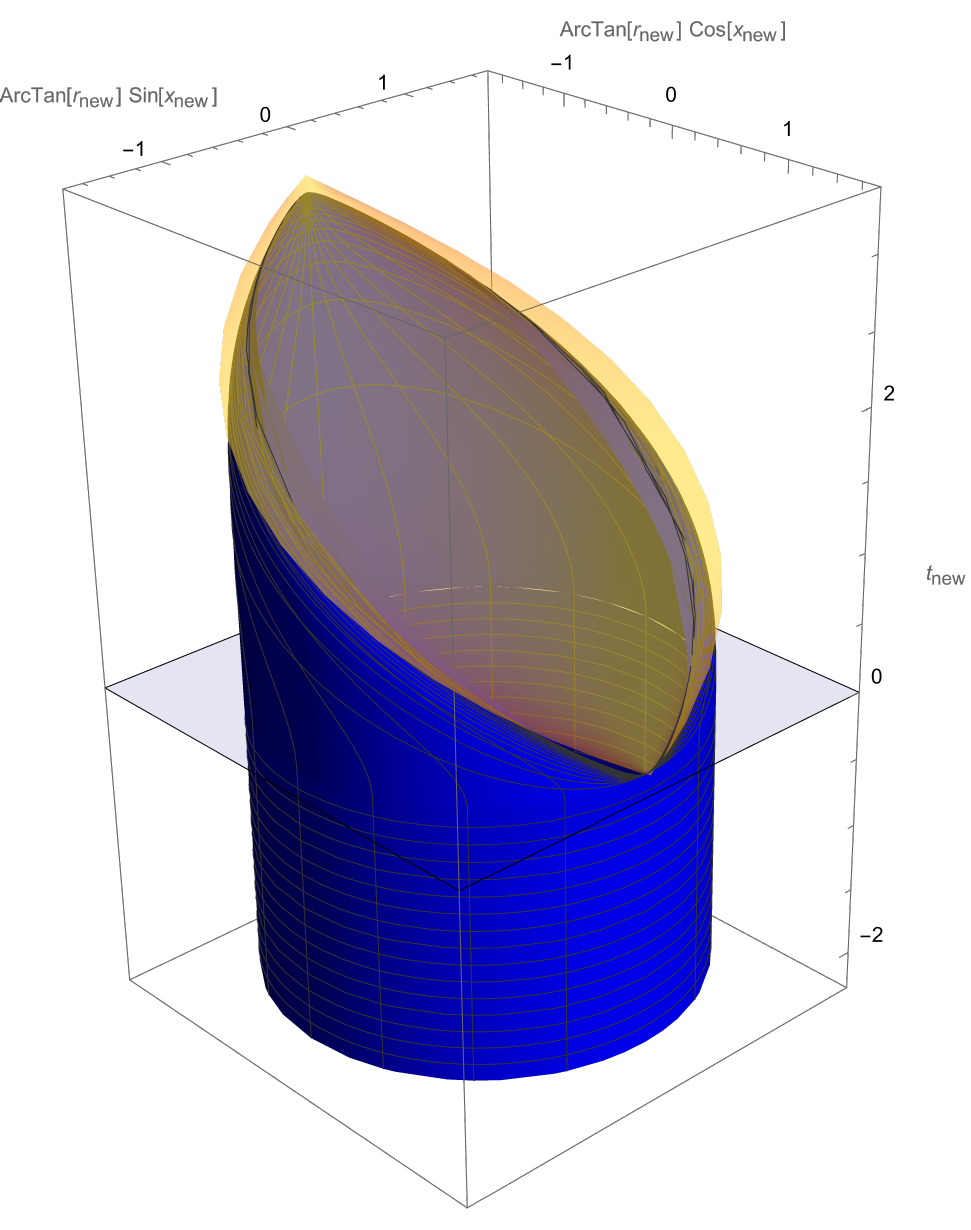}
\includegraphics[width=.37\textwidth]{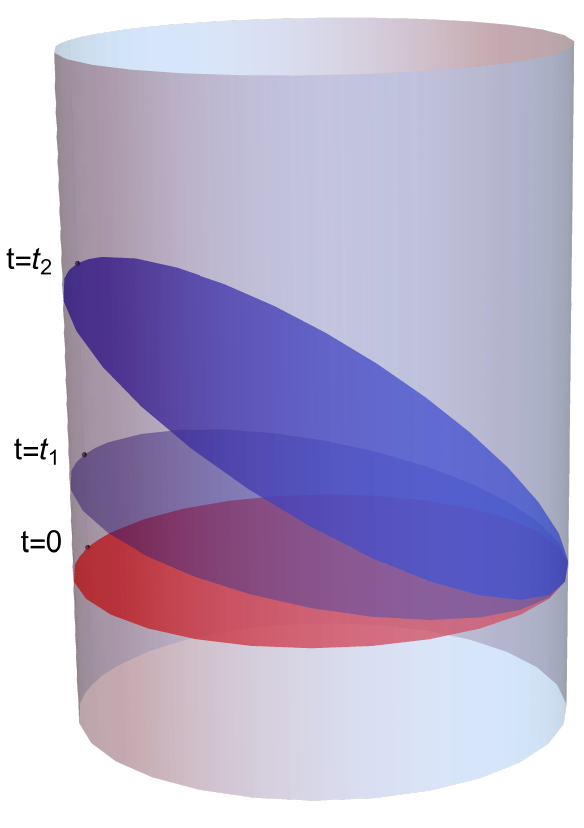}
\qquad
\caption{Left: Plot of the cutoff surface in $(t_{{\rm new}},r_{{\rm new}},x_{{\rm new}})$ spacetime. The light blue surface denotes the $t_{{\rm new}}=t=0$ slice, during which we suddenly switch the Hamiltonian from $H_0$ to $H_1$. At $t<0$ when the driving Hamiltonian is $H_0$, the uniform cutoff $R_\text{new}=\epsilon^{-1}$ is shown in blue. At $t>0$ the driving Hamiltonian becomes $H_1$, and the cutoff surface $R_{{\rm new}}(t_\text{new},x_\text{new})$ is no longer uniform according to \eqref{cutoffpresc}. This significant change leads to the presence of a new horizon, which is depicted above in light yellow. This new horizon is also plotted in Figure \ref{fig:AdSHor} along with its generators. Note that the cutoff surface ends on the horizon (up to finite cutoff corrections). Right: Cartoon of constant $t$ slices in $(t_{{\rm new}},r_{{\rm new}},x_{{\rm new}})$ spacetime in global AdS$_3$. \label{fig:CutoffSurface3d}}
\end{figure}

Besides the movement of the insertion points of operators, there are explicit factors showing up in formulas \eqref{Heisenberg} and \eqref{3desc} multiplying equilibrium correlators. These correspond to taking the holographic cutoff at 
\es{cutoffpresc}{
\epsilon={1\over \sqrt{\frac{\partial w_{{\rm new}}}{\partial w}\frac{\partial\overline{w}_{{\rm new}}}{\partial\overline{w}}}\, R_{{\rm new}}(t_\text{new},x_\text{new})}\,,
}
resulting in a $\emph{nonuniform}$ cutoff surface in the radial coordinate $R_{{\rm new}}(t_\text{new},x_\text{new})$. Note that the formula is applicable for $t>0$, for $t\leq 0$ we have $R_{{\rm new}}=1/\epsilon$. A plot of the cutoff surface \eqref{cutoffpresc} can be found in Figure \ref{fig:CutoffSurface3d}. This formula has appeared in the literature before in~\cite{Goto:2021sqx} based on different arguments; in Appendix~\ref{app:WigglyCutoff} we give a detailed new argument that this  nonuniform cutoff produces the appropriate $\big(\frac{\partial w_{{\rm new}}}{\partial w}\big)^h$ factors in correlators.

We have seen that in the boundary theory a horizon appears, this in the holographic dual leads to a new horizon in the bulk. As constant $t$-slices in the bulk are anchored on the constant $t$-lines on the boundary, this induces a change in the $\emph{foliation}$ of the bulk, see Figure \ref{fig:CutoffSurface3d}. (There is a natural extension of $H_1$ into the bulk in the pure AdS case, but not in BTZ.) As $t\to\infty$, constant-$t$ slices asymptote the bulk null hypersurface anchored on the boundary triangle \eqref{triangle} instead of moving towards $t_{{\rm new}}\to\infty$. As this hypersurface separates bulk points that are connected to the boundary spacetime explored by the CFT from those that are not, it is a new $\emph{horizon}$ in the bulk. 

An alternative approach is provided by the CFT in curved spacetime perspective: we set the boundary metric to agree with \eqref{SSDMetric3}, and use Fefferman-Graham coordinates \cite{Skenderis:1999nb,deHaro:2000vlm} to build a coordinate system adapted to the boundary spacetime. The uniform cutoff surface in the Fefferman-Graham radial coordinate will coincide with the cutoff surface found from \eqref{cutoffpresc}. In 3d pure gravity, the Fefferman-Graham expansion is known to terminate. By solving the Einstein equations order by order in $r$ we obtain the metric~\cite{Skenderis:1999nb}:
\es{FGconcrete}{
ds^2={dr^2\ov r^2}+r^2\big(-v_1(x)^2dt^2+dx^2\big)+g^{(1)}_{ab}(x^c) \,dx^a dx^b+{g^{(2)}_{ab}(x^c) \,dx^a dx^b\ov r^2}\,,
}
where $g^{(j)}_{ab}(x^c)$ is given in Appendix~\ref{app:FGBanados} for the most general solution. There we also give the explicit form of the metric for the case of the vacuum and thermal states. Of course, \eqref{FGconcrete} is just the reslicing of the AdS$_3$ \eqref{AdSMetric} and BTZ \eqref{BTZMetric} spacetimes. (The  reslicing of AdS$_3$ was written down in \cite{MacCormack:2018rwq} before studying the vacuum physics of the SSD CFT.) One advantage of these coordinates is that now the cutoff is at $R=1/\epsilon$ instead of the nonuniform cutoff in $r_{{\rm new}}$ coordinates. 

Note that as discussed in Appendix~\ref{SSDcurvedreview}, the curved spacetime \eqref{SSDMetric3} is obtained by a Weyl transformation from the flat spacetime $ds^2=dw_{{\rm new}} d\overline{w}_{{\rm new}}$. It has been understood from the early days of AdS/CFT~\cite{Witten:1998qj,Henningson:1998gx} that changing the conformal factor of the boundary metric corresponds to choosing a different cutoff prescription. 

\subsection{New horizon in AdS$_3$}\label{New Horizon on AdS_3}

\begin{figure}[htbp]
\centering
\includegraphics[width=.55\textwidth]{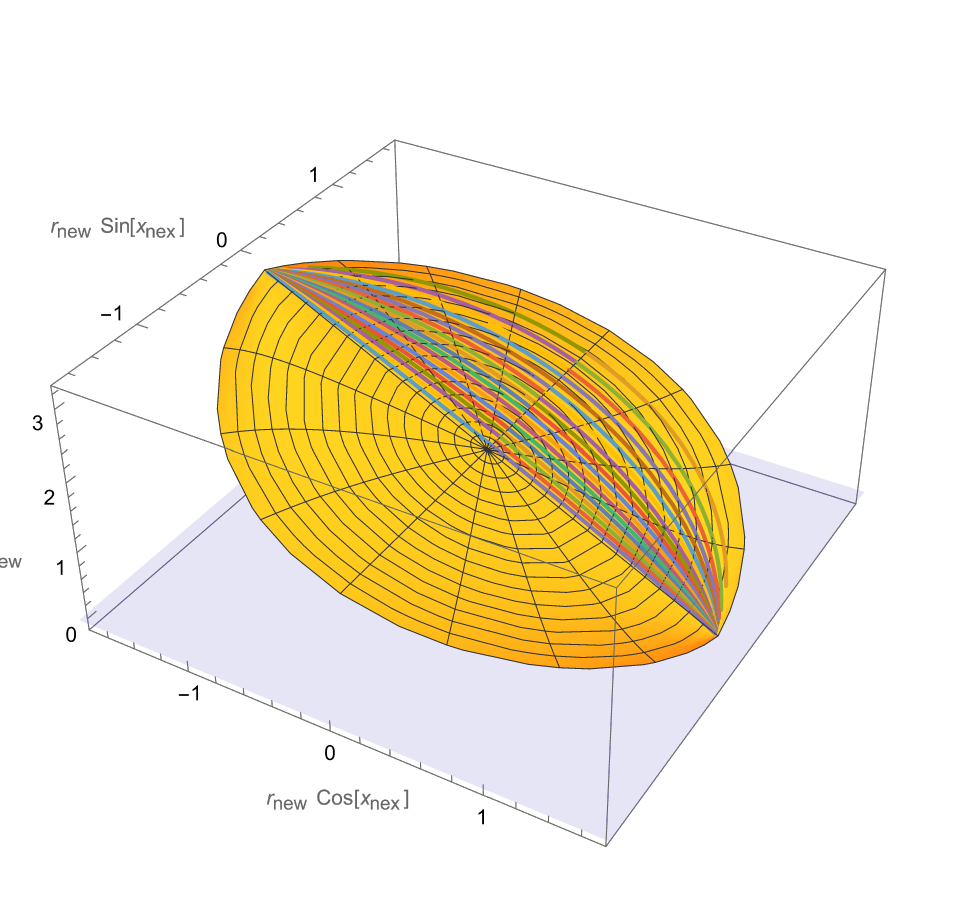}
\qquad
\caption{A plot of the new Killing horizon \eqref{AdSKilling2} in global AdS$_3$. The coloured lines are null geodesics emanating from the top tip of the triangle $(\pi, \pi)$ that generate the hypersurface \eqref{AdSKilling2}; for visualization, we only plotted them on half of the horizon. The light blue surface is the quench surface $t_{{\rm new}}=t=0$. \label{fig:AdSHor}}.
\end{figure}

The isometries of AdS$_3$ form two copies of $SL(2)$ algebras \cite{Brown:1986nw}. The Killing vectors $\zeta_{i}$ and $\overline{\zeta_{i}}$ ($i=-1,0,1$) of AdS$_3$ are given in Appendix \ref{AsymptoticSymmetries}. From these Killing vectors, we can find a linear combination that asymptotes to the conformal Killing vector $k_1^{\mu}$ \eqref{ConfKillingVec}. This provides the natural extension of the $H_1$ evolution into the bulk:
\es{BulkKillingVec}{
K_1&=\zeta_{0}+\overline{\zeta_{0}}-\frac{\zeta_{1}+\overline{\zeta_{1}}}{2}-\frac{\zeta_{-1}+\overline{\zeta_{-1}}}{2}\\
&=\Big(1-\frac{r_{{\rm new}}}{\sqrt{r_{{\rm new}}^2+1}}\cos t_{{\rm new}}\ \cos x_{{\rm new}}\Big)\partial_{t_{{\rm new}}}-\sqrt{r_{{\rm new}}^2+1}\ \sin t_{{\rm new}}\ \cos x_{{\rm new}}\partial_{r_{{\rm new}}}\\
&+\frac{\sqrt{r_{{\rm new}}^2+1}}{r_{{\rm new}}}\sin t_{{\rm new}}\ \sin x_{{\rm new}}\partial_{x_{{\rm new}}}
}
In the near-boundary $r_{{\rm new}}\to\infty$ limit, the components of $K_1^{\mu}$ \eqref{BulkKillingVec} along the boundary indeed go to the boundary conformal Killing vector $k_1^{\mu}$ \eqref{ConfKillingVec}, 
\begin{align}
    \lim_{r_{{\rm new}}\to\infty}K_1^{\mu}=(1-\cos t_{{\rm new}}\ \cos x_{{\rm new}}), \sin t_{{\rm new}}\ \sin x_{{\rm new}}\big)=k_1^{\mu}
\end{align}
The Killing horizon is located where a Killing vector becomes null. From \eqref{BulkKillingVec}, we have for the norm square:
\begin{align}
  |K_1|^2&=\big(\sqrt{r^2+1}\cos t_{{\rm new}}-r\cos x_{{\rm new}}\big)^2=0\label{VNorm}\\
    &\Rightarrow \quad t_{{\rm new}}(r_{{\rm new}},x_{{\rm new}})=\arccos\Big[\frac{r}{\sqrt{r_{{\rm new}}^2+1}}\cos x_{{\rm new}}\Big]\,.\label{AdSKilling}
\end{align}
\eqref{AdSKilling} is the new horizon in pure AdS$_3$. Since $|K_1|^2$ has a double zero, this is an extremal Killing horizon with zero associated temperature.\footnote{A direct calculation shows that this is indeed the case: 
\begin{align}
    T=\frac{1}{2\pi}\sqrt{\nabla_{\mu} |K_1|\nabla^{\mu} |K_1|}\Big|_{{\rm hor.}}=\sqrt{r_{{\rm new}}^2+1}\cos t_{{\rm new}}-r\cos x_{{\rm new}}=0\label{temperature}
\end{align}
where $|K_1|$ given by \eqref{VNorm}. } Just like on the CFT side, the intuition behind this result is that $H_0$ and $H_1$ share the same vacuum state. A plot of the Killing horizon \eqref{AdSKilling} along with its null generators can be found in Figure \ref{fig:AdSHor}. 

We can find the horizon \eqref{AdSKilling} in an alternative manner, which will be a useful method in the following.
To find the horizon as a null hypersurface, we need to investigate the null geodesics emanating from a point $(t_{{\rm new}},x_{{\rm new}})=(y,y)$ on the boundary triangle. The congruence of null geodesics that maximise $t_{{\rm new}}$ in the bulk then generates the null hypersurface. Here, for simplicity, we study only half of the triangle, i.e.~$y\in[0, \pi]$; the result of the other half follows from reflection. 

We will start with the metric \eqref{AdSMetric} in the $(t_{{\rm new}},r_{{\rm new}},x_{{\rm new}})$ coordinate. The null geodesics anchored on a point $(t_{{\rm new}},x_{{\rm new}})=(y,y)$ on the boundary triangle  (see Appendix \ref{subsubsec:PureAdS} for details) is 
\begin{align}
    t_{{\rm new}}(r_{{\rm new}},x_{{\rm new}},y)={\rm arctan}\frac{-r_{{\rm new}}\ \cot (x_{{\rm new}}-y)}{\sqrt{r_{{\rm new}}^2+\csc ^2(x_{{\rm new}}-y)}}+y-\frac{\pi}{2}\label{AdSNullGeod}
\end{align}
where $y\in[0, \pi]$. \eqref{AdSNullGeod} increases monotonically with $y$ (see \eqref{AdSMono}). Hence, the maximal value of $t$ is at $y=\pi$, which is the top tip of the boundary triangle. After some work detailed in Appendix \ref{subsubsec:PureAdS}, one arrives at 
\begin{align}
    t_{{\rm new}}(r_{{\rm new}},x_{{\rm new}})=\arccos\Big[\frac{r_{{\rm new}}}{\sqrt{r_{{\rm new}}^2+1}}\cos x_{{\rm new}}\Big]\label{AdSKilling2}
\end{align}
agreeing with \eqref{AdSKilling} obtained by the Killing vector method. 

\subsection{New horizon on BTZ}\label{New Horizon on BTZ}
The $SL(2)$ isometry of AdS$_3$ is broken to $\mathbb{R}\times U(1)$ in BTZ geometry \cite{Keski-Vakkuri:1998gmz}. Hence, the new horizon is no longer a Killing horizon, nor can we associate temperature with it. To find the explicit expression of the horizon, one needs to investigate the null geodesic equation using the second method of the previous section. The null geodesics anchored at a point on the boundary
triangle $(t_{{\rm new}},x_{{\rm new}})=(y,y)$ (see Appendix \ref{subsubsec:BTZ} for details) is 
\begin{align}
    t_{{\rm new}}(r_{{\rm new}},x_{{\rm new}},y)=\frac{1}{2r_+}\log \frac{r_{{\rm new}}\sqrt{1+r_+^2\sinh ^2(x_{{\rm new}}-y)}-r_+\sqrt{1+r_{{\rm new}}^2\sinh ^2(x_{{\rm new}}-y)}}{r_{{\rm new}}\sqrt{1+r_+^2\sinh ^2(x_{{\rm new}}-y)}+r_+\sqrt{1+r_{{\rm new}}^2\sinh ^2(x_{{\rm new}}-y)}}+y \label{BTZNullGeod1}
\end{align}
where $y\in[0, \pi]$. \eqref{BTZNullGeod1} increases monotonically with $y$ (see \eqref{BTZMono}). Hence, the maximal value of $t$ is at $y=\pi$, which is the top tip of the boundary triangle. After some work detailed in Appendix \ref{subsubsec:BTZ}, one arrives at 
\begin{align}
    t_{{\rm new}}(r,x)=\frac{1}{2r_+}\log \frac{r_{{\rm new}}\sqrt{1+r_+^2\sinh ^2(x_{{\rm new}}-\pi)}-r_+\sqrt{1+r_{{\rm new}}^2\sinh ^2(x_{{\rm new}}-\pi)}}{r_{{\rm new}}\sqrt{1+r_+^2\sinh ^2(x_{{\rm new}}-\pi)}+r_+\sqrt{1+r_{{\rm new}}^2\sinh ^2(x_{{\rm new}}-\pi)}}+\pi\label{BTZNullGeod2}
\end{align}
\begin{figure}[htbp]
\centering
\includegraphics[width=.55\textwidth]{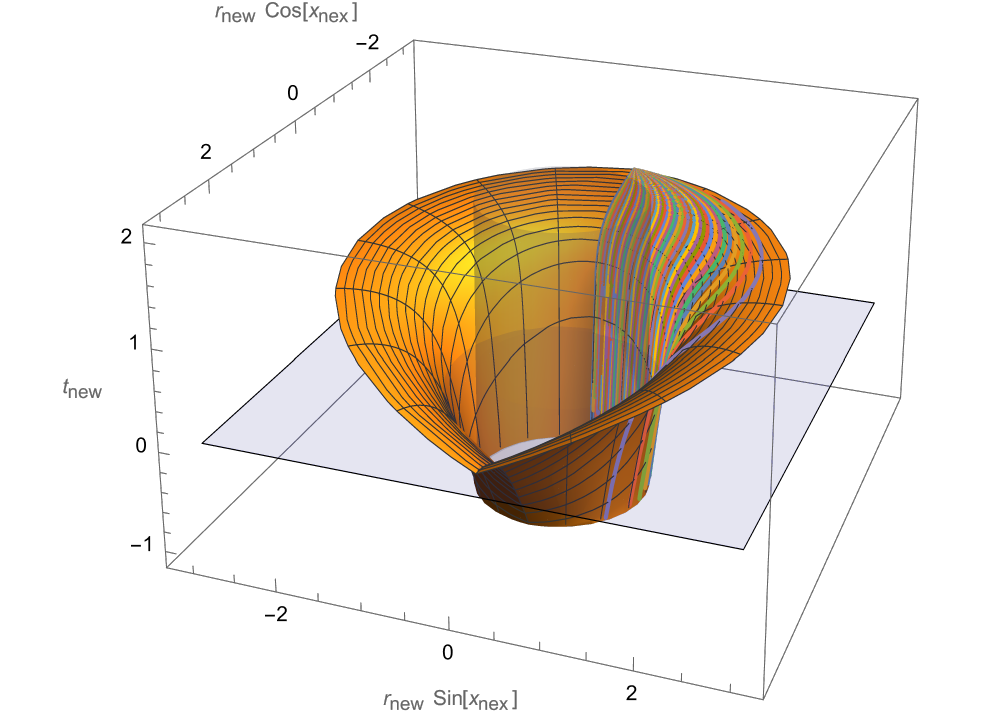}
\qquad
\caption{Plot of the new horizon \eqref{BTZNullGeod2} in the BTZ geometry. The coloured lines are null geodesics emanating from the top tip of the triangle $(\pi, \pi)$ that generate the hypersurface \eqref{BTZNullGeod2}; for visualization, we only plotted them on half of the horizon. The light blue surface is the quench surface $t_{{\rm new}}=t=0$. The BTZ black hole horizon is set to $r_+=1$ and plotted in light gray. The new horizon \eqref{BTZNullGeod2} is larger than the BTZ black hole horizon. \label{fig:BTZHor}}. 
\end{figure}
 A plot of the new horizon \eqref{BTZNullGeod2} along with its null generators can be found in Figure \ref{fig:BTZHor}. 

\begin{figure}[htbp]
\centering
\includegraphics[width=.45\textwidth]{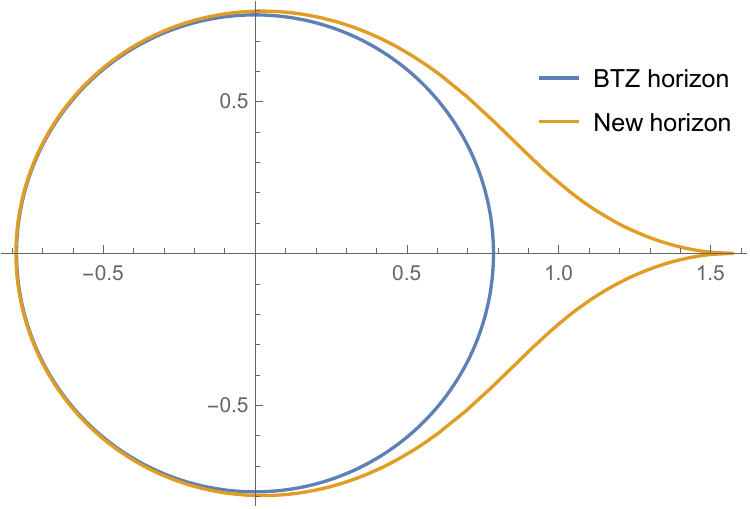}
\qquad
\includegraphics[width=.37\textwidth]{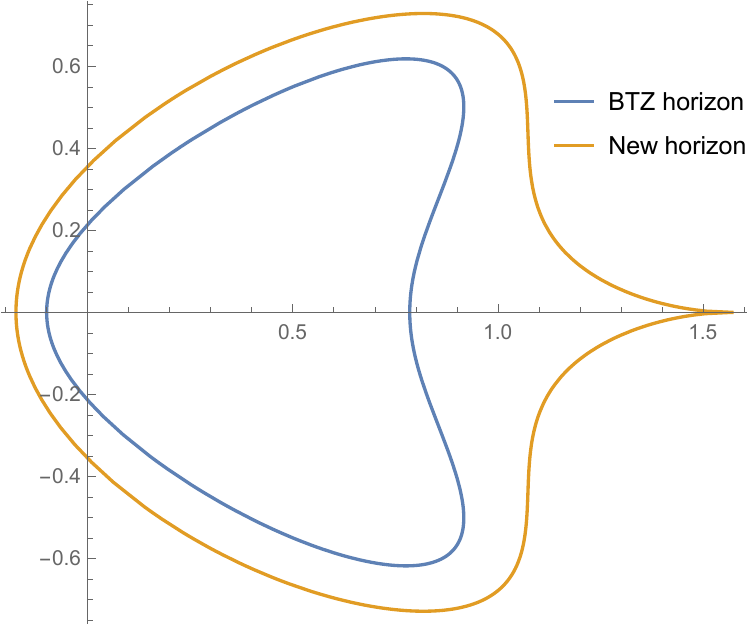}
\caption{Comparison of the new horizon \eqref{BTZNullGeod2} and the BTZ black hole horizon in the coordinates employed by~\cite{Goto:2021sqx}, at $t=0$ (left) and $t=3$ (right), respectively. The blue line denotes the BTZ horizon, whereas the orange line 
 represents the new horizon \eqref{BTZNullGeod2}. \label{fig:HorCompare}}
\end{figure}

For comparison with ref.~\cite{Goto:2021sqx},
we use the coordinate system they were working in to plot time slices of both the new horizon \eqref{BTZNullGeod2} and the BTZ horizon.\footnote{The Fefferman-Graham coordinates discussed around \eqref{FGconcrete} and Appendix~\ref{app:FGBanados} generically break down before encountering horizons, as was recently discussed in~\cite{Abajian:2023bqv}. Hence we cannot use them for addressing this question.} (See Appendix \ref{ComparisonRyu} for details.) From Figure~\ref{fig:BTZHor} it is clear that the new horizon will be slightly $\emph{larger}$ than the BTZ black hole horizon, and indeed this is what we find in Figure \ref{fig:HorCompare}. Note also that the new horizon \eqref{BTZNullGeod2} touches the boundary at the fixed point $x=0$.

The horizon \eqref{BTZNullGeod2} in Figure \ref{fig:HorCompare} develops two peaks that move towards the fixed point $x=0$, as is observed for the BTZ horizon in \cite{Goto:2021sqx}. It was also pointed out in \cite{Goto:2021sqx} that the horizon deformation mimics the energy peaks of the boundary stress tensors. We note that the two peaks of the horizons (both the BTZ horizon and the new horizon) originate from the Weyl factor $\frac{\partial w_{{\rm new}}}{\partial w}\frac{\partial\overline{w}_{{\rm new}}}{\partial\overline{w}}$ in \eqref{r'}; in \cite{Goto:2021sqx}, the peaks of the boundary stress tensors also originate from these prefactors $\frac{\partial w_{{\rm new}}}{\partial w}$ and $\frac{\partial\overline{w}_{{\rm new}}}{\partial\overline{w}}$ in $\langle T(w)\rangle$ and $\langle \overline{T}(\overline{w})\rangle$, respectively. This then explains the qualitative\footnote{As the metric \eqref{r'} is not in Fefferman-Graham gauge \cite{Skenderis:1999nb,deHaro:2000vlm}, it is hard to draw a quantitative connection.} correlations between horizon deformations and energy peaks were first observed in~\cite{Goto:2021sqx}. 

\subsection{Comments on bulk reconstruction}\label{bulk reconstructions}

Horizons make information inaccessible.
The presence of the horizons \eqref{AdSKilling2} and \eqref{BTZNullGeod2} in Figure \ref{fig:AdSHor} and \ref{fig:BTZHor} makes us wonder if an information puzzle can arise in our setup: e.g.~a bulk excitation prepared on the $t=0$ would fall behind the horizon and become inaccessible leading to tension with unitary time evolution. Equivalently, one asks how to reconstruct local bulk operators (dressed to make them diffeomorphism invariant) inside or outside the new horizons, respectively. 

For bulk fields $\emph{inside}$ the causal wedge, i.e.~in the bulk region enclosed by the null hypersurfaces in Figure \ref{fig:AdSHor} or \ref{fig:BTZHor} and the quench surface at $t=0$, we can recover the information using HKLL-like reconstructions \cite{Banks:1998dd,Hamilton:2006az,Witten:2023qsv}. In our case, the $t_\text{new}>0$ part of this subregion is the triangle. (The $t_\text{new}<0$ part is a time strip of width $\pi/2$ for pure AdS and infinite for BTZ.)   The reconstruction formula is given by   
\begin{align}
    \phi(X)=\int_{{\rm triangle + strip}}dx\, K(X|x) O(x)
\end{align}
where $X=(t_{{\rm new}},r_{{\rm new}},x_{{\rm new}})$ and $x=(t_{{\rm new}},x_{{\rm new}})$ stand for the coordinates on the bulk and boundary, respectively. $K(X|x)$ is a smearing function adapted to the subregion. 

For bulk operators $\emph{outside}$ the causal wedge, i.e.~behind the horizon \eqref{AdSKilling2} and \eqref{BTZNullGeod2}, the aforementioned HKLL-like reconstruction does not apply anymore. However, information falling through the horizon \eqref{AdSKilling2} or \eqref{BTZNullGeod2} (but not through the BTZ black hole horizon in the case with \eqref{BTZNullGeod2}) remains reconstructable, by considering $\emph{modular}$ reconstruction \cite{Jafferis:2015del,Dong:2016eik}. In our case, the modular Hamiltonian is simply $H_0$ up to a coefficient $\frac{\beta}{2\pi}$,\footnote{In the vacuum since the density matrix is a projector, the modular Hamiltonian is not well defined, so we need to regulate the vacuum with a small temperature.} so the modular evolution is just $H_0$ time evolution with the modular parameter given by $t_{{\rm new}}=t$.\footnote{Note that when the Hamiltonian is $H_0$, we have $t_{{\rm new}}=t$; $t_{{\rm new}}$ and $t$ are different only when the driving Hamiltonian is $H_1$. } We would like to then consider the $\emph{modular\ flow}$ of local operators $O$ on the boundary, $O(x,t)=U^{-1}(t)O(x,0)U(t)$, where $U(t)=e^{-i \beta H_0 t/ 2\pi}$. Modular reconstruction then states that $\emph{modular\ flows\ equal\ bulk\ modular\ flows}$ \cite{Jafferis:2015del}. This conclusion allows us to recover the information of all bulk operators within the $\emph{entanglement\ wedges}$, which in our case is the entire spacetime for the ground state case and all the spacetime regions outside the BTZ black hole horizon for the thermal state case. In each case, the entanglement wedge is larger than the causal wedge and, importantly, includes the spacetime region that is behind the horizon \eqref{AdSKilling2} or \eqref{BTZNullGeod2} (but outside the BTZ horizon in the thermal state case). 

\section{Quenches governed by other $SL(2)$ Virasoro subalgebras}\label{SSDq}

It is straightforward to generalise the SSD story, by replacing $L_{\pm1}$ in the expression of $H_1$ in \eqref{1} with $L_{\pm q}$ with $q>1$.
This corresponds to the envelope function \cite{Fan:2019upv}
\begin{align}
    v_q(x)=2\sin ^2\Big(q\cdot \frac{x}{2}\Big) 
\end{align}
 and the resulting deformed Hamiltonian $H_q$ is given by
\begin{align}
    H_q&=2\int_0^{2\pi} dx\ \sin ^2\Big(q\cdot \frac{x}{2}\Big)\,T_{00}(x)\\
    &=L_0-\frac{L_{-q}+L_q}{2}+\overline{L}_0-\frac{\overline{L}_{-q}+\overline{L_q}}{2}-q\cdot\frac{c}{12}
\end{align}
In order for $\{L_0, L_q, L_{-q}\}$ to form the same algebra as $\{L_0, L_1, L_{-1}\}$, we have to $\emph{shift}$ the Virasoro generators $\{L_0, L_q, L_{-q}\}$,\footnote{There are subtleties regarding the difference between $L_q$ ($q>1$) and the global conformal charges $\{L_0, L_1, L_{-1}\}$, as the former generate transformations that are not globally well defined and are associated to conformal Killing vectors with poles. But these poles appear at $0$ or $\infty$ on the complex $z$-plane or equivalently $\pm \infty$ on the cylinder, where we will not perform calculations. Therefore, our calculation is unaffected by this subtlety. }
\begin{align}
    \mathcal{L}_0=\frac{1}{q}\big(L_0-\frac{c}{12}q(q^2-1)\big) && \mathcal{L}_{\pm q}=\frac{1}{q}L_{\pm q}
\end{align}
Note that such shifts also define a new vacuum $|0\rangle_q$ for each $H_q$. $H_q$ is now given by 
\begin{align}
    H_q=q\Big(\mathcal{L}_0-\frac{\mathcal{L}_{-q}+\mathcal{L}_q}{2}+\overline{\mathcal{L}}_0-\frac{\overline{\mathcal{L}}_{-q}+\overline{\mathcal{L}}_q}{2}-\frac{c}{12}\Big)\label{Hq}
\end{align}
which follows the same algebraic commutation rules as that of $H_1$. The relevant coordinate transformations then take the same form as \eqref{wnew} and \eqref{wnewbar}, with the circumference $2\pi$ replaced by $2\pi/q$. The system can then be regarded as $q$ copies of identical theories driven by $H_1$, each with a shrunk length scale $2\pi/q$ living in a subregion on the cylinder (see Appendix \ref{SSDGeneralq} for more details).  
Hence the constant $t$ lines approach a $q$-fold triangle, i.e.~$q$ triangles, each living in a subregion $\big[(p-1)\frac{2\pi}{q},p\frac{2\pi}{q}\big]\ (p=1,2,...,q)$. The boundary conformal Killing vector is 
\begin{align}
    k_q^{\mu}=\Big(\cosh  (q\tau_{{\rm new}})\ \cos (qx_{{\rm new}})-1, \sinh (q\tau_{{\rm new}})\ \sin (qx_{{\rm new}})\Big)\label{qTriangles}
\end{align}
The boundary conformal Killing horizon is then the aforementioned $q$-fold triangle. 

\begin{figure}[htbp]
\centering
\includegraphics[width=.55\textwidth]{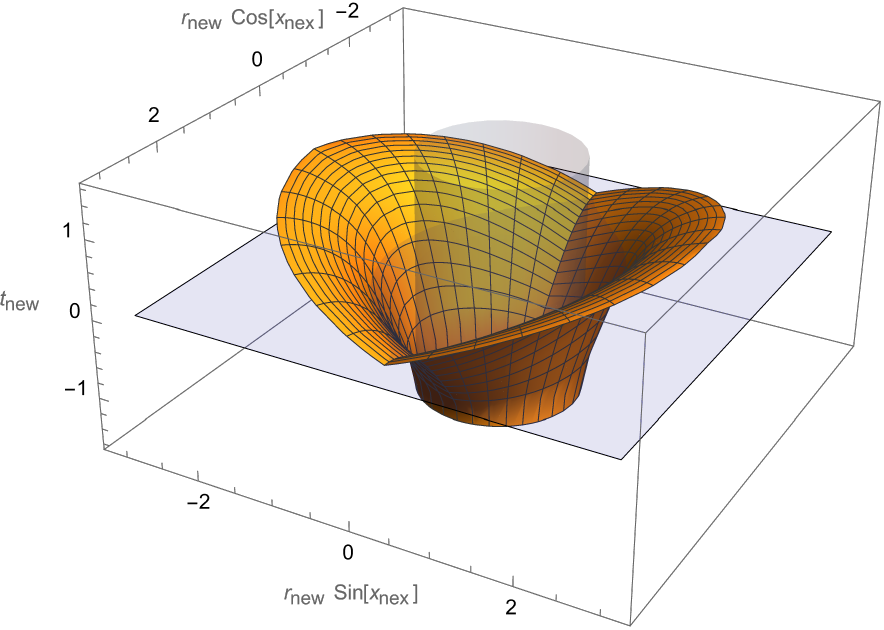}
\qquad
\caption{A plot of the new horizon in BTZ geometry with $H_2$ as driving Hamiltonian. The light blue surface is the quench surface $t_{{\rm new}}=t=0$. The BTZ black hole horizon is set to $r_+=1$ and plotted in light gray. The new horizon is larger than the BTZ black hole horizon. \label{fig:BTZHorq}}
\end{figure}

Generalization of the new horizon with $H_q\ (q>1)$ as Hamiltonian is straightforward. From equation \eqref{qTriangles}, the boundary horizons when $H_q$ drives are $q$-triangles each in a subregion; according to the same maximisation procedure of null geodesics in \eqref{AdSNullGeod} and \eqref{BTZNullGeod1}, the new horizon then consists of $q$-copies of \eqref{AdSKilling2} or \eqref{BTZNullGeod2} when the CFT is in the ground and thermal state, respectively, each anchoring on one of the $q$ triangles on the boundary. An example for $q=2$ in the thermal state is plotted in Figure \ref{fig:BTZHorq}. Note that the bulk horizons for $H_q\ (q>1)$ in the ground state are no longer Killing horizons. (The naive generalizations of \eqref{BulkKillingVec} to $q>1$ do not obey the Killing equation \eqref{KillingEqn}. This is as anticipated as the bulk does not have an $SL(2)$ isometry for each $q$.)

\section{Floquet CFT}\label{Floquetsetup}

\subsection{Floquet CFT at stroboscopic times}
\begin{figure}[htbp]
\centering
\includegraphics[width=.4\textwidth]{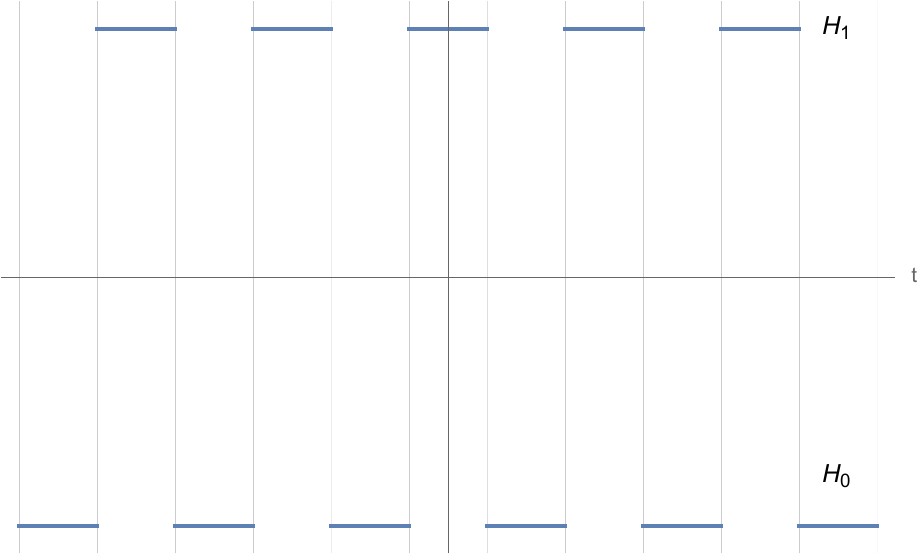}
\qquad
\caption{Floquet CFT. In this example, we have $T_0=T_1$, and choose $t=0$ in the middle of a $H_0$-driving region. \label{fig:FloquetFig}}
\end{figure}

Floquet CFT is defined by $\emph{periodically}$ driving the CFT$_2$ with $H_0$ and $H_1$, for times $T_0$ and $T_1$, respectively. See Figure \ref{fig:FloquetFig}.

We will again work in the Heisenberg picture \eqref{Heisenberg} and study the coordinate transformation $w_{{\rm new}}$ and $\overline{w}_{{\rm new}}$. Previous works \cite{Wen:2018agb,Fan:2019upv,Wen:2020wee,Han:2020kwp,Fan:2020orx,Wen:2022pyj} have been focusing on Floquet CFT at $\emph{stroboscopic}$ time  $t=n(T_0+T_1)$ $(n\in\mathbb{Z}^+)$, i.e.~at the end of each driving cycle.\footnote{Again, in \cite{Wen:2018agb,Fan:2019upv}, $\emph{open}$ boundary conditions are adopted for this set up.} The time evolution operator in \eqref{Heisenberg} (after analytic continuation) is of the form 
\begin{align}
    ...\underbrace{e^{-i H_0 T_0}e^{-i H_1 T_1}\overbrace{e^{-i H_0 T_0}e^{-i H_1 T_1}}^{{\rm 1\ cycle}}e^{-i H_0 T_0}e^{-i H_1 T_1}...e^{-i H_0 T_0}e^{-i H_1 T_1}}_{n\ {\rm times}}...\label{FloquetTimeEvoOp}
\end{align}
We can start at any time during the driving. For technical simplicity, we adopt the setup in \cite{Lapierre:2019rwj} and start in the middle of a $H_0$-driving region, see Figure \ref{fig:FloquetFig}. Then the system possesses $\emph{time-reversal}$ symmetry, i.e.~it is invariant under $t\to -t$. The new coordinates $w_{{\rm new}}$ and $\overline{w}_{{\rm new}}$ on the cylinder at stroboscopic times are given by \cite{Wen:2018agb,Fan:2019upv,Wen:2020wee,Han:2020kwp,Fan:2020orx,Wen:2022pyj, Lapierre:2019rwj} (see Appendix \ref{FloquetReview} for more details)
\begin{align}
    w_n&=w_{{\rm new}}|_{t=n(T_0+T_1)}=\log \Big(\frac{A_n e^{i x}+B_n}{C_ne^{i x}+D_n}\Big)\label{wn}\\
    \overline{w}_n&=\overline{w}_{{\rm new}}|_{t=n(T_0+T_1)}=\log \Big(\frac{A_n e^{-i x}+B_n}{C_ne^{-i x}+D_n}\Big)\label{wnbar}
\end{align}
and
\begin{align}
     t_n=t_{{\rm new}}|_{t=n(T_0+T_1)}= \frac{1}{2i}(w_n+\overline{w}_n) &&
    x_n=x_{{\rm new}}|_{t=n(T_0+T_1)}= \frac{1}{2i}(w_n-\overline{w}_n)\label{tnxnFloquet}
\end{align}
where 
\begin{align}
    A_n=\gamma_1-\eta^n\gamma_2 && B_n= -(1-\eta^n) && C_n=1-\eta^n && D_n=-(\gamma_2-\eta^n\gamma_1)\label{2.19}
\end{align}
and $\gamma_1$, $\gamma_2$ and $\eta$ are determined by the two parameters $T_0$, and $T_1$ of the system, and are given in Appendix \ref{FloquetReview}. The terms inside the logarithm are again $SL(2)$ transformations of the variables $z=e^{i x}$ and $\overline{z}=e^{-i x}$ on the unit circle. Here, $\gamma_1$ and $\gamma_2$ are two $\emph{fixed\ points}$ of the $SL(2)$ transformation, i.e.~they are $\emph{invariant}$ under the $SL(2)$ transformation. In the time-reversal symmetric case \cite{Lapierre:2019rwj}, $\gamma_1\gamma_2=1$. Note that as we are only considering stroboscopic, i.e.~discrete time, we lose the ability to use continuity to choose the appropriate sheet of the logarithm. We simply use the principal sheet, which results in 
$t_n$ and $x_n$ not containing full information about the spacetime. We will remedy this situation in Section \ref{FloquetGenT}.

Depending on the values of $T_0$, and $T_1$, the system ends up in different phases, which are known as the heating phase, non-heating phase, and phase transition, respectively \cite{Wen:2018agb,Fan:2019upv,Wen:2020wee,Han:2020kwp,Fan:2020orx,Wen:2022pyj}. Their properties are summarised as follows: 
\begin{itemize}
    \item In the $\emph{heating}$ phase, $\eta$ is $\emph{real}$ and $0<\eta<1$. $\gamma_1$ and $\gamma_2$ are on the $\emph{unit\ circle}$, $|\gamma_1|=|\gamma_2|=1$. Together with $\gamma_1\gamma_2=1$, $\gamma_2$ is then the $\emph{complex\ conjugate}$ of $\gamma_1$: $\gamma_2^*=\gamma_1$. They are known as the $\emph{stable}$ and $\emph{unstable}$ fixed point, respectively: for $z\neq \gamma_2$, $w_n$ flows to $\log \gamma_1$ at late time; whereas for $z= \gamma_2$, $w_n$ stays at $\log \gamma_2$. The same works for $\overline{w}_n$.\footnote{Namely, for $\overline{z}\neq \gamma_2$, $\overline{w}_n$ flows to $\log \gamma_1$; while for $\overline{z}= \gamma_2$, $\overline{w}_n$ stays at $\log \gamma_2$. Since at $t=0$ the coordinates $z$ and $\bar z$ are complex conjugates of each other equal to $e^{i x}$ and $e^{-i x}$ respectively, the only way the system can end up at an unstable fixed point is if $z=e^{i x}=\gamma_2$, but then $\bar z=\gamma_1$ (the stable fixed point for $z$), since $\gamma_1$ and $\gamma_2$ are conjugate to each other, or vice versa.}

    \item In the $\emph{non-heating}$ phase, $\eta$ is a $\emph{phase}$, $|\eta|=1$. $\gamma_1$ and $\gamma_2$ are both $\emph{real}$ and, in the time-reversal case, inverses of each other. They stay on different sides of the unit circle. $w_n$ and $\overline{w}_n$ show oscillatory behavior. 

    \item At the $\emph{phase\ transition}$, \eqref{2.19} does not apply anymore, and $A_n, B_n, C_n, D_n$ are instead given by \eqref{ABCDphaseTransition}. In this case, the two fixed points $\gamma_1$ and $\gamma_2$ merge into one fixed point $\gamma$, which is on the unit circle, $|\gamma|=1$; in the time-reversal case, we also have $\gamma=1$ (see \eqref{gammaPhaseTransition}). $w_n$ flows to $\log \gamma$, albeit at a slower rate. 
\end{itemize}

In the heating phase, it is shown in \cite{Fan:2019upv}  that there are $\emph{energy\ peaks}$ forming at the $\emph{unstable}$ fixed points;\footnote{Again note that results in \cite{Fan:2019upv} are derived with systems prepared in the ground state with $\emph{open}$ boundary conditions.} moreover, the entanglement entropy of a single interval $A$ $\emph{grows\ linearly}$ if $A$ includes one of the two peaks, i.e.~includes one $\emph{unstable}$ fixed point. These phenomena indicate that at late time, nearly all degrees of freedom are localised at the two fixed points.  At the phase transition, the two peaks merge into one \cite{Fan:2019upv}. In the non-heating phases, energy and entanglement oscillate without a well-defined period \cite{Fan:2019upv}. More recently, the work \cite{Wen:2022pyj} studied Floquet CFT prepared in a $\emph{thermal}$ initial state. They discovered that the system $\emph{heats up}$ in the "heating region" in the small neighborhood of the energy peaks, where energy and local temperatures grow exponentially and entanglement grows linearly; the rest of the system, nevertheless $\emph{cools down}$, with energy decreasing exponentially and entanglement entropy as well as local temperatures become that of the ground state. 

In this work, we primarily focus on the heating phases mainly because we are interested in understanding the meaning of the fixed points for continuous times (beyond the stroboscopic times) and the role they play in the holographic dual.
Also, it was shown in \cite{Han:2020kwp} that heating phases always exist in more generic Floquet CFT setups than we analyse here, whereas non-heating phases might be absent.

\subsection{Floquet CFT at general time}\label{FloquetGenT}
The field theoretic study of Floquet CFT~\cite{Wen:2018agb,Fan:2019upv,Wen:2020wee,Han:2020kwp,Fan:2020orx,Wen:2022pyj} focuses primarily on stroboscopic time $t=n(T_0+T_1)$. To build the dual bulk spacetime, however, one needs to study the time evolution of $t_{{{\rm new}}}$ and $x_{{{\rm new}}}$ at general $t>0$. This was first explored in \cite{Andersen:2020xvu}. Our treatment here will be similar but more comprehensive. It is convenient to explore Floquet CFT at arbitrary $t$ on the complex $z$-plane, related to the $w$-cylinder \eqref{wnew}, \eqref{wnewbar} via the logarithmic map $w=\log z$ and $\overline{w}=\log \overline{z}$. On the $z$-plane, the effect of $H_1$ is an $SL(2)$ transformation of the $z$ and $\overline{z}$ coordinates; the explicit expressions of $z_n$ and $\overline{z}_n$ are given by \eqref{zn} and \eqref{znbar}, respectively. 

To obtain the expression of $z_{{{\rm new}}}$ at general time, one can start with \eqref{zn} at a stroboscopic time step, and then study how $z_n$ time evolves in the $(n+1)$-th cycle, i.e.~$n(T_0+T_1)\leq t\leq (n+1)(T_0+T_1)$. The full time evolution then follows inductively. In each cycle, the time evolution of $z_{{{\rm new}}}$ depends on whether the Hamiltonian is $H_0$ or $H_1$. As the effect of $H_0$ and $H_1$ are ordinary time translations and $SL(2)$ transformations according to \eqref{wnew} (see also \eqref{znew}), respectively, the expression of $z_{{{\rm new}}}$ in the $(n+1)$-th cycle is: 
\begin{equation}
\label{znewGeneralTime}
z_{{{\rm new}}}=
\left\{
\begin{aligned}
& e^{i\mathfrak{t}_{n+1}}z_n,\ \ &&0\leq \mathfrak{t}_{n+1}\leq \frac{T_0}{2}\\
& \frac{\Big(1+\frac{i(\mathfrak{t}_{n+1}-T_0/2)}{2}\Big)\le(e^{\frac{i T_0}{2}}z_n\ri)-\frac{i(\mathfrak{t}_{n+1}-T_0/2)}{2}}{\frac{i(\mathfrak{t}_{n+1}-T_0/2)}{2}\le(e^{\frac{i T_0}{2}}z_n\ri)+\Big(1-\frac{i(\mathfrak{t}_{n+1}-T_0/2)}{2}\Big)},\ \ &&\frac{T_0}{2}< \mathfrak{t}_{n+1}\leq \frac{T_0}{2}+T_1\\
& e^{i(\mathfrak{t}_{n+1}-(T_0+T_1))}z_{n+1},\ \ &&\frac{T_0}{2}+T_1\leq \mathfrak{t}_{n+1}\leq T_0+T_1\\
\end{aligned}
\right.
\end{equation}
where $\mathfrak{t}_{n+1}\equiv t-n(T_0+T_1)\in[0,T_0+T_1]$ is the time in the $(n+1)$-th cycle, and $z_n$ is replaced by $z=e^{ i x}$ in the first cycle, i.e.~when $n=0$. For $\overline{z}_{{{\rm new}}}$ we simply have $z_n\to\overline{z}_n$ in \eqref{znewGeneralTime}, where $\overline{z}_n$ \eqref{znbar} is given by $z\to \overline{z}=e^{- i x}$ in \eqref{zn}, see \eqref{znewbarGeneralTime} for the full expression. A proof of the formula \eqref{znewGeneralTime} using mathematical induction can be found in Appendix \ref{znewGeneralTimeProof}. 

The rather complicated equation \eqref{znewGeneralTime} simplifies greatly in the heating phase due to properties of the fixed points. As $0<\eta<1$, at late stroboscopic  time \eqref{zn} becomes 
\begin{align}
    \lim_{n\to\infty}z_n=\lim_{n\to\infty}\frac{(\gamma_1-\eta^n\gamma_2)z-(1-\eta^n)}{(1-\eta^n)z-(\gamma_2-\eta^n\gamma_1)}=\frac{\gamma_1 z-1}{z-\frac{1}{\gamma_1}}=\gamma_1 && (z\neq\gamma_2)\label{zngamma1}
\end{align}
where we have used $\gamma_1\gamma_2=1$. For $z=\gamma_2$, $z_n=\gamma_2$ by the definition of fixed point. Henceforth, we have 
\begin{equation}
\label{znewLateTime}
    \lim_{n\to\infty}z_n=
    \left\{
    \begin{aligned}
        &\gamma_1,\ \ z\neq\gamma_2\\
        &\gamma_2,\ \ z=\gamma_2
    \end{aligned}
    \right.
\end{equation}
From the middle equation of \eqref{znewGeneralTime}, $H_1$ sends $e^{\frac{i T_0}{2}}z_n$ to $e^{-\frac{i T_0}{2}}z_{n+1}$ (see Appendix \eqref{EffOfH1} for details), and the Floquet dynamics can be summarised as the following $\emph{iteration}$ 
\begin{align}
    ...\overset{H_1}{\longrightarrow}e^{-\frac{i T_0}{2}}z_n\overset{H_0}{\longrightarrow}e^{\frac{i T_0}{2}}z_n\overset{H_1}{\longrightarrow}e^{-\frac{i T_0}{2}}z_{n+1}\overset{H_0}{\longrightarrow}e^{\frac{i T_0}{2}}z_{n+1}\overset{H_1}{\longrightarrow}...\label{recursion}
\end{align}
As $z_n$ either approaches the stable fixed points $\gamma_1$ or stays at the unstable fixed point $\gamma_2$ at late time \eqref{znewLateTime}, the recursion \eqref{recursion} eventually becomes the $\emph{repetition}$ of the two fixed points
\begin{align}
    &...\overset{H_1}{\longrightarrow}e^{-\frac{i T_0}{2}}\gamma_1\overset{H_0}{\longrightarrow}e^{\frac{i T_0}{2}}\gamma_1\overset{H_1}{\longrightarrow}e^{-\frac{i T_0}{2}}\gamma_1\overset{H_0}{\longrightarrow}e^{\frac{i T_0}{2}}\gamma_1\overset{H_1}{\longrightarrow}\label{iterGamma1}...\\
    &...\overset{H_1}{\longrightarrow}e^{-\frac{i T_0}{2}}\gamma_2\overset{H_0}{\longrightarrow}e^{\frac{i T_0}{2}}\gamma_2\overset{H_1}{\longrightarrow}e^{-\frac{i T_0}{2}}\gamma_2\overset{H_0}{\longrightarrow}e^{\frac{i T_0}{2}}\gamma_2\overset{H_1}{\longrightarrow}...\label{iterGamma2}
\end{align}
where we have used the definition of fixed point as points that leave \eqref{zn} invariant. The analysis \eqref{zngamma1}-\eqref{iterGamma2} also work for the anti-holomorphic coordinate with $z\to\overline{z}$ and $z_n\to\overline{z}_n$. 

From \eqref{znewGeneralTime}, \eqref{iterGamma1} \eqref{iterGamma2}, we see that the term "fixed points" is meaningful only in the stroboscopic sense; in general time, they undergo $\emph{repetitive}$ motions on the unit circle in the complex plane and return to the same location after completing each cycle. Thus, after the transient dies out, the late-time behavior of Floquet CFT boils down to that of the two fixed points $\gamma_1$ and $\gamma_2$ (or even $\emph{one}$ point, considering that $\gamma_2$ is the complex conjugate of $\gamma_1$ in the time-reversal case). 

\begin{figure}[htbp]
\centering
\includegraphics[width=.55\textwidth]{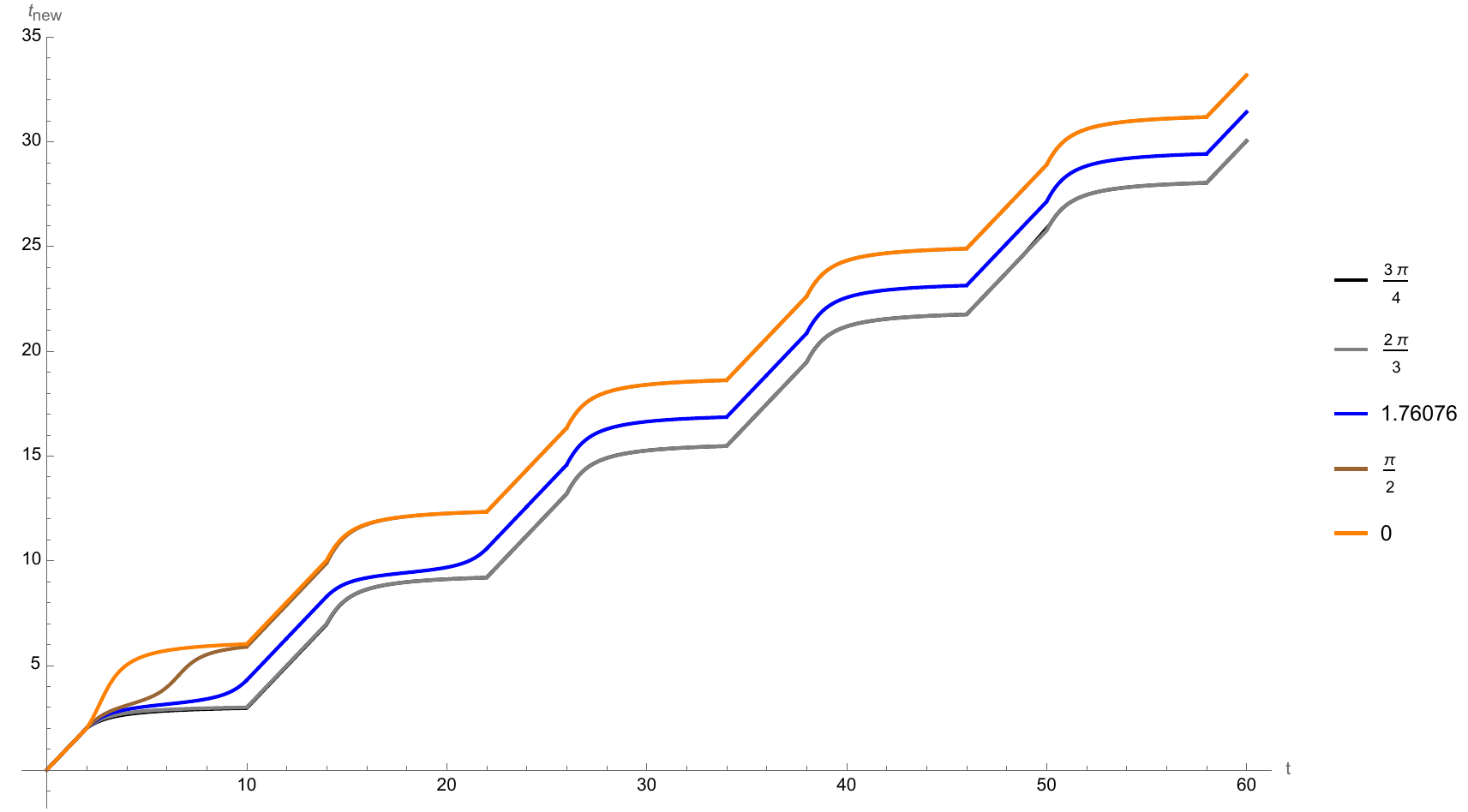}\\[4pt]
\includegraphics[width=.55\textwidth]{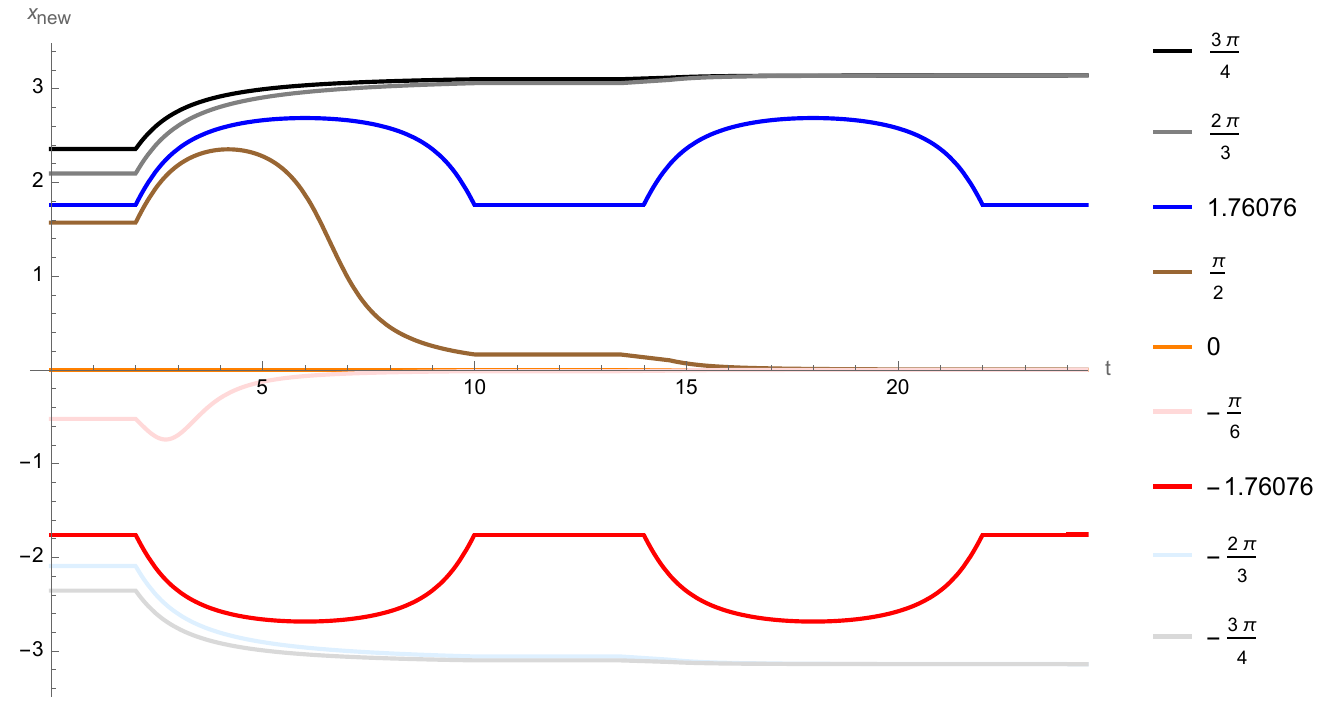}
\qquad
\caption{Plot of $t_{{{\rm new}}}$ and $x_{{{\rm new}}}$ at general $t$ in the heating phase. Here we set $T_0=4$ and $T_1=8$. We plotted for $0, \frac{\pi}{2}, \frac{1}{ i}\log \gamma_1\approx 1.76076, \frac{2\pi}{3}, \frac{3\pi}{4}$ for the first 5 cycles of $t_{{{\rm new}}}$ (as $t_{{{\rm new}}}$ is symmetric under $x\to-x$) and for  $x=-\frac{3\pi}{4},-\frac{2\pi}{3}, \frac{1}{i}\log \gamma_2\approx -1.76076, -\frac{\pi}{6}, 0, \frac{\pi}{2}, \frac{1}{ i}\log \gamma_1\approx 1.76076, \frac{2\pi}{3}, \frac{3\pi}{4}$ for the first two cycles of $x_{{{\rm new}}}$. \label{fig:i}}
\end{figure}

\begin{figure}[htbp]
\centering
\includegraphics[width=.68\textwidth]{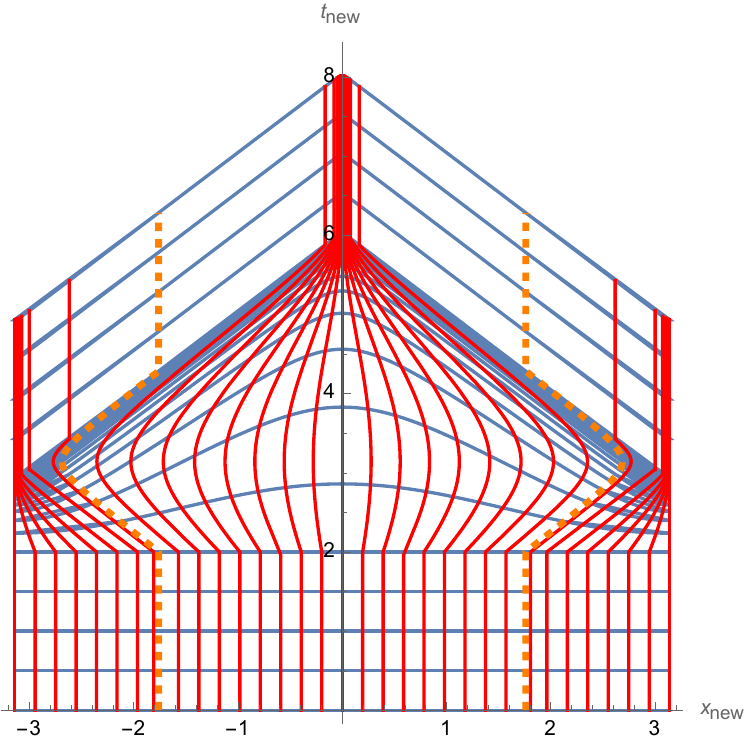}
\qquad
\caption{Constant $t$ lines (blue) and constant-$x$ lines (red) in the $(t_{{\rm new}}, x_{{\rm new}})$-plane. Here we set $T_0=4$ and $T_1=8$, and plot for the first cycle. At $0<t<2$ and $10<t<12$, the driving Hamiltonian is $H_0$, constant-$t$ lines and constant-$x$ lines are uniformly distributed straight lines. At $2<t<10$ the driving Hamiltonian is $H_1$, constant-$t$ lines asymptote $\emph{two}$ triangles, whose tips are at $x_{{\rm new}}=0$ and $x_{{\rm new}}=\pm \pi$, respectively (notice that $x_{{\rm new}}$ is periodic: $x_{{\rm new}}\sim x_{{\rm new}}+2\pi$); constant-$x$ lines converge to these two tips, depending on the range of $x$. The dashed orange lines are the trajectories of the two fixed points. \label{fig:FloquetGrid}}
\end{figure}

There is a second sense that the "fixed points" are not really fixed: so far we have used a principal sheet prescription for the logarithm, but continuity in $t$ leads to additions of multiples of $\pi$ to $t_\text{new}$ compared to $t_n$, as also noted in \cite{Andersen:2020xvu}. In preparation for the investigation of the holographic dual, we need to study the continuous time evolution of $t_{{\rm new}}$ and $x_{{\rm new}}$. Using \eqref{znewGeneralTime} and \eqref{znewbarGeneralTime}, we plot $t_{{{\rm new}}}$ and $x_{{{\rm new}}}$ as functions of $t$ for different values of $x$ in Figure \ref{fig:i}, and further plot constant-$t$ lines and constant-$x$ lines on the $(t_{{\rm new}}, x_{{\rm new}})$ plane in the first cycle in Figure \ref{fig:FloquetGrid}. To better understand $t_{{{\rm new}}}$ and $x_{{{\rm new}}}$, it is instructive to first plot $\frac{1}{i}w_{{{\rm new}}}=\frac{1}{i}\log z_{{{\rm new}}}=t_{{{\rm new}}}+x_{{{\rm new}}}$ and $\frac{1}{i}\overline{w}_{{{\rm new}}}=\frac{1}{i}\log \overline{z}_{{{\rm new}}}=t_{{{\rm new}}}-x_{{{\rm new}}}$, which we include in Appendix \ref{Plotwnewwnewbar} with discussions. 

From  Figure \ref{fig:i} and Figure \ref{fig:FloquetGrid}, it is evident that $t_{{{\rm new}}}$ and $x_{{{\rm new}}}$ indeed converge to different curves as $t$ evolves. In particular, 
\begin{itemize}
    \item At $|x|<\frac{1}{ i}\log \gamma_1$, $x_{{{\rm new}}}$ converges to $x_{{{\rm new}}}=0$, and $t_{{{\rm new}}}$ converges to the curve whose stroboscopic value is $\frac{1}{ i}\log \gamma_1$. This is because both the $\frac{1}{i}w_{{{\rm new}}}$ and the $\frac{1}{i}\overline{w}_{{{\rm new}}}$ curve converge to that of the stable fixed point $\gamma_1$.
    
    \item At $|x|=\frac{1}{i}\log \gamma_1$, $x_{{{\rm new}}}$ follows repetitive motions with stroboscopic values $x_{{{\rm new}}}=\pm\frac{1}{i}\log \gamma_1$; $t_{{{\rm new}}}$ follows a curve whose stroboscopic value is $0$. This is because $\emph{either}$ the $\frac{1}{i}w_{{{\rm new}}}$ $\emph{or}$ the $\frac{1}{i}\overline{w}_{{{\rm new}}}$ curve corresponds to the unstable fixed point $\gamma_2$.\footnote{$z_n$ and $\overline{z}_n$ cannot both be at the unstable fixed point $\gamma_2$, since initially they are complex conjugates. }
    
    \item At $|x|>\frac{1}{ i}\log \gamma_1$, $x_{{{\rm new}}}$ converges to $x_{{{\rm new}}}=\pm\pi$ (these are the same point); $t_{{{\rm new}}}$ again converges to the curve whose stroboscopic value is $\frac{1}{i}\log \gamma_1$, but is shifted down by $\pi$ compared to the first case above. This is because either the $\frac{1}{i}w_{{{\rm new}}}$ or the $\frac{1}{i}\overline{w}_{{{\rm new}}}$ curve is on another Riemann sheet and is therefore $2\pi$ lower, 
    see Appendix \ref{Plotwnewwnewbar} for details. These subtle $\pi$ shifts were not spelled out in previous work that focused on the stroboscopic dynamics~\cite{Wen:2018agb,Fan:2019upv,Wen:2020wee,Han:2020kwp,Fan:2020orx,Wen:2022pyj}.
\end{itemize}

\section{The gravity dual of Floquet CFT}\label{Gravity dual of Floquet CFT 1}

\subsection{Gravity dual of Floquet CFT in heating phases}\label{Gravity dual of Floquet CFT}
The dual gravitational picture can be established similarly to that of the SSD quench case. The gravity dual for the ground and thermal state is pure AdS$_3$ and BTZ black hole, respectively, whose metrics are given by \eqref{AdSMetric} and \eqref{BTZMetric}, with $t_{{\rm new}}$ and $x_{{\rm new}}$ obtained from \eqref{znewGeneralTime} and \eqref{znewbarGeneralTime}. The cutoff surface is also nonuniform in the radial coordinate, like in \eqref{cutoffpresc} in the SSD quench case. 

From Figure \ref{fig:i}, the gravity dual of Floquet CFT in the $(t_{{\rm new}},x_{{\rm new}},r_{{\rm new}})$ coordinate can thus be deduced as follows: as $t$ evolves, constant $x$ lines on the $(t_{{\rm new}}, x_{{\rm new}})$ cylinder converge to $x_{{\rm new}}=0$ or $x_{{\rm new}}=\pi$, or stay at $\pm\frac{1}{i}\log \gamma_1$ as evolving forward in $t_{{\rm new}}$, depending on whether $|x|$ is smaller than, greater than or equal to $\frac{1}{i}\log \gamma_1$. Since $t_{{\rm new}}\to\infty$ as $t\to\infty$ for any $x\in[-\pi,\pi]$, there are no new horizons in the bulk anymore. 

\begin{figure}[htbp]
\centering
\includegraphics[width=.55\textwidth]{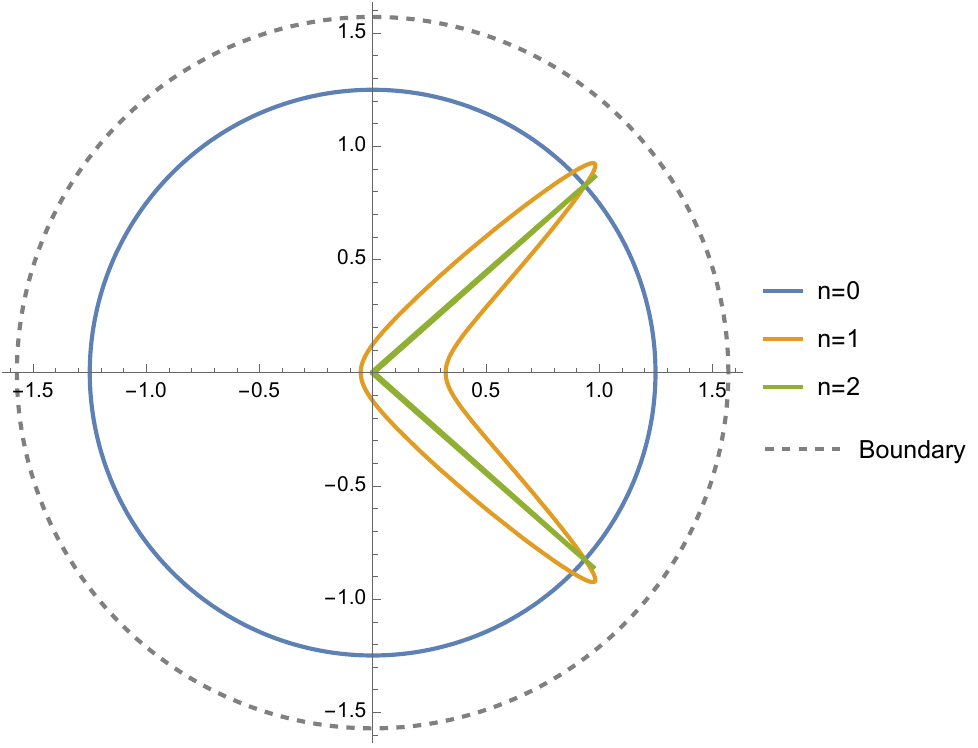}
\qquad
\caption{Time evolution of the BTZ black hole horizon. We chose $r_+=3$, $T_0=2$ and $T_1=8$. We plotted the cutoff surface at $n=0,1,2$. We used the ${\rm arctan}$ function to make the plot, so the boundary is at $r=\frac{\pi}{2}$, which is plotted in a dashed gray line.\label{fig:FloquetHor}}
\end{figure}

Next, we study the evolution of the BTZ horizon $r_{{\rm new}}=r_+$ in the $(t,x,r)$ coordinates. Using the foliation in Appendix \ref{ComparisonRyuFloquet},  the stroboscopic dynamics of the horizon are shown in Figure \ref{fig:FloquetHor}. 

We observe that the horizon approaches the boundary at the two fixed points, i.e.~$x=\pm \frac{1}{ i}\log \gamma_1$, developing two peaks. These peaks originate from the prefactor $\sqrt{\frac{\partial w_n}{\partial w}\frac{\partial \overline{w}_n}{\partial \overline{w}}}$ in \eqref{r'Floquet}, which grows exponentially at the two heating points, see \eqref{r'Floquet}. In \cite{deBoer:2023lrd}, the evolution of BTZ horizon in the slow-driving limit (i.e.~when $v(x)= 1+O(\epsilon)$ in \eqref{dsprime}) for gravity dual of generally deformed CFT  was studied,\footnote{Assuming that these driven CFTs obey the slow-driving condition at least within a range of parameters.} and it was concluded that the horizon approaches the conformal boundary at the heating point(s).  Here, our Floquet setup is not in the slow-driving limit.
This provides evidence that the phenomenon of BTZ horizons evolving toward the boundary at the fixed points is likely generic in heating phases. 

For CFT prepared in the thermal states on a circle, there are energy peaks forming at the unstable fixed points. To see this, note first that in a holographic CFT the thermal stress tensor can be approximated by that of a high energy pure state with dimension $h$ as in the thermal SSD quench case \cite{Goto:2021sqx}, which is $\langle T(z_n)\rangle=\frac{h}{z_n^2}$ on the complex $z$-plane. Transforming back to the $w$-cylinder, 
\begin{align}
    \langle T(w)\rangle = \Big(\frac{\partial z}{\partial w}\Big)^2 \Big(\frac{\partial z_n}{\partial z}\Big)^2 \frac{h}{z_n^2}-\frac{c}{24}\label{T(w)}
\end{align}
where the last term comes from ${\rm Sch}(z,w)$. The remaining analysis parallels that in \cite{Fan:2019upv}: the Jacobian factor $\big(\frac{\partial z_n}{\partial z}\big)^2$ determines the chiral energy peak. The anti-chiral peak is specified by $\big(\frac{\partial \overline{z}_n}{\partial \overline{z}}\big)^2$ in $\langle \overline{T}(\overline{w})\rangle$. The two energy peaks locate at $x=\pm\frac{1}{i}\log \gamma_1$, which are positions at which the black hole horizon approaches the boundary in Figure \ref{fig:FloquetHor}. This is because both the horizon peaks and energy peaks originate from $\frac{\partial z_n}{\partial z}$ or equivalently $\frac{\partial w_n}{\partial w}$ as well as their anti-holomorphic counterparts in the $SL(2)$ transformations. They are big at the unstable fixed point, since in their neighborhood there is a strong sensitivity to the initial data. This is in contrast with the vicinity of the stable fixed point that cools down because it is insensitive to small changes in the initial data.

\subsection{Non-heating phases and phase transition}
At the phase transition, the two fixed points $\gamma_1$ and $\gamma_2$ merge into one, which is $\gamma=1$ in the time-reversal symmetric case.\footnote{Although the $SL(2)$ transformation~\eqref{ABCDphaseTransition} in the phase transition case belongs to the same (parabolic) class of $SL(2)$ transformation of SSD~\eqref{znew}, the two cases are importantly different: \eqref{znew} is for continuous time, whereas the transformation~\eqref{ABCDphaseTransition} only encodes the $\emph{stroboscopic}$ information of the Floquet time evolution. In the phase transition case of Floquet CFT, the general time expression for $z_{{\rm new}}$ takes a piecewise form similar to~\eqref{znewGeneralTime} in the heating phase case. The stroboscopic fixed point specified by~\eqref{ABCDphaseTransition} likewise undergoes repetitive motions on the unit circle within each cycle. } In the non-heating phase, $z_{{\rm new}}$ shows oscillatory behavior instead of approaching the fixed points (which are no longer on the unit circle \cite{Fan:2019upv}). Therefore, one would expect $t_{{\rm new}}$ to show stronger $x$-dependence. In both cases, $t_{{\rm new}}$ increases for different $x\in[0,\pi]$ just as in the heating phases: when $H_0$ acts as Hamiltonian, $t_{{\rm new}}$ increases linearly; when $H_1$ drives, it is also straightforward to show from \eqref{tnewxnew} that $t_{{\rm new}}$ is monotonically increasing. Therefore, $t_{{\rm new}}\to\infty$ at any initial $x\in[0,\pi]$ as $t\to\infty$. Consequently, there are no new horizons in the bulk in non-heating phases or at the phase transition either. 

\section{Dynamics of entanglement entropy}\label{Holographic Entanglement Entropy}
\begin{figure}[htbp]
\centering
\includegraphics[width=.35\textwidth]{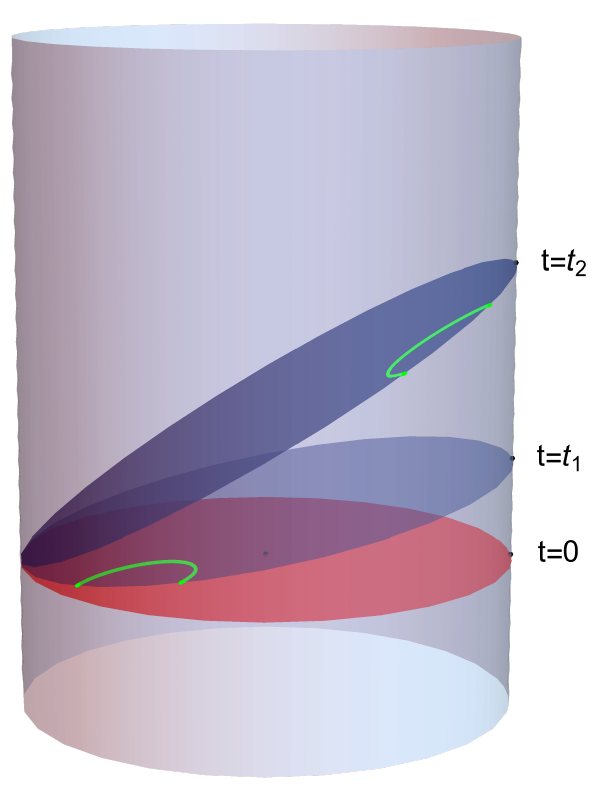}
\qquad
\caption{A demonstration of non-equal-$t_{{\rm new}}$ spacelike geodesics, plotted as green curves in the bulk anchoring on the two endpoints of the subregion $A$ on the boundary constant-$t$ lines.  These spacelike geodesics compute the holographic entanglement entropy. \label{fig:EEGroundState}}
\end{figure}

In this section, we study the holographic entanglement entropy of a single interval $A=[x_1, x_2]$ at a constant $t$-slice. To this end, we again work in the $(t_{{\rm new}},x_{{\rm new}},r_{{\rm new}})$ geometry, and then perform a coordinate transformation back to $(t,x,r)$. At time $t$, the endpoints of $A=[x_1, x_2]$ are obtained from \eqref{wnew}, \eqref{wnewbar}, \eqref{tnewxnew},
\begin{align}
    w_{{{\rm new}},j}=\frac{1}{2i}(t_{{{\rm new}},j}+x_{{{\rm new}},j}) && \overline{w}_{{{\rm new}},j}=\frac{1}{2i}(t_{{{\rm new}},j}-x_{{{\rm new}},j}) && (j=1,2) \label{7.1}
\end{align}
The holographic entanglement entropy of a single interval $A$ is given by the length of the spacelike geodesic anchored on $\partial A=\{(t_{{{\rm new}},1},x_{{{\rm new}},1}), (t_{{{\rm new}},2},x_{{{\rm new}},2})\}$ on the boundary \cite{Ryu:2006bv,Ryu:2006ef}: 
\begin{align}
    S_A=\frac{\mathcal{L}}{4G_N}=\frac{c}{6}\mathcal{L}
\end{align}
where we have used the Brown-Henneaux central charge, $c=3/2G_N$ \cite{Brown:1986nw}. Therefore, to calculate the holographic entanglement entropy, one needs to obtain the geodesic length $\mathcal{L}$, using the temporal and angular separation $\Delta t_{{{\rm new}}}=t_{{{\rm new}},2}-t_{{{\rm new}},1}$ and $\Delta x_{{{\rm new}}}=x_{{{\rm new}},2}-x_{{{\rm new}},1}$ as inputs, and then transform back to $(t,x,r)$. Note that in general $\Delta t_{{{\rm new}}}\neq0$, so our calculation is $\emph{not}$ of equal time ("time" as $t_{{\rm new}}$). See Figure \ref{fig:EEGroundState} for a demonstration. 

The length of the spacelike geodesic anchoring on the boundary is infinite, therefore, we need to regulate it by imposing a $\emph{cutoff}$ at some large radius $r_{{\rm new}}=R_{{\rm new}}$, and adopt the $\emph{regulated}$ length $\mathcal{L}_{{\rm reg}}$. As is discussed in Subsection \ref{4.1}, the cutoff surface is non-uniform in the radial direction, with $R_{{\rm new}}(t_\text{new},x_\text{new})$ given by \eqref{cutoffpresc}. See Figure \ref{fig:CutoffSurface3d} for a plot of the cutoff surface.

\subsection{Gravity dual of SSD: ground state}
\subsubsection{Holographic computation}\label{sec7.1.1}
The metric \eqref{AdSMetric} leads to two conserved quantities for geodesics, $E=(r_{{\rm new}}^2+1) \frac{dt_{{\rm new}}}{d\lambda}$ and $L=r_{{\rm new}}^2 \frac{dx_{{\rm new}}}{d\lambda}$, where $\lambda$ is the affine parameter along the geodesic. In \cite{Hubeny:2012ry}, it was found that in pure AdS$_3$, the temporal and angular distances are given in terms of the two conserved quantities by 
\begin{align}
    \cos (\Delta t_{{\rm new}})&=\frac{L^2-E^2+1}{\sqrt{(L^2-E^2+1)^2+4E^2}}\label{delta_tAdS}\\
    \cos (\Delta x_{{\rm new}})&=\frac{L^2-E^2-1}{\sqrt{(L^2-E^2-1)^2+4L^2}}\label{delta_xAdS}
\end{align}
From which one can solve for $E$ and $L$. Using these $E$ and $L$, we compute the regulated length of the spacelike geodesic \cite{Hubeny:2012ry} in Appendix \ref{AdSSpaceLikeGeod}
\begin{align}
    \mathcal{L}_{{\rm reg}}&=-\log \sqrt{(E^2-L^2+1)^2+4L^2}+\log (4R_{{\rm new,1}}R_{{\rm new,2}})\label{geodLenAdS}\\
    &=\log \le[4 \Big|\sinh 
    \big(\frac{1}{2}(w_{{{\rm new}},1}-w_{{{\rm new}},2})\big)\ \sinh \big(\frac{1}{2}(\overline{w}_{{{\rm new}},1}-\overline{w}_{{{\rm new}},2})\big)\Big|\ \ri]+\log (R_{{\rm new,1}}R_{{\rm new,2}})\,,\label{7.7}
\end{align}
where $\log (4R_{{\rm new,1}}R_{{\rm new,2}})$ is the universal diverging piece. From \eqref{cutoffpresc}, we have for the large-radius cutoff
\begin{align}
    R_{{\rm new},i}=\frac{R_i}{\sqrt{\frac{\partial w_{{\rm new},i}}{\partial w_i}\frac{\partial\overline{w}_{{\rm new},i}}{\partial\overline{w}_i}}}\label{regCutoff}
\end{align}
where $i=1,2$ labels the two large-radius cutoffs near the two endpoints of the geodesic. With \eqref{regCutoff}, the holographic entanglement entropy is:\footnote{Note that as $w_{{\rm new}}$ is purely imaginary, the terms inside the logarithm are in fact $\sin $ functions of $t_{{\rm new}}$ and $x_{{\rm new}}$.}
\begin{align}
    S_A=\frac{\mathcal{L}_{{\rm reg}}}{4G_N}=\frac{c}{6} \log \le[4\frac{\Big|\sinh 
    \big(\frac{1}{2}(w_{{{\rm new}},1}-w_{{{\rm new}},2})\big)\ \sinh \big(\frac{1}{2}(\overline{w}_{{{\rm new}},1}-\overline{w}_{{{\rm new}},2})\big)\Big|}{\prod_{i=1}^2\sqrt{\frac{\partial w_{{\rm new},i}}{\partial w_i}\frac{\partial\overline{w}_{{\rm new},i}}{\partial\overline{w}_i}}\epsilon_i}\ri]\label{SSDEEVacAdS}
\end{align}
where we have introduced the regularization parameter $\epsilon_i=R_i^{-1}$. One can easily verify that $S_A=S_{A^c}$, which is a property of pure states. An example of the spacelike geodesic computing the holographic entanglement entropy is in Figure \ref{fig:EEGroundState}.\footnote{One may wonder what general statements we can make about geodesics computing entropies in these coordinates.  Due to the Killing symmetry of AdS$_3$ discussed in Appendix~\ref{AsymptoticSymmetries}, a spacelike geodesic whose endpoints are on a constant-$t$ line on the boundary entirely lies on a constant-$t$ slice in the bulk. As the new horizon~\eqref{AdSKilling} is the $t=\infty$ slice and a Killing horizon, any spacelike geodesic is upper bounded by the new horizon. These statements do not continue to hold in the thermal state.} 

\subsubsection{Match with field theory result}\label{SSDGSEE}
In 1+1 dimensional CFT, the entanglement entropy of a single interval $A=[x_1, x_2]$ can be evaluated from the two-point function of twist operators \cite{Calabrese:2004eu},
\begin{align}
    S_A=-\lim_{n\to 1}\frac{\partial}{\partial n}{\rm Tr}\rho_A^n && {\rm Tr}\rho_A^n&	\propto \langle \mathcal{T}_{n}(w_1,\overline{w}_1)\mathcal{T}_{-n}(w_2,\overline{w}_2)\rangle\label{EEfund}
\end{align}
with dimension $h=\overline{h}=\frac{c}{24}(n-\frac{1}{n})$. The time evolution of the two-point function is captured by the $SL(2)$ transformation $(w,\overline{w})\to (w_{{\rm new}}, \overline{w}_{{\rm new}})$: 
\begin{align}
    \langle \mathcal{T}_{n}(w_1,\overline{w}_1)\mathcal{T}_{-n}(w_2,\overline{w}_2)\rangle=\prod_{i=1}^2\Big(\frac{\partial w_{{\rm new},i}}{\partial w}\Big)^h\Big(\frac{\partial\overline{w}_{{\rm new},i}}{\partial\overline{w}}\Big)^h\langle \mathcal{T}_{n}(w_{{{\rm new}},1},\overline{w}_{{{\rm new}},1})\mathcal{T}_{-n}(w_{{{\rm new}},2},\overline{w}_{{{\rm new}},2})\rangle\label{7.12}
\end{align}
The two-point function $\langle \mathcal{T}_{n}(w_{1,{{\rm new}}},\overline{w}_{1,{{\rm new}}})\mathcal{T}_{-n}(w_{2,{{\rm new}}},\overline{w}_{2,{{\rm new}}})\rangle$ is evaluated following the methods in \cite{Calabrese:2004eu} but paying close attention to the difference between holomorphic and antiholomorphic parts (see Appendix \ref{Calc2pt} for details), 
\begin{align}
    &\langle \mathcal{T}_{n}(w_{{{\rm new}},1},\overline{w}_{{{\rm new}},1})\mathcal{T}_{-n}(w_{{{\rm new}},2},\overline{w}_{{{\rm new}},2})\rangle\nonumber\\
    =&\Big[4\sinh \Big(\frac{1}{2}(w_{{{\rm new}},1}-w_{{{\rm new}},2})\Big)\ \sinh \Big(\frac{1}{2}(\overline{w}_{{{\rm new}},1}-\overline{w}_{{{\rm new}},2})\Big)\Big]^{-2h}\,,\label{2ptFuncNew}
\end{align}
where
\begin{align}
    w_{{\rm new}}=\tau_{{\rm new}}+i x_{{\rm new}} && \overline{w}_{{\rm new}}=\tau_{{\rm new}}-i x_{{\rm new}}\,.
\end{align}
Substituting \eqref{2ptFuncNew} into \eqref{7.12} and then into \eqref{EEfund} and performing the analytic continuation $\tau_{{\rm new}}\to it_{{\rm new}}$, the CFT entanglement entropy is then
\begin{align}
    S_A&=\frac{c}{6} \log \le[\frac{4 \Big|\sinh \big(\frac{1}{2}(w_{{{\rm new}},1}-w_{{{\rm new}},2})\big)\ \sinh \big(\frac{1}{2}(\overline{w}_{{{\rm new}},1}-\overline{w}_{{{\rm new}},2})\big)\Big|}{\prod_{i=1}^2\sqrt{\frac{\partial w_{{\rm new},i}}{\partial w_i}\frac{\partial\overline{w}_{{\rm new},i}}{\partial\overline{w}_i}}\epsilon_i}\ri]\label{SSDEEVacCFT}
\end{align}
where we have introduced the UV cutoff $\epsilon_i$. The terms multiplying $\epsilon_i$ in the denominator in \eqref{SSDEEVacCFT} come from the Jacobian factors in \eqref{7.12}. The field theory calculation \eqref{SSDEEVacCFT} precisely matches our earlier holographic result \eqref{SSDEEVacAdS}.\footnote{Note, however, that the field theory result is valid for $\emph{generic}$ CFTs, while our holographic result is limited to $\emph{holographic}$, i.e.~large $c$, large gap CFTs only.}

\subsection{Gravity dual of SSD: thermal state}
\subsubsection{Holographic computations}\label{sec7.2.1}
The metric \eqref{BTZMetric} has two conserved quantities, $E=(r_{{\rm new}}^2-r_+^2) \frac{dt_{{\rm new}}}{d\lambda}$ and $L=r_{{\rm new}}^2 \frac{dx_{{\rm new}}}{d\lambda}$, where $\lambda$ is the affine parameter along the geodesic. We will focus on the region outside the BTZ horizon, $r>r_+$. In Appendix \ref{BTZSpaceLikeGeod}, we calculated the temporal and angular separation of the two endpoints of the spacelike geodesic
\begin{align}
    \cosh  (r_+\Delta t_{{\rm new}})&=\frac{L^2-E^2-r_+^2}{\sqrt{(L^2-E^2-r_+^2)^2-4r_+^2E^2}}\label{delta_tBTZ}\\
    \cosh  (r_+\Delta x_{{\rm new}})&=\frac{L^2-E^2+r_+^2}{\sqrt{(L^2-E^2+r_+^2)^2-4r_+^2L^2}}\label{delta_xBTZ}
\end{align}
where $L\ge E+r_+$ so that the spacelike geodesic does not enter the BTZ horizon (see Appendix \ref{BTZSpaceLikeGeod} for details).\footnote{Note that if one sets the BTZ black hole radius $r_+^2=-1$ and hence replaces $\cosh$ with $\cos$  in \eqref{delta_tBTZ} and \eqref{delta_xBTZ}, one recovers \eqref{delta_tAdS} and \eqref{delta_xAdS} in pure AdS$_3$, respectively.} Therefore one can solve for $E$ and $L$. Using these $E$ and $L$, we computed the regulated length of the spacelike geodesic in Appendix \ref{BTZSpaceLikeGeod}.
\begin{align}
    \mathcal{L}_{{\rm reg}}&=-\log \sqrt{(L^2-E^2+r_+^2)^2-4r_+^2L^2}+\log (4R_{{\rm new,1}}R_{{\rm new,2}})\label{geodLenBTZ}\\
    &=\log \le[\Big(\frac{2}{r_+}\Big)^2\Big|\sin 
    \Big(r_+\frac{w_{{{\rm new}},1}-w_{{{\rm new}},2}}{2}\Big)\ \sin \Big(r_+\frac{\overline{w}_{{{\rm new}},1}-\overline{w}_{{{\rm new}},2}}{2}\Big)\Big|\ \ri]+\log (R_{{\rm new,1}}R_{{\rm new,2}})\label{7.18}
\end{align}
For the BTZ black hole, the radius and inverse temperature are related via $r_+={2\pi/\beta}$. The large-radius cutoff is the same as \eqref{regCutoff}. The holographic entanglement entropy is:\footnote{Note that as $w_{{\rm new}}$ is purely imaginary, the terms inside the logarithm are in fact $\sinh $ functions of $t_{{\rm new}}$ and $x_{{\rm new}}$.}
\begin{align}
    S_A^{(1)}=\frac{\mathcal{L}_{{\rm reg}}}{4G_N}=\frac{c}{6} \log \le[\frac{\big(\frac{\beta}{\pi}\big)^2 \Big|\sin \big(\frac{\pi}{\beta}(w_{{{\rm new}},1}-w_{{{\rm new}},2})\big)\ \sin \big(\frac{\pi}{\beta}(\overline{w}_{{{\rm new}},1}-\overline{w}_{{{\rm new}},2})\big)\Big|}{\prod_{i=1}^2\sqrt{\frac{\partial w_{{\rm new},i}}{\partial w_i}\frac{\partial\overline{w}_{{\rm new},i}}{\partial\overline{w}_i}}\epsilon_i}\ri]\label{HEEfiniteT}
\end{align}
Where the meaning of the superscript $(1)$ will be clear in later subsections. $S_A^{(1)}$ and $S_{A^c}^{(1)}$ are not equal, as the state is no longer pure but thermal. An example of the spacelike geodesic computing the holographic entanglement entropy is in the left plot of Figure \ref{fig:EEThermalState}. 

\subsubsection{Non-equal-time entanglement plateau}\label{Non-equal-time entanglement plateaux}

\begin{figure}[htbp]
\centering
\includegraphics[width=.35\textwidth]{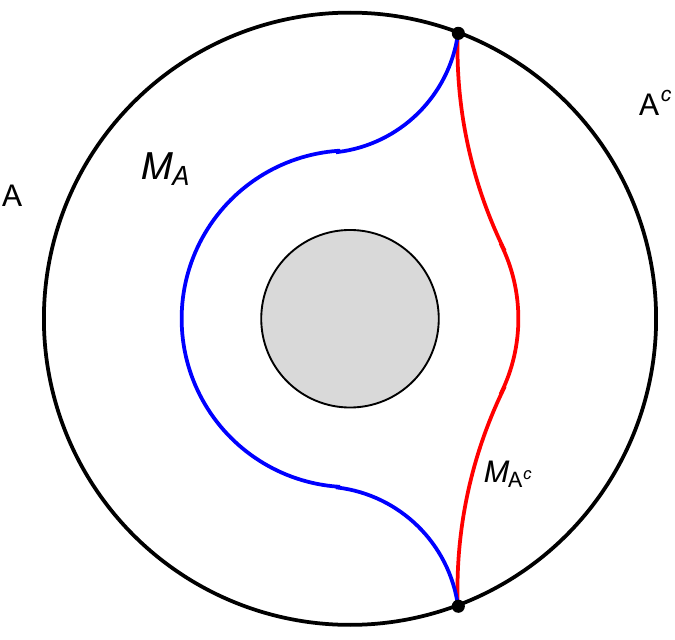}
\qquad
\caption{A demonstration of the minimal surface for a subregion $A$ and its complementary $A^c$. \label{fig:plateaux}}
\end{figure}

 When the state of the total system is a thermal density matrix, $S_A$ and $S_{A^c}$ are not equal, as the minimal surfaces $M_A$ and $M_{A^c}$ computing them encircle complementary parts of the black hole horizon \cite{Ryu:2006bv,Ryu:2006ef} (see Figure \ref{fig:plateaux}).
 Their difference $\Delta S_A$ is bounded from above by the Araki-Lieb inequality \cite{Araki:1970ba}, 
\begin{align}
    |\Delta S_A|=|S_A-S_{A^c}|\leq S_{A\cup A^c}=S_{\rho}\label{Araki-Lieb}
\end{align}
where $S_{\rho}$ is the entropy of the total density matrix, which in our case is the entropy of the BTZ black hole $S_{\rho}=\frac{c}{3}\pi r_+=\frac{2\pi^2c}{3\beta}$.
If we used the geodesics on Figure \ref{fig:plateaux} to determine the entanglement entropy of the CFT at finite temperature using the static BTZ geometry, for large $A$ we would get a violation of \eqref{Araki-Lieb}. The way out is to consider another minimal surface besides $M_A$ that also satisfies the homology constraint \cite{Headrick:2007km}, $M_{A^c}\cup \mathbb{S}_{\rm hor}$ with $\mathbb{S}_{\rm hor}$ the bifurcation surface of the black hole. Hence
\es{2minsurfs}{
    S_A=\frac{1}{4G_N}{\rm min}\{\mathcal{L}(M_A),\mathcal{L}(M_{A^c})+ \mathcal{L}(\mathbb{S}_{\rm hor})\}\,,
}
where $\mathcal{L}(\mathbb{S}_{\rm hor})=2\pi r_+$. When $M_{A^c}\cup \mathbb{S}_{\rm hor}$ is minimal in \eqref{2minsurfs}, the Araki-Lieb inequality is saturated, which ref.~\cite{Hubeny:2013gta} called the $\emph{plateau}$ phenomenon.

To study whether our non-equal time holographic entanglement entropy \eqref{HEEfiniteT} shows the plateau phenomenon, let us check if \eqref{HEEfiniteT} obeys the Araki-Lieb inequality \eqref{Araki-Lieb},
\begin{align}
    \Delta S_A^{(1)}=|S_A^{(1)}-S_{A^c}^{(1)}|=\frac{c}{6} \log \le[\sinh ^2\Big(\frac{\pi}{\beta}\Delta t_{{\rm new}}\Big)\ {\rm csch}\Big(\frac{\pi}{\beta}(\Delta t_{{\rm new}}+2\pi)\Big)\ {\rm csch}\Big(\frac{\pi}{\beta}(\Delta t_{{\rm new}}-2\pi)\Big)\ri]\label{DeltaS}
\end{align}
Recall that in \eqref{tnewxnewLimit}, as $t\to\infty$, the two endpoints $(t_{{\rm new},1},x_{{\rm new},1})$ and $(t_{{\rm new},2},x_{{\rm new},2})$ flow to the top tip of the triangle, $(t_{{\rm new}},x_{{\rm new}})=(\pi, \pi)$. Therefore, $\Delta t_{{\rm new}}\to 0$ in \eqref{DeltaS} as $t$ evolves. In this limit $\Delta S_A^{(1)}$ $\emph{diverges}$, which means that \eqref{DeltaS} $\emph{violates}$ the Araki-Lieb inequality \eqref{Araki-Lieb}. Consequently, our non-equal time holographic entanglement entropy \eqref{HEEfiniteT} also possesses the plateau phenomenon.  The non-equal-time generalization of \cite{Hubeny:2013gta} is straightforward: in addition to the connected minimal surface (see the red curve in the right plot of Figure \ref{fig:EEThermalState}) that computes \eqref{HEEfiniteT}, there is also a $\emph{disconnected}$ minimal surface, given by the $\emph{union}$ of the minimal surface for $A^c$ (i.e.~$\Delta t_{{\rm new}}$ stays the same while $\Delta x_{{\rm new}}$ becomes the complementary value $2\pi-\Delta x_{{\rm new}}$, see the green curve in the right plot of Figure \ref{fig:EEThermalState}) and the bifurcation surface of the event horizon.\footnote{The bifurcation surface is minimal in all static black hole geometries \cite{Hubeny:2013gta}.} The holographic entanglement entropy is then computed by,
\begin{align}
    S_A={\rm min}\{S_A^{(1)},S_A^{(2)}\}
\end{align}
where $S_A^{(1)}$ is given by \eqref{HEEfiniteT}, and 
\begin{align}
    S_A^{(2)}&=\frac{2\pi^2c}{3\beta}+\frac{c}{6} \log \le[\Big(\frac{\beta}{\pi}\Big)\frac{\big|\sinh \big[\frac{\pi}{\beta}
    ((t_{{{\rm new}},1}-t_{{{\rm new}},2})+(2\pi-(x_{{{\rm new}},1}-x_{{{\rm new}},2}))\big]\Big|}{\prod_{i=1}^2\sqrt{\frac{\partial w_{{\rm new},i}}{\partial w_i}\epsilon_i}}\ri]\nonumber\\[4pt]
    &+\frac{c}{6}\log \le[\Big(\frac{\beta}{\pi}\Big)\frac{\big|\sinh \big[\frac{\pi}{\beta}
    ((t_{{{\rm new}},1}-t_{{{\rm new}},2})-(2\pi-(x_{{{\rm new}},1}-x_{{{\rm new}},2}))\big]\Big|}{\prod_{i=1}^2\sqrt{\frac{\partial \overline{w}_{{\rm new},i}}{\partial \overline{w}_i}\epsilon_i}}\ri]\,.\label{HEEfiniteT2}
\end{align}
When $S_A^{(2)}$ is minimal, the Araki-Lieb inequality \eqref{Araki-Lieb} is saturated automatically. 

The plateau phenomenon in thermal SSD quench was first proposed in \cite{Goto:2021sqx}. We remark that while our results are qualitatively similar, our holographic construction is quantitatively different, as the holographic entanglement entropy is computed by spacelike geodesics anchoring on the boundary at non-equal times. 

\subsubsection{Holographic cooling/heating effect}\label{sec:HoloCoolHeat}
\begin{figure}[htbp]
\centering
\includegraphics[width=.38\textwidth]{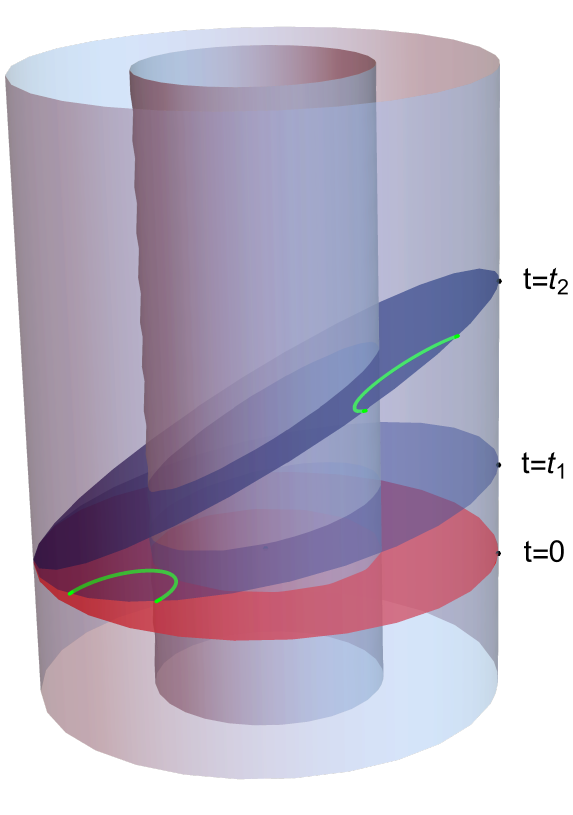}
\includegraphics[width=.38\textwidth]{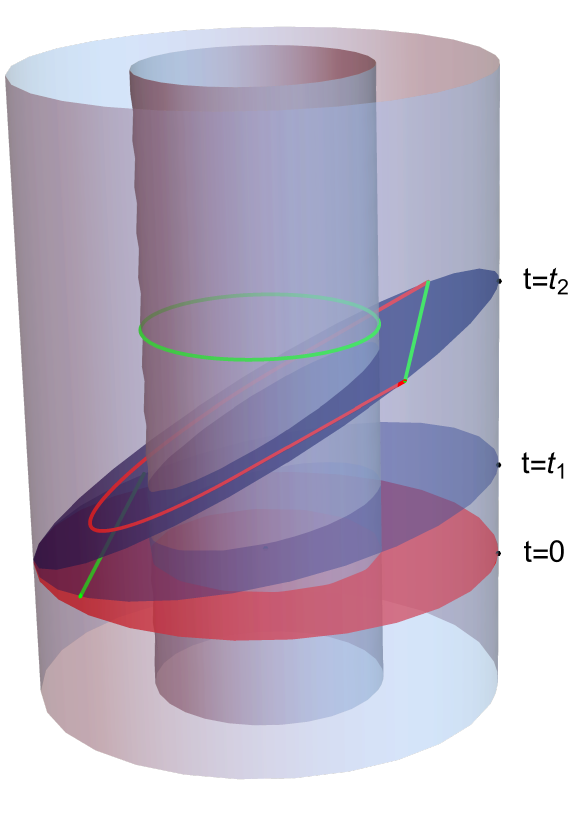}
\qquad
\caption{Non-equal time spacelike geodesics in BTZ geometry. The two geodesic endpoints flow towards the top tip of the triangle $(\pi,\pi)$ as $t$ evolves, from either the same (left) or different (right) side. In the latter case, at late time, the minimal surface computing holographic entanglement entropy switches from the red surface homologous to $A$ to the union of the green surface homologous to $A^c$ and the bifurcation surface of the black hole horizon (also coloured in green). \label{fig:EEThermalState}}.
\end{figure}

To further understand the time-evolution of holographic entanglement entropy, let us first note that the endpoints of the interval $t_{{\rm new},i}$ and $x_{{\rm new},i}$ $(i=1,2)$ can flow to the top tip of the triangle \eqref{tnewxnewLimit} from the same or different side, depending on whether $A$ includes the fixed point $x=0$ or not.
\begin{itemize}
    \item If $A$ does not include $x=0$, $(t_{{\rm new},1},x_{{\rm new},1})$ and $(t_{{\rm new},2},x_{{\rm new},2})$ flow to the top tip of the triangle from the $\emph{same}$ side, and $|x_{{{\rm new}},1}-x_{{{\rm new}},2}|\to 0$ in \eqref{HEEfiniteT}. See the left plot of Figure \ref{fig:EEThermalState}. Therefore, $S_A^{(1)}$ is always smaller than $S_A^{(2)}$, so we have $S_A=S_A^{(1)}$ at all $t>0$.  At late-time, one can show \cite{Goto:2021sqx} that $S_A$ is given by the $\emph{vacuum}$ entanglement entropy, 
\begin{align}
        S_A=S_A^{(1)}\to\frac{c}{3}\log \Big[2\Big(\sin \frac{1}{2}(x_2-x_1)\Big)\Big]\label{SSDCoolingEq}
\end{align}
    \item If $A$ includes $x=0$, $(t_{{\rm new},1},x_{{\rm new},1})$ and $(t_{{\rm new},2},x_{{\rm new},2})$ flow to the top tip of the triangle from $\emph{different}$ sides, making $|x_{{{\rm new}},1}-x_{{{\rm new}},2}|\to 2\pi$ in \eqref{HEEfiniteT}. At some point, therefore, the entanglement entropy switches from $S_A^{(1)}$ to $S_A^{(2)}$. See the right plot of Figure \ref{fig:EEThermalState}. At late time, the second and third terms in \eqref{HEEfiniteT2} become much smaller than the first term,\footnote{For this the time has to compensate the smallness of the cutoff $\epsilon_i$.} so that $S_A$ is dominated by the $\emph{thermal}$ entanglement entropy, 
\begin{align}
    S_A=S_A^{(2)}\to S_{\rho}=\frac{2\pi^2c}{3\beta}\,. \label{SSDHeatingEq}
\end{align}
\end{itemize}

We conclude that whether the entanglement entropy cools down to the vacuum value or heats up and eventually reaches thermal value has a simple geometric interpretation in the bulk, which is whether the spacelike geodesic computing the holographic entanglement entropy winds around the BTZ black hole or not at late time.  Again, \eqref{SSDCoolingEq} and \eqref{SSDHeatingEq} match the results in \cite{Goto:2021sqx} qualitatively, albeit the setup is quantitatively different.  

\subsubsection{Comparison with field theory result on a line}\label{Comparison with field theory result on a line}
Computing the entanglement entropy of 1+1 dimensional CFT on a finite interval at finite temperature $\beta$ is difficult. However, one can compare the holographic result \eqref{HEEfiniteT} with that of the thermal state of CFT$_2$ on a line.  This CFT result can be obtained by Wick rotating $w_{{\rm new}}\to iw_{{\rm new}}$ and rescaling $2 \pi\to\beta$ the cylinder that computes \eqref{SSDEEVacCFT} (see Appendix \ref{Calc2pt} for details), 
\begin{align}
    S_A=\frac{\mathcal{L}_{{\rm reg}}}{4G_N}=\frac{c}{6} \log \le[\frac{\big(\frac{\beta}{\pi}\big)^2 \Big|\sin \big(\frac{\pi}{\beta}(w_{{{\rm new}},1}-w_{{{\rm new}},2})\big)\ \sin \big(\frac{\pi}{\beta}(\overline{w}_{{{\rm new}},1}-\overline{w}_{{{\rm new}},2})\big)\Big|}{\prod_{i=1}^2\sqrt{\frac{\partial w_{{\rm new},i}}{\partial w_i}\frac{\partial\overline{w}_{{\rm new},i}}{\partial\overline{w}_i}}\epsilon_i}\ri]\,.\label{HEEfiniteTCFT}
\end{align}
This result matches with the holographic result $S_A^{(1)}$ before the switch in the dominant extremal surface \eqref{HEEfiniteT2}. In \cite{Asplund:2014coa}, it is explained that the coincidence of \eqref{HEEfiniteT} and \eqref{HEEfiniteTCFT} is due to properties of large-$c$ Virasoro identity blocks arising from a CFT calculation obtaining entanglement entropy from four point functions with two heavy and two light operators of the cyclic orbifold CFT$^n/\mathbb{Z}_n$. In particular, the identity block in the dominant channel corresponds to the spacelike geodesic computing \eqref{HEEfiniteT}. 

\subsection{Gravity dual of Floquet CFT}
We have seen from Figure \ref{fig:FloquetGrid} that points with $x\neq \pm \frac{1}{i}\log \gamma_1$ quickly converge to $x_{{\rm new}}=0$ or $\pm\pi$ on the $(t_{{\rm new}},x_{{\rm new}})$ cylinder while evolving forward in $t_{{\rm new}}$. Therefore, to study the time evolution of holographic entanglement entropy, it is convenient to work in the stroboscopic time, and the general-time results follow. Adopting the same methods in Subsection \ref{sec7.1.1} and \ref{sec7.2.1}, we find the holographic entanglement entropy at $t=n(T_0+T_1)$ for ground state 
\begin{align}
    S_A=\frac{c}{6} \log \le[\frac{4 \Big|\sinh \big(\frac{1}{2}(w_{{{ n}},1}-w_{{{ n}},2})\big)\ \sinh \big(\frac{1}{2}(\overline{w}_{{{ n}},1}-\overline{w}_{{{ n}},2})\big)\Big|}{\prod_{i=1}^2\sqrt{\frac{\partial w_{{ n},i}}{\partial w_i}\frac{\partial\overline{w}_{{ n},i}}{\partial\overline{w}_i}}\epsilon_i}\ri]\label{FloquetEEGround}
\end{align}
and thermal state 
\begin{align}
    S_A^{(1)}=\frac{c}{6} \log \le[\frac{\big(\frac{\beta}{\pi}\big)^2 \Big|\sin \big(\frac{\pi}{\beta}(w_{{{ n}},1}-w_{{{ n}},2})\big)\ \sinh \big(\frac{\pi}{\beta}(\overline{w}_{{{ n}},1}-\overline{w}_{{{ n}},2})\big)\Big|}{\prod_{i=1}^2\sqrt{\frac{\partial w_{{ n},i}}{\partial w_i}\frac{\partial\overline{w}_{{ n},i}}{\partial\overline{w}_i}}\epsilon_i}\ri]\label{FloquetEEThermal}
\end{align}
Using the approach in Subsection \ref{SSDGSEE}, one can find agreements between the ground state entanglement entropy \eqref{FloquetEEGround} and the corresponding CFT result. The thermal state entanglement entropy \eqref{FloquetEEThermal} matches with that of field theory on a $\emph{line}$ at finite temperature $\beta^{-1}$ in \cite{Wen:2022pyj} $\emph{exactly}$, for reasons explained in \cite{Asplund:2014coa} and re-iterated in Subsection \ref{Comparison with field theory result on a line}. Therefore, the cooling/heating effects in \cite{Wen:2022pyj} can be reproduced using holography. In what follows, we discuss the geometric understanding of the cooling/heating effects like in Subsection \ref{sec:HoloCoolHeat}. 

\subsubsection{Holographic cooling/heating effect}
For a given subregion $A=[x_1,x_2]$, the two endpoints are mapped to two distinct points $(w_{n,1},\overline{w}_{n,1})$ and $(w_{n,2},\overline{w}_{n,2})$ or equivalently $(t_{n,1},x_{n,1})$ and $(t_{n,2},x_{n,2})$ on the $(t_{{\rm new}},x_{{\rm new}})$ cylinder at stroboscopic time, and the holographic entanglement entropy is computed by the non-equal-time spacelike geodesic anchoring at these two points, resulting in \eqref{FloquetEEThermal}. The time evolution of holographic entanglement entropy then boils down to those of $(t_{n,1},x_{n,1})$ and $(t_{n,2},x_{n,2})$, which was discussed in Section \ref{Gravity dual of Floquet CFT}: for $|x|\neq \frac{1}{i}\log \gamma_1$, $(t_n,x_n)$ flows to either $\big(\frac{1}{ i}\log \gamma_1,0\big)$ or $\big(\frac{1}{ i}\log \gamma_1-\pi,\pi\big)$, depending on whether $|x|<\frac{1}{i}\log \gamma_1$ or $\frac{1}{i}\log \gamma_1<|x|<\pi$. This is already evident in the first cycle, see Figure \ref{fig:FloquetGrid}. At general time, points $\big(\frac{1}{ i}\log \gamma_1,0\big)$ and $\big(\frac{1}{ i}\log \gamma_1-\pi,\pi\big)$ evolve forward in $t_{{\rm new}}$ along the lines $x_{{\rm new}}=0$ or $x_{{\rm new}}=\pi$. There are three different possible cases: 
\begin{itemize}
    \item $A=[x_1,x_2]$ includes $\emph{none}$ of the two fixed points $x=\frac{1}{i}\log \gamma_1$ or $x=\frac{1}{ i}\log \gamma_2$. In that case, $x_{n,1}$ and $x_{n,2}$ will $\emph{both}$ flow to $\big(\frac{1}{ i}\log \gamma_1,0\big)$ or $\big(\frac{1}{ i}\log \gamma_1-\pi,\pi\big)$, depending on whether $|x_j|<\frac{1}{i}\log \gamma_1$ or $|x_j|>\frac{1}{ i}\log \gamma_1$ $(j=1,2)$. The spacelike geodesic computing the holographic entanglement entropy then shrinks to the neighborhood of $\big(\frac{1}{ i}\log \gamma_1,0\big)$ or $\big(\frac{1}{ i}\log \gamma_1-\pi,\pi\big)$ at stroboscopic time, resulting in $\Delta w_n\to0$ and $\Delta\overline{w}_n\to 0$. Plugging in \eqref{FloquetEEThermal}, it can be shown that the entanglement entropy equals that of the $\emph{vacuum}$ entanglement entropy \cite{Wen:2022pyj}, and the system is said to be in the $\emph{cooling}$ region.   
    
    \item $A=[x_1,x_2]$ includes $\emph{one\ of}$ the two fixed points $x=\frac{1}{i}\log \gamma_1$ or $x=\frac{1}{i}\log \gamma_2$. In that case, one of the two $x_{n,j}$ will flow to the point $\big(\frac{1}{ i}\log \gamma_1,0\big)$, while the other instead to the point $\big(\frac{1}{ i}\log \gamma_1-\pi,\pi\big)$. The holographic entanglement entropy is then computed by the non-equal-time spacelike geodesic ending at point $\big(\frac{1}{ i}\log \gamma_1,0\big)$ and $\big(\frac{1}{ i}\log \gamma_1-\pi,\pi\big)$, thus $\Delta t_n = \pi$, and $\Delta x_n = \pm\pi$, leading to 
    \begin{align}
        \frac{1}{i}\Delta w_n = \Delta t_n+\Delta x_n = 2\pi && \frac{1}{i}\Delta \overline{w}_n = \Delta t_n-\Delta x_n= 0
    \end{align}
    or
     \begin{align}
        \frac{1}{i}\Delta w_n = \Delta t_n+\Delta x_n = 0 && \frac{1}{i}\Delta \overline{w}_n = \Delta t_n-\Delta x_n= 2\pi
    \end{align}
    Substituting in \eqref{FloquetEEThermal}, and using~\eqref{wn} and~\eqref{wnbar} one can show that the entanglement entropy grows $\emph{linearly}$ with slope $\frac{c}{6}\log \frac{1}{\eta}$ \cite{Wen:2022pyj}, and the system is said to be in the $\emph{heating}$ region.   

    Note that it is impossible to have $\frac{1}{i}\Delta w_n=\frac{1}{i}\Delta \overline{w}_n=2\pi$, as the $-\pi$ in $\big(\frac{1}{ i}\log \gamma_1-\pi,0\big)$ originates from $\emph{either}$ $\frac{1}{i}w_{{\rm new}}$ $\emph{or}$ $\frac{1}{i}\overline{w}_{{\rm new}}$ being $2\pi$ lower. There does not exist an $x$ where both would be $2\pi$ lower (see Appendix \ref{Plotwnewwnewbar}). Equivalently speaking, $x_1, x_2$ cannot be at the two different sides of both $\frac{1}{i}\log \gamma_1$ and $\frac{1}{i}\log \gamma_2$. Thus, the heating effect is induced by $\emph{either}$ the chiral $\emph{or}$ the anti-chiral component of the CFT; these two parts cannot both heat.  

    Another interesting feature of the heating case is that as $\Delta x_n$ encloses $\emph{half}$ of the $(t_{{\rm new}},x_{{\rm new}})$ cylinder, the extremal geodesic computing holographic entanglement entropy will not switch to the disconnected one, and $S_A$ will always be computed by $S_A^{(1)}$. This is in contrast to the thermal SSD quench case in Subsection \ref{sec:HoloCoolHeat}, where the spacelike geodesic computing holographic entanglement entropy winds the black hole at late time. 
    
    \item $A=[x_1,x_2]$ includes $\emph{both}$ fixed points. In that case, again, $x_{n,1}$ and $x_{n,2}$ will both flow to $\big(\frac{1}{ i}\log \gamma_1,0\big)$ or $\big(\frac{1}{ i}\log \gamma_1-\pi,\pi\big)$ depending on $|x_j|$, and the system is again in the $\emph{cooling}$ region. 
\end{itemize}
We conclude that whether the entanglement entropy cools down to the vacuum value or heats up and grows linearly has a simple geometric interpretation in the bulk, which is whether the spacelike geodesic computing the holographic entanglement entropy encloses $\emph{half}$ of the BTZ black hole or not at late time. The difference between heating in Floquet CFT and in SSD quench \ref{sec:HoloCoolHeat} is because in the latter case, one injects energy into the CFT only once at $t=0$, and then allows it to equilibrate. That the extremal surface winds the black hole at late time and the entanglement entropy is computed by the thermal entropy is a signature of this equilibration; this could perhaps be made more precise.
In Floquet CFT, however, one keeps adding energy to the CFT at each cycle by switching Hamiltonian from $H_0$ to $H_1$ (therefore the energy increases exponentially \cite{Wen:2022pyj}), so that the CFT will never reach equilibrium, and entanglement keeps growing.   

\section{Conclusions and outlook}\label{conclusion}  

In this work, we studied the gravitational dual of SSD CFT and Floquet CFT prepared in both vacuum and thermal states with periodic boundary conditions. We found that the holographic dual of the non-trivial time evolution of SSD CFT in the ground and thermal states can be understood as pure AdS$_3$ and BTZ black hole spacetime with novel foliations, respectively. We further discovered that new horizons are present in the bulk in both cases, which are given by equations \eqref{AdSKilling} and \eqref{BTZNullGeod2}, and plotted in Figure \ref{fig:AdSHor} and \ref{fig:BTZHor}, respectively. We then argued that bulk operators outside the BTZ horizon can be reconstructed regardless of whether they are inside or outside the new horizons. Subsequently, we studied the gravity dual of Floquet CFT after working out Floquet dynamics in general time, and concluded that there are no new horizons in the bulk anymore. Lastly, we computed the holographic entanglement entropy in the gravity dual of SSD and Floquet CFT, respectively. We found agreement with CFT results in all cases that are comparable and reproduced the cooling/heating effects holographically. 

There are two important technical results of the paper from which most of the results follow. First, the expressions  $w_{{\rm new}}$ \eqref{wnew} and $\overline{w}_{{\rm new}}$ \eqref{wnewbar} lead to the causal structure of spacetime, such as the horizon \eqref{AdSKilling2} and \eqref{BTZNullGeod2}, and the Floquet dual gravitational picture. Second, the Jacobians $\frac{\partial {\rm new}}{\partial {\rm old}}$, $\frac{\partial \overline{{\rm new}}}{\partial \overline{{\rm old}}}$ determine the curved spacetime CFT that the driving is equivalent to and lead to the new slicing of the bulk in holography; they
lead to the energy peaks in SSD and Floquet CFT, hence they result in the exponential growth of energy \cite{Fan:2019upv,Wen:2022pyj}; they determine how  the twist operator two-point function  transforms under the change of coordinates \eqref{7.12}, hence they result in the linear growth of entanglement entropy \cite{Fan:2019upv,Wen:2022pyj}.

Throughout our work, we focused only on one of the simplest $SL(2)$ deformations: the sine-square deformation as well as the Floquet CFT constructed from it, both with periodic boundary conditions. There are several direct generalizations. First of all, it would be interesting to explore holographic aspects of more general $SL(2)$ deformed CFTs such as Mobius deformations \eqref{MobiusDeformedCFT}, using the approaches in this work. Time evolution of entanglement entropy after a Mobius quench was studied in \cite{Wen:2018vux}, Floquet CFT with ordinary and Mobius deformed CFT were constructed in \cite{Wen:2020wee}, and a version of holographic dynamics of Mobius quench with thermal initial states was investigated in \cite{Goto:2021sqx}, where it was pointed out that entanglement entropy for subsystems after a Mobius quench shows periodic oscillations. It would also be interesting to go beyond the subgroup $SL(2)$ and explore holographic aspects of deformations involving the full Virasoro algebra \cite{moosavi2021inhomogeneous,Fan:2020orx,Lapierre:2020ftq,Erdmenger:2021wzc,deBoer:2023lrd}. In studying 2d CFTs with general deformations, the presence of a new horizon is \emph{not} a generic feature. For example, the Mobius-deformed CFT \eqref{MobiusDeformedCFT} does not have any horizon in its Heisenberg time evolution.\footnote{To see this, one can follow the procedures in section \ref{SSD: Boundary Horizons} and obtain the conformal Killing vector corresponding to $H_{\theta}$ \eqref{MobiusDeformedCFT}
\begin{align}
    k_{\theta}=(1-\tanh{2\theta}\cos{t}\cos{x})\partial_t+\tanh{2\theta}\sin{t}\sin{x}\partial_x
\end{align}
which becomes null only in the $\theta\to\infty$, i.e. SSD limit at the triangle \eqref{triangle}.} Henceforth, a particularly meaningful question is: under what conditions does a deformed 2d CFT possess a new horizon?\footnote{We thank Bastien Lapierre for discussions on this topic.} Another interesting direction is to study holographic constructions of deformed CFT with $\emph{open}$ (instead of periodic) boundary conditions. While $H_0$, $H_1$ and the Mobius deformed Hamiltonian $H_\theta$ \eqref{MobiusDeformedCFT} share the same ground states with periodic boundary conditions and hence lead to a very simple quench dynamics, imposing open boundary conditions can set up a non-trivial quantum quench problem for CFTs prepared in the ground state by suddenly switching the CFT Hamiltonian from $H_0$ to $H_1$ or $H_\theta$ at $t=0$. This is indeed the construction adopted in \cite{Wen:2018vux,Wen:2018agb,Fan:2019upv,Bernamonti:2024fgx}. Holographic studies of SSD or Mobius deformed CFT with open boundary condition are carried out in \cite{Bernamonti:2024fgx,Kudler-Flam:2023ahk}. It would be interesting to generalise our holographic constructions in section \ref{4.1} to deformed CFTs with open boundary conditions, and make comparisons with their results. Furthermore, in addition to Floquet, i.e. periodically driven CFT, one can also drive the CFT quasi-periodically
\cite{Wen:2020wee,Lapierre:2020roc}. It is of interest to study the holographic duals of these CFTs.

One of the most interesting discoveries in our work is the presence of new horizons \eqref{AdSKilling2} and \eqref{BTZNullGeod2}. While we have confirmed the entanglement wedge reconstructability of bulk operators behind them, it would be interesting to explore whether any trace of the easy (HKLL) and hard (entanglement wedge) reconstruction dichotomy persists  away from the large $c$ limit, in systems more relevant for condensed matter. 

The recently proposed BKM metric \cite{deBoer:2023lrd} provides an interesting information-theoretic viewpoint of driven CFTs, which links SSD and Floquet CFT with circuit complexity. Therefore, our holographic construction offers a possible testing ground for holographic complexity. In addition to the BKM information metric, there are also other complexity measures of relevance. For example, it was shown that Krylov complexity for $\emph{states}$ (a.k.a.~spread complexity) \cite{Balasubramanian:2022tpr} can be defined for the Hamiltonian
\begin{align}
    H=\alpha(L_{-1}+L_1)+\gamma L_0 +\delta \mathbb{I}\,,
\end{align}
which includes (the holomorphic part of) $H_1$ as a special case with $\alpha=-1/2$, $\gamma=1$ and $\delta=0$. Another possible theory of relevance is the path integral optimization formalism \cite{Caputa:2017urj,Caputa:2017yrh}, where foliations of (Euclidean) AdS$_3$ can be obtained from optimizing the Liouville action \cite{Polyakov:1981rd}. The path integral optimization formalism was extended to Lorentzian time in \cite{Boruch:2021hqs}, and to the bulk metric with SSD boundary \cite{MacCormack:2018rwq} in \cite{Caputa:2020mgb}. It would be interesting to explore the possibility of reproducing our results in Figure \ref{fig:CutoffSurface3d} from path integral optimization. 

Another interesting direction is to extend the formalism we developed to "quantum scars" based on the Virasoro algebras \cite{Caputa:2022zsr,Liska:2022vrd}. In these works, generalised coherent states \cite{Perelomov:1974yw}  of the form 
\begin{align}
    |\Psi_n(\xi)\rangle = e^{\xi L_{-n}-\overline{\xi}L_n}|h\rangle\label{scar}
\end{align}
 are interpreted as scarred states, where $|h\rangle$ is the highest weight state in 2d CFT and $\xi$ is a complex parameter. In \cite{Liska:2022vrd}, the effects of displacement operators corresponding to scarred states \eqref{scar} are also elucidated as inducing coordinate transformations on the complex plane, similar to that of $H_1$ in our work. Therefore, it would be interesting to explore the bulk causal structures of the gravity dual of scarred states of the form \eqref{scar}.

\section*{Acknowledgments}

We thank Yiming Chen, Ruihua Fan, Zohar Komargodski, Conghuan Luo, Matthew Roberts, Kotaro Tamaoka, Mao Tian Tan, Apoorv Tiwari, Jie-qiang Wu, Jingxiang Wu, Zhou Yang, and especially Xueda Wen for discussions and/or correspondence. We thank Bastien Lapierre and Chen Bai for valuable feedback on the first version of the preprint, and the anonymous referee for constructive questions. HJ is supported in part by Lady Margaret Hall, University of Oxford. MM is supported in part by the STFC grant ST/X000761/1.

\appendix
\section{Review of SSD CFT}\label{Review of SSD CFT}
In the main text, we introduced the cylinder $\mathbb{R}\times S^1$ with coordinate \eqref{Eucledian_w}.  
To study CFT$_2$, we map the cylinder to the $\emph{complex\ plane}$ $\mathbb{C}$ via the exponential map, 
\begin{equation}
\label{expMap}
    \left\{
    \begin{aligned}
            z&=e^{w}=e^{\tau+i x}\\
    \overline{z}&=e^{\overline{w}}=e^{\tau-i x}
    \end{aligned}
    \right.
\end{equation}
$H_0$ is $\emph{dilation}$ on the $z$-plane. Analytically continuing to real time $\tau\to it$, we have
\begin{equation}
\label{Lorentzianz}
\left\{
    \begin{aligned}
    z&=\lambda e^{w}=e^{i(t+x)}\\
    \overline{z}&=\lambda e^{\overline{w}}=e^{i(t-x)}
\end{aligned}
\right.
\end{equation}
where $\lambda=e^{\tau}$ in Euclidean signature. In Lorentzian signature, $z_{{\rm new}}$ and $\overline{z}_{{\rm new}}$ are on the $\emph{unit\ circle}$. 

\subsection{SSD CFT}\label{SSDreview}
\subsubsection{The action of $H_1$}
Under the exponential map \eqref{expMap}, the SSD Hamiltonian \eqref{1} becomes 
\begin{align}
    H_1&=2\int_0^{2\pi} dx\ \sin ^2\Big(\frac{x}{2}\Big)T_{00}(x)\\
    &=\oint \frac{dz}{2\pi i} \Big(-\frac{1}{2}+z-\frac{z^2}{2}\Big)T(z)-(z\to\bar{z})-\frac{c}{12}\label{H_1_z}\\
    &=L_0-\frac{L_{-1}+L_1}{2}+\bar{L}_0-\frac{\bar{L}_{-1}+\bar{L_1}}{2}-\frac{c}{12}
\end{align}
While $H_0$ acts as dilation, the action of $H_1$ is a more general $SL(2)$ transformation. To see this, we perform a coordinate transformation $(z,\overline{z})\to(\chi,\overline{\chi})$: 
\es{H1start}{
    H_1&=\oint\frac{dz}{2\pi i}\ \Big(-\frac{1}{2}(z-1)^2\Big)T(x)+h.c\\
    &=\oint\frac{d\chi}{2\pi i}\ \frac{dz}{d\chi}\Big(-\frac{1}{2}(z-1)^2\Big)\Big[\Big(\frac{d\chi}{dz}\Big)^2 T(\chi)+\frac{c}{12}{\rm Sch}(\chi,z)\Big]+h.c\,, 
}
    where ${\rm Sch}(\chi,z)$ is the Schwarzian. By requiring that $H_1$ is dilation on the $\chi$-plane \cite{Fan:2019upv}, i.e.~that the  prefactor of $T(\chi)$ is equal to $\chi$, we obtain a differential equation with the solution:\footnote{The most general solutions take the form $\chi(z)=c \,e^{\frac{z+1}{z-1}}$ and $\overline{z}=\overline{c}\,e^{\frac{\overline{z}+1}{\overline{z}-1}}$; the constants $c$ and $\overline{c}$ have no effect on the coordinate transformations.} 
    \begin{align}
    \chi(z)=e^{\frac{z+1}{z-1}}\,, && \overline{\chi}(\overline{z})=e^{\frac{\overline{z}+1}{\overline{z}-1}}\,,\label{chi}
\end{align}
Plugging in this map into  \eqref{H1start}, we get
   \begin{align} 
      H_1 &=\oint\frac{d\chi}{2\pi i}\Big(\chi T_{00}(\chi)-\frac{c}{24}\frac{1}{\chi}\Big)+h.c\,,
\end{align}
and evaluating the contour integral (together with an identical contribution coming from the anti-holomorphic term) gives the Casimir energy $-c/12$.

To our knowledge, the map \eqref{chi} has not appeared in the literature, but we show around \eqref{MobiusDeformedCFT} that it can be obtained as a limit of maps known from prior work. Note that the maps contain essential singularities at $z=1$ and $\overline{z}=1$, respectively, which means that we need to exclude them in defining \eqref{chi}.  These are the fixed points of the $SL(2)$ transformations \eqref{znew} and \eqref{znewbar}, so we simply have $z_{{\rm new}}=z=1$ and $\overline{z}_{{\rm new}}=\overline{z}=1$ in these two excluded points. On the $\chi$ plane, we then have the Heisenberg evolution 
\begin{align}
    e^{\tau H_1 }O(\chi,\bar{\chi})e^{-\tau H_1 }=O(\lambda\chi,\lambda\bar{\chi})
\end{align}
where $\lambda=e^{\tau}$. Back to the $z$-plane, the effect of $H_1$ is to shift the operator $O$ from $z$ to $z_{{\rm new}}$, where $z_{{\rm new}}$ is related to $z$ via \cite{Wen:2018vux}
\begin{align}
    \chi(z_{{\rm new}})&=\lambda\chi(z)
\end{align}
from which one can solve for $z_{{\rm new}}$: 
\begin{align}
    e^{\frac{z_{{\rm new}}+1}{z_{{\rm new}}-1}}&=e^{\tau}\cdot e^{\frac{z+1}{z-1}}\\
    \Rightarrow z_{{\rm new}}&=\frac{(1+\frac{\tau}{2})z-\frac{\tau}{2}}{\frac{\tau}{2}z+(1-\frac{\tau}{2})}\label{znew}
\end{align}
For the anti-holomorphic part, we repeat the above calculation with $z\to\overline{z}$, arriving at
\begin{align}
    \overline{z}_{{\rm new}}&=\frac{(1+\frac{\tau}{2})\overline{z}-\frac{\tau}{2}}{\frac{\tau}{2}\overline{z}+(1-\frac{\tau}{2})}\label{znewbar}
\end{align}
\eqref{znew} and \eqref{znewbar} are $SL(2)$ transformations on the complex plane. Transforming back to the cylinder via the inverse of \eqref{expMap}, and performing analytic continuation $\tau=it$ then yields \eqref{wnew} and \eqref{wnewbar} in the main text. 

The relation among the cylinder (left), complex plane (middle), and the $SL(2)$ transformed plane (right) are summarised as follows:
\begin{align}
    \begin{CD}
w,\overline{w}@>z=e^{w}>\overline{z}=e^{ \overline{w}}> z, \overline{z}@>\chi=e^{\frac{z+1}{z-1}}>\overline{\chi}=e^{\frac{\overline{z}+1}{\overline{z}-1}}>\chi, \overline{\chi}\\
@VH_1VV @VH_1VV @VH_1V{\rm Dilations}V\\
w_{{\rm new}}, \overline{w}_{{\rm new}} @>z_{{\rm new}}=e^{ w_{{\rm new}}}>\overline{z}_{{\rm new}}=e^{\overline{w}_{{\rm new}}}> z_{{\rm new}}, \overline{z}_{{\rm new}}@>\lambda\chi=e^{\frac{z_{{\rm new}}+1}{z_{{\rm new}}-1}}>\lambda\overline{\chi}=e^{\frac{\overline{z}_{{\rm new}}+1}{\overline{z}_{{\rm new}}-1}}> \lambda\chi, \lambda\overline{\chi}
\end{CD}
\label{CoordTab}
\end{align}

\subsubsection{Trigonometric forms of the coordinate transformations}
In Lorentzian signature, $|z_{{\rm new}}|=|\overline{z}_{{\rm new}}|=1$ as Mobius transformations preserve the unit circles \eqref{Lorentzianz} $z_{{\rm new}}$ and $\overline{z}_{{\rm new}}$ lives on. For technical conveniences in making Figure \ref{fig:Triangle} and \ref{fig:CutoffSurface3d}, we follow \cite{Goto:2021sqx} and write $z_{{\rm new}}$ and $\overline{z}_{{\rm new}}$ in trigonometric  forms: \eqref{znew} and \eqref{wnew} can be written as 
\begin{align}
    z_{{\rm new}}=\cos \ 2\varphi+i\ \sin \ 2\varphi && w_{{\rm new}}=2i\varphi\,,
\end{align}
where 
\begin{align}
    \sin \varphi=\frac{\sin \big(\frac{x}{2}\big)}{\sqrt{1-t\ \sin x+t^2 \sin ^2\big(\frac{x}{2}\big)}} && \cos \varphi=\frac{\cos \big(\frac{x}{2}\big)-t\ \sin \big(\frac{x}{2}\big)}{\sqrt{1-t\ \sin x+t^2 \sin ^2\big(\frac{x}{2}\big)}}\,.\label{znewphi}
\end{align}
We then have $\varphi\in[\frac{ x}{2},\pi]\in [0,\pi]$ \cite{Goto:2021sqx} as $t$ ranges from $[0,\infty)$. Similarly, \eqref{znewbar} and \eqref{wnewbar} can be written as  
\begin{align}
    \overline{z}_{{\rm new}}=\cos \ 2\overline{\varphi}+i\ \sin \ 2\overline{\varphi} && \overline{w}_{{\rm new}}=2i\overline{\varphi}\,,
\end{align}
where 
\begin{align}
    \sin \overline{\varphi}=\frac{-\sin \big(\frac{x}{2}\big)}{\sqrt{1+t\ \sin x+t^2 \sin ^2\big(\frac{x}{2}\big)}} && \cos \overline{\varphi}=\frac{\cos \big(\frac{x}{2}\big)+t\ \sin \big(\frac{x}{2}\big)}{\sqrt{1+t\ \sin x+t^2 \sin ^2\big(\frac{x}{2}\big)}}\,.\label{overlineznewphi}
\end{align}
We then have $\overline{\varphi}\in[-\frac{ x}{2},0]\in [-\pi,0]$ \cite{Goto:2021sqx} as $t$ ranges from $[0,\infty)$. Note that $\varphi$ and $\overline{\varphi}$ are in general $\emph{not}$ complex conjugates. 

The explicit expressions of $t_{{\rm new}}(t,x)$ and $x_{{\rm new}}(t,x)$ can be found from \eqref{wnew}, \eqref{wnewbar}, \eqref{tnewxnew}, \eqref{znewphi} and \eqref{overlineznewphi}
\begin{align}
    t_{{\rm new}}(t,x)&=\arccos\Big[\frac{\cos \big(\frac{x}{2}\big)-t\ \sin \big(\frac{x}{2}\big)}{\sqrt{1-t\ \sin x+t^2 \sin ^2\big(\frac{x}{2}\big)}}\Big]-\arccos\Big[\frac{\cos \big(\frac{x}{2}\big)+t\ \sin \big(\frac{x}{2}\big)}{\sqrt{1+t\ \sin x+t^2 \sin ^2\big(\frac{x}{2}\big)}}\Big]\label{tnewtrig}\\
    x_{{\rm new}}(t,x)&=\arccos\Big[\frac{\cos \big(\frac{x}{2}\big)-t\ \sin \big(\frac{x}{2}\big)}{\sqrt{1-t\ \sin x+t^2 \sin ^2\big(\frac{x}{2}\big)}}\Big]+\arccos\Big[\frac{\cos \big(\frac{x}{2}\big)+t\ \sin \big(\frac{x}{2}\big)}{\sqrt{1+t\ \sin x+t^2 \sin ^2\big(\frac{x}{2}\big)}}\Big]\,.\label{xnewtrig}
\end{align}

\subsubsection{SSD CFT as a CFT on curved spacetime}\label{SSDcurvedreview}

Under Weyl transformations, primary correlators transform as 
\es{WeylTf}{
\le\langle O(z,\bar z) \dots \ri\rangle_{ds'^2=\Om \bar{\Om} \,ds^2}=\Om(z)^{-h}\,\bar{\Om}(\bar z)^{-\bar h}\dots\le\langle O(z,\bar z) \dots \ri\rangle_{ds^2}
}
Now let us choose $\Om=\big(\frac{\partial w_{{\rm new}}}{\partial (i x)}\big)^{-1}$ and similarly for $\bar \Om$. Then the second quantity in \eqref{3desc} becomes the RHS of the above equation. It only remains to compute metric $ds'^2$ in $\tau,x$ coordinates. To this end, we write:

\es{dsprime}{
ds'^2&=\Om \bar \Om\ dw_{{\rm new}}d\overline{w}_{{\rm new}}\\
&=\Big(\frac{\partial w_{{\rm new}}}{\partial (i x)}\Big)^{-1}\Big(\frac{\partial\overline{w}_{{\rm new}}}{\partial(-ix))}\Big)^{-1}\Big(\frac{\partial w_{{\rm new}}}{\partial \tau}d\tau+\frac{\partial w_{{\rm new}}}{\partial x}dx\Big)\Big(\frac{\partial \overline{w}_{{\rm new}}}{\partial \tau} d\tau+\frac{\partial \overline{w}_{{\rm new}}}{\partial x} dx\Big)\\
&=\Big({\partial_\tau w_{{\rm new}}\over \partial_x w_{{\rm new}}}\, d\tau+dx\Big)\Big({\partial_\tau \overline{w}_{{\rm new}}\over \partial_x \overline{w}_{{\rm new}}} \, d\tau+ dx\Big)\\
&=\Big(-iv_1(x)\, d\tau+dx\Big)\Big(iv_1(x)\, d\tau+ dx\Big)\\
&=v_1(x)^2d\tau^2+dx^2\,,
}
which is the curved space metric \eqref{SSDMetric3}.
Thus the Weyl transformation indeed directly gives the third quantity in \eqref{3desc}.

\subsubsection{$H_q$ with $q>1$}\label{SSDGeneralq}
As for $H_q$ with $q>1$, the conformal map $z=e^{qw}$ takes the cylinder to a $q$-sheet Riemann surface, where $H_q$ takes the same form as $H_1$
\begin{align}
     H_q&=2\int_0^{2\pi} dx\, \sin ^2\Big(q\cdot \frac{x}{2}\Big)T_{00}(x)\\
    &=q \oint \frac{dz}{2\pi i} \Big(-\frac{1}{2}+z-\frac{z^2}{2}\Big)T(z)-(z\to\bar{z})-q\frac{c}{12}
\end{align}
The calculation then follows the same as above, with $2\pi\to 2\pi/q$. Namely, 
\begin{align}
    z_{{\rm new}}&=\frac{(1+q\frac{\tau}{2})z-q\frac{\tau}{2}}{q\frac{\tau}{2}z+(1-q\frac{\tau}{2})}\\
    ds^2&=q^2  \frac{e^{2q\tau}}{(1-\cos \ qx)^2}\Big[(1-\cos \ qx)^{2}d\tau^2+dx^2\Big]
\end{align}
Note that now $z$ and $\overline{z}$ are coordinates on the $q$-sheeted Riemann surface and there is a branch cut in the $z$-plane. The fact that the system is $q$ copies of identical theories transcribes to the crossing of sheets on the $z$-plane. 

\subsection{General Cases}
\cite{Wen:2018vux,Goto:2021sqx} also studied quantum quench with the one-parameter generalization of the SSD Hamiltonian $H_1$, the Mobius Hamiltonian, 
The Mobius deformed Hamiltonian $H_\theta$ is given by 
\begin{align}
    H_\theta=L_0-{\rm tanh}\ 2\theta\ \frac{L_{-1}+L_1}{2}+\overline{L}_0-{\rm tanh}\ 2\theta\ \frac{\overline{L}_{-1}+\overline{L}_1}{2}-\frac{c}{12}\label{MobiusDeformedCFT}
\end{align}
The $\theta\to 0$ and $\theta\to\infty$ limits of $H_\theta$ are $H_0$ and $H_1$, respectively. This family of Hamiltonians act as dilatations on the $\xi$ plane as $H_\theta^{(\xi)}={1\ov \cosh(2\theta)}\le(L^{(\xi)}_0+\bar{L}_0^{(\xi)}\ri)-{c\ov 12}$ with
\es{xiplane}{
\xi=-{\cosh \theta\, z-\sinh \theta\ov \sinh \theta\, z-\cosh \theta}\,.
}
By raising this map to the power $\cosh(2\theta)$, we obtain a properly normalised dilatation operator:\footnote{Through the exponential map we can see that raising the map to some power corresponds to rescaling time on the cylinder.} 
\es{chiThetaplane}{
\chi_\theta=\le(-\xi\ri)^{\cosh(2\theta)}=\le({\cosh \theta\, z-\sinh \theta\ov \sinh \theta\, z-\cosh \theta}\ri)^{\cosh(2\theta)}\,,
}
which has the limit
\es{chiThetaLimit}{
\chi=\lim_{\theta\to \infty}\chi_\theta=e^{\frac{z+1}{z-1}} \,.
}
This agrees with \eqref{chi}.

The methodology of describing the non-trivial time evolution of inhomogeneous non-equilibrium CFT as coordinate transformations reviewed in Section \ref{SSDsetup} is valid beyond SSD (where only $SL(2)$ deformations are considered) and extends to deformations involving the $\emph{full}$ Virasoro algebras. In these cases, it is shown \cite{moosavi2021inhomogeneous} that the projective unitary representation of the universal cover of the orientation-preserving diffeomorphisms of the circle $\widetilde{{\rm Diff}}_+(S^1)$ allows one to flatten out the deformation $v(x)$ and map the expectations to ordinary equilibrium CFT under coordinate transformations. Like in the SSD case, these coordinate transformations capture the full information of the time evolution of the deformed CFT. In \cite{deBoer:2023lrd}, it is further demonstrated that just as in SSD CFT, these generally deformed CFTs can be understood as CFTs on curved spacetime as well. 

\section{Changing the cutoff in AdS/CFT}

\subsection{AdS/CFT with a nonuniform cutoff}\label{app:WigglyCutoff}

We will show on the example of a scalar field how the extrapolate dictionary \cite{Banks:1998dd} of AdS/CFT changes under changing the cutoff prescription. In the absence of a source of a scalar field, we have
\es{Falloff}{
\phi(\rho,y^\mu)=(2\Delta-d)\rho^{-\Delta}\, O(y^\mu)+\text{(subleading)}\,,
}
where $\rho$ is the bulk radial direction and $O(y^\mu)$ is the boundary CFT primary field. It is clear that if we choose two different bulk radial coordinates related by $r=\Om(y^\mu) r_\text{new}$ then the same bulk physics will give rise to boundary correlators that differ by some factors
\es{BulkFactors}{
\le\langle O(y^\mu)\dots\ri\rangle_\text{CFT, flat}&=\lim_{r_\text{new}\to \infty} {r_\text{new}^{\Delta}\ov2\Delta-d} \dots\le\langle \phi(r_\text{new},y^\mu) \dots \ri\rangle_\text{AdS}\\[4pt]
&=\Om(y^\mu)^{-\Delta}\dots \le\langle O(y^\mu)\dots\ri\rangle_\text{wiggly bdy}\,,
}
where the last object is the boundary correlation function that is computed from the nonuniform cutoff prescription on the gravity side
\es{cutoffagain}{
\epsilon={1 \ov R}={1\ov \Om(y^\mu) R_\text{new}} \,,
}
where in our application the choice $\Om=\sqrt{\frac{\partial w_{{\rm new}}}{\partial w}\frac{\partial\overline{w}_{{\rm new}}}{\partial\overline{w}}}$ is made.
$\le\langle O(y^\mu)\dots\ri\rangle_\text{wiggly bdy}$ is hence equal to the second object in \eqref{3desc}, as stated in the main text. 

It is also equal to the curved space CFT correlation function, as we show on the CFT side from applying a Weyl transformation and on the gravity side by adapting Fefferman-Graham coordinates to the nonuniform cutoff surface.

\subsection{The Fefferman-Graham expansion with a curved boundary}\label{app:FGBanados}

Let us change the spatial coordinate as
\es{xX}{
x=2\arccot X\,, 
}
then the boundary metric \eqref{SSDMetric3} takes the conformally flat form:
\es{confflat}{
ds_\text{bdy}^2={4\ov(1+X^2)^2}\le(-dt^2+dX^2\ri)\,.
}
The Fefferman-Graham expansion with this boundary metric takes the form:
\es{BanadosFG}{
ds^2=&{dr^2\ov r^2}-{(4r^2+X^2-(1+X^2)^2\,L(t+X))(4r^2+X^2-(1+X^2)^2\,R(t-X))\ov 4r^2(1+X^2)^2}\,dt^2\\
&+{(4r^2-1+(1+X^2)^2\,L(t+X))(4r^2-1+(1+X^2)^2\,R(t-X))\ov 4r^2(1+X^2)^2}\,dX^2\\
&+{(8r^2-1+X^2)(L(t+X)-R(t-X))\ov 4r^2}\,dt\,dX\,,
}
where $L(t+X)$ and $R(t-X)$ are arbitrary left and right moving functions corresponding to the holomorphic and anti-holomorphic stress tensor. The geometry \eqref{BanadosFG} first considered in~\cite{Skenderis:1999nb} is the generalisation of the well-known Banados geometries \cite{Banados:1998gg} appropriate for a flat boundary, and reduce to them in the limit of small $X$. We obtained these expressions by solving Einstein's equations in a $1/r$ series expansion, a method reviewed in \cite{Skenderis:1999nb,deHaro:2000vlm}. We can easily convert the metric to $x$ coordinates using the relation \eqref{xX}.

If we set $t=0$ and convert $X$ back to $x$, the resulting metric on the $t=0$ slice should match with the metric on the $t_\text{new}=0$ slice of the empty AdS$_3$ or BTZ metrics. Converting them to Fefferman-Graham form allows us to read off the functions $L$ and $R$. For the AdS$_3$ case, we get that $L=R=0$. In this case, a redefinition of $r$ reproduces the metric presented in \cite{MacCormack:2018rwq}.  For BTZ we find:
\es{LR}{
L(y)=R(y)={1+r_+^2\ov (1+y^2)^2}\,.
}

\section{Asymptotic symmetries of AdS$_3$}\label{AsymptoticSymmetries}
In this appendix we present the Killing vectors of global AdS$_3$ in \cite{Hartman}: 
\begin{align}
    \zeta_{-1}&=\frac{1}{2}e^{-i(t+x)}\Big(\frac{r}{\sqrt{r^2+1}}\partial_t+\frac{\sqrt{r^2+1}}{r}\partial_x+i\sqrt{r^2+1}\partial_r\Big)\\
    \zeta_{0}&=\frac{1}{2}(\partial_t+\partial_x)\\
    \zeta_{1}&=\frac{1}{2}e^{i(t+x)}\Big(\frac{r}{\sqrt{r^2+1}}\partial_t+\frac{\sqrt{r^2+1}}{r}\partial_x-i\sqrt{r^2+1}\partial_r\Big)\\
    \overline{\zeta_{-1}}&=\frac{1}{2}e^{-i(t-x)}\Big(\frac{r}{\sqrt{r^2+1}}\partial_t-\frac{\sqrt{r^2+1}}{r}\partial_x+i\sqrt{r^2+1}\partial_r\Big)\\
    \overline{\zeta_{0}}&=\frac{1}{2}(\partial_t-\partial_x)\\
    \overline{\zeta_{1}}&=\frac{1}{2}e^{i(t-x)}\Big(\frac{r}{\sqrt{r^2+1}}\partial_t-\frac{\sqrt{r^2+1}}{r}\partial_x-i\sqrt{r^2+1}\partial_r\Big)
\end{align}
where we dropped the "new" scripts for notation simplicity. These vectors are indeed obey the Killing equations
\begin{align}
    \nabla_{(\mu}(\zeta_i)_{\nu)}=0 && \nabla_{(\mu}(\overline{\zeta_i})_{\nu)}=0 && (i=0,\pm 1)\,.\label{KillingEqn}
\end{align}
The Killing vectors obey the $SL(2,\mathbb{R})_L\times SL(2,\mathbb{R})_R$ algebra:
\begin{align}
    i\{\zeta_{1},\zeta_{-1}\}=2\zeta_{0} && i \{\zeta_{1},\zeta_{0}\}=\zeta_{1} && i\{\zeta_{-1},\zeta_{0}\}=-\zeta_{-1}\label{4.9}\\    
    i \{\overline{\zeta_{1}},\overline{\zeta_{-1}}\}=2\overline{\zeta_{0}} && i \{\overline{\zeta_{1}},\overline{\zeta_{0}}\}=\overline{\zeta_{1}} && 
    i \{\overline{\zeta_{-1}},\overline{\zeta_{0}}\}=-\overline{\zeta_{-1}}\label{4.10}
\end{align}
where $\{.\,,.\}$ here denotes the Lie bracket and hence we have extra $i$'s on the LHS compared to when writing commutators.

\section{Geodesic equations}
In this appendix, we study null and spacelike geodesic equations, on pure AdS$_3$ and BTZ black holes, respectively. Our method is largely based on \cite{Hubeny:2012ry}. We will be working in geometries in the $(t_{{\rm new}},x_{{\rm new}},r_{{\rm new}})$ coordinate throughout this appendix, so we will drop the "new" subscript for notation simplicity. 

We consider Lorentzian (2+1)-dimensional metric: 
\begin{align}
    ds^2=-f(r)dt^2+\frac{dr^2}{f(r)}+r^2dx^2
\end{align}
where $\lim_{r\to\infty}f(r)=r^2 $ so that the spacetime is asymptotically AdS$_3$. As the metric is independent of $t$ and $x$, there are two conserved quantities: 
\begin{align}
    E=f(r) \frac{dt}{d\lambda} && L=r^2 \frac{dx}{d\lambda} \label{conservedQuantities}
\end{align}
which correspond to the energy and angular momentum, respectively. The norm of the tangent vector is: 
\begin{align}
    \kappa&=g_{\mu\nu}\frac{dx^{\mu}}{d\lambda}\frac{dx^{\nu}}{d\lambda}=-\frac{1}{f(r)}E^2+\frac{1}{f(r)}\Big(\frac{dr}{d\lambda}\Big)^2+\frac{1}{r^2}L^2\label{normTan}
\end{align}
where $\kappa=+1, 0,-1$ for spacelike, null, and timelike geodesics, respectively. We will focus on null and spacelike geodesics. Equation \eqref{normTan} can be put into the general form 
\begin{align}
    \Big(\frac{dr}{d\lambda}\Big)^2+V(r)=0 && V(r)=-E^2-\kappa f(r)+L^2\frac{f(r)}{r^2}\label{effPotential}
\end{align}
where $V(r)$ can be thought of as an effective potential. Geodesic motion requires $\big(\frac{dr}{d\lambda}\big)^2\ge 0$. Consequently, the condition for a geodesic to reach the boundary is $\lim_{r\to\infty}V(r)\leq 0$; for a geodesic emanating from a point on the boundary, the deepest distance, i.e.~the smallest $r$ it can reach $r_*$ is given by the largest positive zero of the effective potential $V(r)$ \cite{Hubeny:2012ry}.  
 
\subsection{Null geodesics}\label{Null Geodesics}
For null geodesic, we have $\kappa=0$ in \eqref{effPotential}, and the two conserved quantities can be combined to a single parameter $\ell=L/E$ by rescaling the affine parameter $\lambda$ \cite{Hubeny:2012ry}. \eqref{effPotential} then becomes  
\begin{align}
    \Big(\frac{dr}{d\lambda}\Big)^2+\ell^2\frac{f(r)}{r^2}-1=0\,.\label{nullGeod}
\end{align}
From \eqref{nullGeod}, it is clear that in order for the null geodesic to reach the boundary, one must have $\ell^2\leq 1$; $\ell^2=1$ corresponds to the case that the null geodesic stays on the boundary \cite{Hubeny:2012ry}. 
\subsubsection{Pure AdS$_3$}
\label{subsubsec:PureAdS}
For pure AdS$_3$, we have $f(r)=r^2+1$ in \eqref{nullGeod}, which integrates to 
\begin{align}
    r(\lambda)=\sqrt{(1-\ell^2)\ \lambda^2+\frac{\ell^2}{1-\ell^2}}  && \lambda\in \big[0,\infty\big)\label{AdSnullGeod}
\end{align}
The null geodesic reaches the boundary, i.e.~$r\to\infty$ when $\lambda\to\infty$. 

Substituting \eqref{AdSnullGeod} into the two conserved quantities \eqref{conservedQuantities} with $f(r)=r^2+1$, one finds 
\begin{align}
    E&=(r^2+1) \frac{dt}{d\lambda}=\frac{(1-\ell^2)^2\ \lambda^2+1}{1-\ell^2} \frac{dt}{d\lambda}=1\\
    L&=r^2 \frac{dx}{d\lambda}=\frac{(1-\ell^2)^2\ \lambda^2+\ell^2}{1-\ell^2} \frac{dx}{d\lambda}=\ell
\end{align}
which integrate to 
\begin{align}
    t(\lambda)&={\rm arctan}[(1-\ell^2)\lambda]+C_t\\
    x(\lambda)&={\rm arctan}\Big[\frac{1}{\ell}(1-\ell^2)\lambda\Big]+C_x
\end{align}
When the null geodesic approaches the boundary 
\begin{align}
    \lim_{\lambda\to\infty}  t(\lambda) = C_t +\frac{\pi}{2} =y&& \lim_{\lambda\to\infty}  x(\lambda) = C_x +\frac{\pi}{2}=y
\end{align}
The other half of the triangle can be studied by reflection, i.e.~by performing a change of variable $x'=2\pi-x\in[0,\pi]$. The general expressions of null geodesics equations are therefore 
\begin{equation}
\left\{
    \begin{aligned}
    r(\lambda)&=\sqrt{(1-\ell^2)\ \lambda^2+\frac{\ell^2}{1-\ell^2}}\\
    t(\lambda)&={\rm arctan}[(1-\ell^2)\lambda]+y-\frac{\pi}{2}\\
    x(\lambda)&={\rm arctan}\Big[\frac{1}{\ell}(1-\ell^2)\lambda\Big]+y-\frac{\pi}{2}
\end{aligned}
\right.
\end{equation}
From the expression of $x(\lambda)$, we can solve for $\ell$ 
\begin{align}
    -\cot (x-y)&=\frac{1}{\ell}(1-\ell^2)\lambda\label{B16}\\
    \ell&=\frac{r}{\sqrt{1+r^2+\cot ^2(x-y)}}\leq 1\label{ell}
\end{align}
Plugging \eqref{B16}, \eqref{ell} and $r(\lambda)$ into $t(\lambda)$
\begin{align}
    t(r,x,y)&={\rm arctan}\Big[\ell\cdot\frac{1}{\ell}(1-\ell^2)\lambda\Big]+y-\frac{\pi}{2}\\
    &={\rm arctan}\Big[\frac{-r\cot (x-y)}{\sqrt{r^2+\csc ^2(x-y)}}\Big]+y-\frac{\pi}{2}
\end{align}
which is repeated  in the main text in \eqref{AdSNullGeod}. It is easy to see that $t$ increases monotonically with respect to $y$, 
\begin{align}
    \frac{dt}{dy}=1-\underbrace{\frac{r}{\sqrt{r^2+\csc ^2(x-y)}}}_{\leq 1}\ge 0\label{AdSMono}
\end{align}
So $t$ maximises at $y=\pi$. 
\begin{align}
    t(r,x,y=\pi)=-{\rm arctan}\Big[\frac{r\cot x}{\sqrt{r^2+\csc ^2x}}\Big]+\frac{\pi}{2}
\end{align}
Further simplifications depend on the range of $x$: 
\begin{itemize}
    \item For $\cot x>0$, i.e.~$0<x<\pi/2$, we have 
    \begin{align}
    t(r,x)={\rm arctan}\Big[\frac{\sqrt{r^2+\csc ^2x}}{r\cot x}\Big]=\arccos\Big[\frac{r}{\sqrt{r^2+1}}\cos x\Big]
\end{align}
    \item For $\cot x<0$, i.e.~$\pi/2<x<\pi$, we have
    \begin{align}
    t(r,x)&=-{\rm arctan}\Big[\frac{r\cot x}{\sqrt{r^2+\csc ^2x}}\Big]-\frac{\pi}{2}+\pi={\rm arctan}\Big[\frac{\sqrt{r^2+\csc ^2x}}{r\cot x}\Big]+\pi\nonumber\\
    &=\arccos\Big[\frac{r}{\sqrt{r^2+1}}\cos x\Big]
\end{align}
\end{itemize}
both of which are \eqref{AdSKilling2} in the main text. 

\subsubsection{BTZ}\label{BTZNullGeod}
\label{subsubsec:BTZ}
For BTZ, we have $f(r)=r^2-r_+^2$ in \eqref{nullGeod}, which integrates to 
\begin{align}
    r(\lambda)=\sqrt{(1-\ell^2)\ \lambda^2-\frac{\ell^2r_+^2}{1-\ell^2}}  && \lambda\in \Big[\frac{\ell r_+}{1-\ell^2},\infty\big)\label{BTZnullGeod}
\end{align}
The null geodesic reaches the boundary, i.e.~$r\to\infty$ when $\lambda\to\infty$. 

Substituting \eqref{BTZnullGeod} into the two conserved quantities \eqref{conservedQuantities} with $f(r)=r^2-r_+^2$, we find 
\begin{align}
    E&=(r^2-r_+^2) \frac{dt}{d\lambda}=\frac{(1-\ell^2)^2\ \lambda^2-r_+^2}{1-\ell^2} \frac{dt}{d\lambda}=1\\
    L&=r^2 \frac{dx}{d\lambda}=\frac{(1-\ell^2)^2\ \lambda^2-\ell^2r_+^2}{1-\ell^2} \frac{dx}{d\lambda}=\ell
\end{align}
which outside the horizon, $r\ge r_+$ for integrate to:
    \begin{align}
        t(\lambda)&=\frac{1}{2 r_+}\log \frac{(1-\ell^2)\ \lambda-r_+}{(1-\ell^2)\ \lambda+r_+}+C_t\\
        x(\lambda)&=\frac{1}{2 r_+}\log \frac{(1-\ell^2)\ \lambda-\ell r_+}{(1-\ell^2)\ \lambda+\ell r_+}+C_x
    \end{align}

When the null geodesic approaches the boundary  
\begin{align}
    \lim_{\lambda\to\infty}  t(\lambda) =y= C_t && \lim_{\lambda\to\infty}  x(\lambda) = y= C_x
\end{align}
The general expressions of null geodesics equations outside the horizon are therefore 
\begin{equation}
\left\{
    \begin{aligned}
    r(\lambda)&=\sqrt{(1-\ell^2)\ \lambda^2-\frac{\ell^2r_+^2}{1-\ell^2}}\\
    t(\lambda)&=\frac{1}{2 r_+}\log \frac{(1-\ell^2)\ \lambda-r_+}{(1-\ell^2)\ \lambda+r_+}+y\\
    x(\lambda)&=\frac{1}{2 r_+}\log \frac{(1-\ell^2)\ \lambda-\ell r_+}{(1-\ell^2)\ \lambda+\ell r_+}+y
\end{aligned}
\right.
\end{equation}
From the expression of $x(\lambda)$, we can solve for $\ell$ 
\begin{align}
    \ell=\frac{(-1+e^{2r_+(x-y)})r}{\sqrt{(-1+e^{2r_+(x-y)})^2r^2+4r_+^2e^{2r_+(x-y)}}}\leq 1\label{B37}
\end{align}
Plugging \eqref{B37} and $r(\lambda)$ into $t(\lambda)$ gives
\begin{align}
    t(r,x,y)=\frac{1}{2r_+}\log \frac{r\sqrt{1+r_+^2\sinh ^2(x-y)}-r_+\sqrt{1+r^2\sinh ^2(x-y)}}{r\sqrt{1+r_+^2\sinh ^2(x-y)}+r_+\sqrt{1+r^2\sinh ^2(x-y)}}+y
\end{align}
which is \eqref{BTZNullGeod1} in the main text. It is easy to see that $t$ increases monotonically with respect to $y$ outside the BTZ horizon
\begin{align}
    \frac{dt}{dy}\propto 4r(r^2-r^2_+)\ \sinh (x-y)\ge0\label{BTZMono}
\end{align}
So $t$ is maximised at $y=\pi$. 
\begin{align}
    t(r,x,y=\pi)=\frac{1}{2r_+}\log \frac{r\sqrt{1+r_+^2\sinh ^2(x-\pi)}-r_+\sqrt{1+r^2\sinh ^2(x-\pi)}}{r\sqrt{1+r_+^2\sinh ^2(x-\pi)}+r_+\sqrt{1+r^2\sinh ^2(x-\pi)}}+\pi
\end{align}
which is \eqref{BTZNullGeod2} in the main text. 

\subsection{Spacelike geodesics}\label{SpacelikeGeodesics}
For spacelike geodesic, we have $\kappa=1$ in \eqref{normTan}
\begin{align}
    \Big(\frac{dr}{d\lambda}\Big)^2+V(r)=0 && V(r)=-f(r)-E^2+L^2\frac{f(r)}{r^2}=0\label{spacelikeGeod}
\end{align}
To compute the holographic entanglement entropy, one needs to compute the length of the spacelike geodesic obeying the appropriate homology constraints. In our case, the spacelike geodesic equation is for non-equal time. The length of the spacelike geodesic is given in terms of the temporal and angular separation of the two points on the boundary on which our geodesic anchors. 

The length of the spacelike geodesic is given by 
\begin{align}
    \mathcal{L}=2\int_{r_*}^{\infty}d\lambda
\end{align}
with the factor of two coming from symmetry. But this is of course infinite; to get a finite result, we impose the large-radius cutoffs $R_1$, $R_2$ at the two geodesic endpoints as in \cite{Hubeny:2012ry}. Note that from \eqref{effPotential}, one can change integration variable $d\lambda=\frac{dr}{\sqrt{-V(r)}}$, which is appropriate since $V(r)<0$ in $r\in(r_*,\infty)$. Thus, we have
\begin{align}
    \mathcal{L}_{{\rm reg}}=\Big(\int_{r_*}^{R_1}+\int_{r_*}^{R_2}\Big)\frac{dr}{\sqrt{-V(r)}}\label{lReg}
\end{align}
and the universal diverging piece is simply $\log 4R_1R_2$, since $V(r)\propto -r^2$ for spacelike geodesics as $r$ gets large in \eqref{spacelikeGeod}. 

The temporal and angular separation are then given by \cite{Hubeny:2012ry}
\begin{align}
    \Delta t&=2\int_{t_*}^{\infty}dt=2\int_{\lambda_*}^{\infty}\frac{dt}{d\lambda}d\lambda=2\int_{\lambda_*}^{\infty}\frac{E}{f(r)} d\lambda=2\int_{r_*}^{\infty}\frac{E}{f(r)\sqrt{-V(r)}}dr\label{Deltat}\\
    \Delta x&=2\int_{x_*}^{\infty}dx=2\int_{\lambda_*}^{\infty}\frac{dx}{d\lambda}d\lambda=2\int_{\lambda_*}^{\infty}\frac{L}{r^2} d\lambda=2\int_{r_*}^{\infty}\frac{L}{r^2\sqrt{-V(r)}}dr\label{Deltax}
\end{align}
where the conserved quantities \eqref{conservedQuantities} have been substituted in. 

\subsubsection{Pure AdS$_3$}\label{AdSSpaceLikeGeod}
For pure AdS$_3$, we have $f(r)=r^2+1$ in \eqref{spacelikeGeod}. The relevant quantities needed in our calculation have been worked out in \cite{Hubeny:2012ry}. The minimal radius the spacelike geodesic can reach is
\begin{align}
    r_*=\sqrt{\frac{1}{2}\Big[-(E^2-L^2+1)+\sqrt{(E^2-L^2+1)^2+4L^2}\Big]}\label{r*AdS}
\end{align}
The regulated length of the spacelike geodesic is
\begin{align}
    \mathcal{L}_{\rm reg}=\Big(\int_{r_*}^{R_1}+\int_{r_*}^{R_2}\Big)\frac{dr}{\sqrt{-V(r)}}=-\log \sqrt{(E^2-L^2+1)^2+4L^2}+\log (4R_1R_2)\label{D45}
\end{align}
where we added the universal diverging term $\log (4R_1R_2)$ that \cite{Hubeny:2012ry} subtracted off, as it will play a role in our analysis later on. The temporal and angular distance between the two geodesic endpoints on the boundary are
\begin{align}
    \Delta t&=2\int_{r_*}^{\infty}\frac{E}{f(r)\sqrt{-V(r)}}dr=\frac{\pi}{2}+\arcsin \Big[\frac{E^2-L^2-1}{\sqrt{(E^2-L^2-1)^2+4E^2}}\Big]\\
    \Delta x&=2\int_{r_*}^{\infty}\frac{L}{r^2\sqrt{-V(r)}}dr=\frac{\pi}{2}+\arcsin \Big[\frac{E^2-L^2+1}{\sqrt{(E^2-L^2+1)^2+4L^2}}\Big]
\end{align}
which are \eqref{delta_tAdS} and \eqref{delta_xAdS} in the main text. Note that $(E^2-L^2-1)^2+4E^2=(E^2-L^2+1)^2+4L^2$. 

From \eqref{delta_tAdS} and \eqref{delta_xAdS}, we can solve for $E$ and $L$:
\begin{align}
    E=\frac{\sin \Delta t}{\cos \Delta t-\cos \Delta x} && 
    L=\frac{\sin \Delta x}{\cos \Delta t-\cos \Delta x}\label{ELAdS}
\end{align}
Plugging \eqref{ELAdS} into \eqref{D45}, we have, 
\begin{align}
    \mathcal{L}_{{\rm reg}}&=\log \Big[\frac{1}{2}\big|\cos \Delta t-\cos \Delta x\big|\Big]+\log (4R_{{1}}R_{{2}})\\
    &=\log \Big[4\Big|\sinh 
    \Big(\frac{w_{1}-w_{2}}{2}\Big)\ \sinh \Big(\frac{\overline{w}_{1}-\overline{w}_{2}}{2}\Big)\Big|\ \Big]+\log (R_{{1}}R_{{2}})
\end{align}
where we have used \eqref{7.1} in getting from the second to the third line. Restoring the "new" subscript then gives \eqref{7.7} in the main text. 

\subsubsection{BTZ}\label{BTZSpaceLikeGeod}
For BTZ, we have $f(r)=r^2-r_+^2$ in \eqref{spacelikeGeod}. Let us consider the cases in which the spacelike geodesic never enters the black hole. In such a case, $V(r)\ge 0$ for some $r>r_+$ \cite{Hubeny:2012ry}. To see what this indicates more concretely, we first note that $V(r)$ peaks at $r=\sqrt{L r_+}$: 
\begin{align}
    \frac{d V(r)}{dr}=-2r+\frac{2L^2 r_+^2}{r^3} && \frac{d V(r)}{dr}=0\Rightarrow r=\sqrt{L r_+} 
\end{align}
where one needs $\sqrt{L r_+}\ge r_+$, i.e.~$L\ge r_+$. The condition for the maximal value of $V(r)$, that is $V(\sqrt{L r_+})$, to be non-negative is 
\begin{align}
    V(\sqrt{L r_+})=-E^2+(L-r_+)^2\ge0 \qquad \Rightarrow \qquad L\ge E+r_+
\end{align}
which is more stringent than $L\ge r_+$. Henceforth, the criterion for the spacelike geodesic not entering the BTZ horizon is  $L\ge E+r_+$. The limiting case $L=r_+$ when $E=0$ corresponds to the known case that the spacelike geodesic gets trapped in a circular orbit at the horizon, which is studied in \cite{Hubeny:2012ry}. At the minimal radius the spacelike geodesic can reach, $dr/d\lambda=0$, so the minimal radius is given by the larger positive zero of $V(r)$ 
\begin{align}
    0&=r^2-r_+^2+E^2-L^2\frac{r^2-r_+^2}{r^2}\\
    \Rightarrow r_*&=\sqrt{\frac{1}{2}\Big[L^2-E^2+r_+^2+\sqrt{(L^2-E^2+r_+^2)^2-4r_+^2L^2}\Big]}\label{r_*BTZ}
\end{align}
One can verify that for $L= E+r_+$, $r_*=\sqrt{r_+(r_++E)}> r_+$; increasing $L$ results in increasing $r_*$. Hence $r_*> r_+$ for $L\ge E+r_+$. By setting $r_+^2=-1$ in \eqref{r_*BTZ}, we recover \eqref{r*AdS} in pure AdS$_3$. 

The regulated geodesic length is then given by integrating \eqref{lReg}
\begin{align}
    \mathcal{L}_{\rm reg}=\Big(\int_{r_*}^{R_1}+\int_{r_*}^{R_2}\Big)\frac{dr}{\sqrt{-V(r)}}=-\log \sqrt{(L^2-E^2+r_+^2)^2-4r_+^2L^2}+\log 4R_1R_2\label{D53}
\end{align}
Note that by setting the BTZ black hole radius $r_+^2=-1$ in \eqref{D53}, we recover \eqref{D45} in pure AdS$_3$. The temporal and angular distance between the two geodesic endpoints on the boundary are given by integrating  \eqref{Deltat} and \eqref{Deltax}, respectively
\begin{align}
    \Delta t&=2\int_{r_*}^{\infty}\frac{E}{f(r)\sqrt{-V(r)}}dr\\
    &=\frac{1}{r_+}\log \Big[\frac{\sqrt{(L^2-E^2+r_+^2)^2-4r_+^2E^2}}{(L-E-r_+)(L+E+r_+)}\big]\\
    &=\frac{1}{r_+}{\rm arccosh}\Big[\frac{L^2-E^2-r_+^2}{\sqrt{(L^2-E^2-r_+^2)^2-4r_+^2E^2}}\Big]\\
    \Delta x&=2\int_{r_*}^{\infty}\frac{L}{r^2\sqrt{-V(r)}}dr\\
    &=\frac{1}{r_+}\log \Big[\frac{\sqrt{(E^2-L^2-r_+^2)^2-4r_+^2L^2}}{(L+E-r_+)(L-E-r_+)}\big]\nonumber\\
    &=\frac{1}{r_+}{\rm arccosh}\Big[\frac{L^2-E^2+r_+^2}{\sqrt{(L^2-E^2+r_+^2)^2-4r_+^2L^2}}\Big]\label{D61}
\end{align}
which are \eqref{delta_tBTZ} and \eqref{delta_xBTZ} in the main text. Note that $(L^2-E^2+r_+^2)^2-4r_+^2L^2=(L^2-E^2-r_+^2)^2-4r_+^2E^2$. One can verify that when $E=0$, \eqref{D61} becomes 
\begin{align}
    \Delta x=\frac{1}{r_+}{\rm arccosh}\Big(\frac{L^2+r_+^2}{L^2-r_+^2}\Big)=\frac{2}{r_+}{\rm arctanh}\Big(\frac{r_+}{L}\Big)
\end{align}
which matches with a known result in section 2.3 of \cite{Hubeny:2012ry}.

From \eqref{delta_tBTZ} and \eqref{delta_xBTZ}, one can solve for $E$ and $L$
\begin{align}
    E=r_+\frac{\sinh (r_+\Delta t)}{\cosh  (r_+\Delta t)-\cosh  (r_+\Delta x)} && 
    L=r_+\frac{\sinh (r_+\Delta x)}{\cosh  (r_+\Delta t)-\cosh  (r_+\Delta x)}\label{ELBTZ}
\end{align}
Plugging \eqref{ELBTZ} into \eqref{D53}, we have,
\begin{align}
    \mathcal{L}_{{\rm reg}}
    &=\log \Big[\Big(\frac{2}{r_+}\Big)^2\frac{1}{2}\big|\cosh  (r_+\Delta t)-\cosh  (r_+\Delta x)\big|\Big]+\log (R_{{1}}R_{{2}})\label{7.22}\\
    &=\log \Big[\Big(\frac{2}{r_+}\Big)^2\Big|\sin 
    \Big(r_+\frac{w_{1}-w_{2}}{2}\Big)\ \sin \Big(r_+\frac{\overline{w}_{1}-\overline{w}_{2}}{2}\Big)\Big|\ \Big]+\log (R_{{1}}R_{{2}})
\end{align}
where we have used \eqref{7.1} in getting from the second to the third line. This then gives \eqref{7.18} in the main text with the "new" subscript restored. 

\section{Time evolution of horizons}\label{ComparisonRyu}
\subsection{SSD CFT}
In this appendix, we show how the coordinates in Figure \ref{fig:HorCompare} are established. In \cite{Goto:2021sqx}, a redefinition of the radial coordinate 
\begin{align}
    r'=\sqrt{\frac{\partial w_{{\rm new}}}{\partial w}\frac{\partial\overline{w}_{{\rm new}}}{\partial\overline{w}}}r_{{\rm new}}=\frac{1}{\sqrt{4t^2\sin ^4\frac{ x}{2}+\Big(1- t^2 \sin ^2\frac{ x}{2}\Big)^2}}r_{{\rm new}}\label{r'}
\end{align}
was introduced, and the BTZ black hole horizon was plotted on constant $t$ slices in the $(t, x, r')$ coordinate. In order to compare with their result, we adopt the same $(t, x, r')$ coordinate when plotting Figure \ref{fig:HorCompare}. Notice that unlike \eqref{BanadosFG}, the metric in the $(t,x,r')$ coordinate is not in the Fefferman-Graham gauge \cite{Skenderis:1999nb,deHaro:2000vlm}, as $r'$ is $t$ and $x$ dependent, leading to non-zero $dr'dt$ and $dr'dx$ terms in the metric. Therefore quantitative comparisons with the boundary stress tensor are in general hard to achieve. 

\subsection{Floquet CFT}\label{ComparisonRyuFloquet}

As for Floquet CFT, we introduce the analogue of \eqref{r'} at stroboscopic time $t=n(T_0+T_1)$ to plot Figure \ref{fig:FloquetHor}. 
\begin{align}
    r'&=\sqrt{\frac{\partial w_n}{\partial w}\frac{\partial \overline{w}_n}{\partial \overline{w}}}r_n=\frac{e^{ix} |\left(\eta ^n-1\right)^2-\left(\gamma _1 \eta ^n-\gamma _2\right) \left(\gamma _2 \eta ^n-\gamma
   _1\right)|}{\sqrt{\Xi(x,t)}}r_n\nonumber\\
   \Xi(x,t)&=\left(e^{ix} \left(\gamma _1 \eta ^n-\gamma _2\right)-\eta ^n+1\right) \left(\gamma _2+e^{ix} \left(\eta
   ^n-1\right)-\gamma _1 \eta ^n\right)\nonumber\\
   &\left(\gamma _1+e^{ix} \left(\eta ^n-1\right)-\gamma _2 \eta ^n\right) \left(e^{ix}
   \left(\gamma _2 \eta ^n-\gamma _1\right)-\eta ^n+1\right)\label{r'Floquet}
\end{align}
where $r_n$=$r_{{\rm new}}|_{t=n(T_0+T_1)}$. One can verify that $\sqrt{\frac{\partial w_n}{\partial w}\frac{\partial \overline{w}_n}{\partial \overline{w}}}$ grows as $\eta^{-n}$ at at $z=\gamma_1$ and $\overline{z}=\gamma_1$, i.e.~$x=\pm \frac{1}{ i}\log \gamma_1$. 

\section{Review of Floquet CFT}\label{FloquetReview}
\subsection{Floquet CFT at stroboscopic time}
\subsubsection{First cycle}\label{Floquet1st}
In the first cycle, the Floquet Hamiltonian is: 
\begin{equation}
\label{eq6}
H(t)=
\left\{
\begin{aligned}
H_1&=2\int_0^{2\pi} dx\, \sin ^2\Big(\frac{x}{2}\Big)\,T_{00}(x),\ \ &&  0<t<T_1\\
H_0&=\int_0^{2\pi} dx\, T_{00}(x),\ \ && T_1<t<T_1+T_0\\
\end{aligned}
\right.
\end{equation}
Each step can be understood separately in Euclidean signature and by working in the complex $z$-plane. We then analytically continue to Lorentzian time.

\begin{itemize}
    \item In the first step, $H_0$ simply acts as $\emph{dilation}$, and 
    \begin{equation}
        \left\{
        \begin{aligned}
            z_{{\rm new},0}&=e^{\tau}z\\
            \overline{z}_{{\rm new},0}&=e^{\tau}\overline{z}
        \end{aligned}
        \right.
    \end{equation}
    \item In the second step, $H_1$ acts as $SL(2)$ $\emph{transformation}$ $\emph{on}$ $z_{{\rm new},0}$ $\emph{and}$ $\overline{z}_{{\rm new},0}$ according to \eqref{znew} and \eqref{znewbar}, 
        \begin{equation}
        \left\{
        \begin{aligned}
            z_{{\rm new},1}&=\frac{(1+\frac{\tau}{2})(e^{\tau}z)-\frac{\tau}{2}}{\frac{\tau}{2}(e^{\tau}z)+(1-\frac{\tau}{2})}\\
            \overline{z}_{{\rm new},1}&=\frac{(1+\frac{\tau}{2})(e^{\tau}\overline{z})-\frac{\tau}{2}}{\frac{\tau}{2}(e^{\tau}\overline{z})+(1-\frac{\tau}{2})}
        \end{aligned}
        \right.
    \end{equation}
    
\end{itemize}
After the completion of one cycle, and analytic continue $\tau=it$, we have \cite{Wen:2018agb,Fan:2019upv}
\begin{align}
    z_1=f(z)=\frac{az+b}{cz+d} && 	\mathfrak{H}=\begin{pmatrix}
a & b \\
c & d 
\end{pmatrix}
\label{z1}
\end{align}
where
\begin{align}
    a=\Big(1+i\frac{T_1}{2}\Big)e^{\frac{i T_0}{2}} && b= -i\frac{ T_1}{2}e^{-i\frac{T_0}{2}} && c=i{\frac{T_1}{2}e^{i\frac{T_0}{2}}} && d=\Big(1-i\frac{T_1}{2}\Big)e^{-i\frac{T_0}{2}}
\end{align}
We further introduce the quantity $\Delta$ through the square of the trace of $\mathfrak{H}$ \cite{Wen:2018agb}, 
\begin{align}
    ({\rm Tr}\mathfrak{H})^2&=(a+d)^2=4(1-\Delta)\\
    \Delta&=\Big(1-\Big(\frac{T_1}{2}\Big)^2\Big)\ \sin ^2\frac{T_0}{2}+\frac{T_1}{2}\ \sin T_0\label{Delta}
\end{align}
The sign of $\Delta$ determines different types of Mobius transformations. 
\subsubsection{$n$-cycle}\label{Floquetnth}
Now let us repeat the first cycle for $n$ times, with the time evolution operator given by \eqref{FloquetTimeEvoOp}. $n$-cycle driving can be represented via a Mobius transformation in the normal form \cite{Wen:2018agb,Fan:2019upv}, 
\begin{align}
    \frac{z_n-\gamma_1}{z_n-\gamma_2}=\eta^n\  \frac{z-\gamma_1}{z-\gamma_2}\label{znOrg}
\end{align}
where $\gamma_{1}$ and $\gamma_{2}$ are the $\emph{fixed points}$ of the Mobius transformation, defined as points that are $\emph{invariant}$ under the $SL(2)$ transformation \eqref{z1}: $f(\gamma_i)=\gamma_i, (i=1,2)$; and $\eta$ is given by $\eta=f'(z)|_{z=\gamma_1}$. From \eqref{znOrg} 
one can solve for $z_n$,  
\begin{align}
    z_n=\frac{(\gamma_1-\eta^n\gamma_2)z-(1-\eta^n)\gamma_1\gamma_2}{(1-\eta^n)z-(\gamma_2-\eta^n\gamma_1)}=:\frac{A_nz+B_n}{C_nz+D_n}\label{AnBnCnDn}
\end{align}
And the two fixed points $\gamma_{1}$ and $\gamma_{2}$ are determined by 
\begin{align}
    f(\gamma_i)=\frac{a\gamma_i+b}{c\gamma_i+d}=\gamma_i\ \Rightarrow\ \gamma_{1,2}=\frac{(a-d)\mp\sqrt{(a-d)^2+4bc}}{2c}=\frac{(a-d)\mp2\sqrt{-\Delta}}{2c}\label{gammai}
\end{align}
where we have used $(a-d)^2+4bc=-4\Delta$. 
$\eta$ can be obtained from 
\begin{align}
    \eta=f'(z)|_{z=\gamma_1}=\frac{(a+d)+\sqrt{(a-d)^2+4bc}}{(a+d)-\sqrt{(a-d)^2+4bc}}=\frac{(a+d)+2\sqrt{-\Delta}}{(a+d)-2\sqrt{-\Delta}}=\frac{c\gamma_2+d}{c\gamma_1+d}
\end{align}
At the phase transition, $\gamma_1=\gamma_2=(a-d)/2c$, and \eqref{AnBnCnDn} does not apply anymore. Instead we have 
\begin{align}
    \frac{1}{z_n-\gamma}+n=\frac{1}{z-\gamma}+n\cdot \mathfrak{b} \label{znOrg2}
\end{align}
where $\mathfrak{b}$ is the translation length, $\mathfrak{b}=\frac{a-d}{2c}$ \cite{Fan:2019upv}. From \eqref{znOrg2}, one can solve for $z_n$,
\begin{align}
    z_n=\frac{(1+n\mathfrak{b}\cdot\gamma)z-n\mathfrak{b}\cdot\gamma^2}{n\mathfrak{b}\cdot z+(1-n\mathfrak{b}\cdot\gamma)}=:\frac{A_nz+B_n}{C_nz+D_n}\label{ABCDphaseTransition}
\end{align}
Different types of Mobius transformations are characterised by the sign of $\Delta$ \eqref{Delta} 
\begin{itemize}
    \item $\Delta>0$: $\eta$ is a phase. This case corresponds to the $\emph{non-heating}$ phase.  
    \item $\Delta<0$: $\eta$ is a $\emph{real}$ number.
    This case corresponds to the $\emph{heating}$ phase. $\eta$ can be either in $(0,1)$ or $(1,\infty)$. From \eqref{znOrg} we see that if $0<\eta<1$, then at large $n$, i.e.~late stroboscopic time, $z_n$ converges to $\gamma_1$ if $z\neq \gamma_2$ and stays at $\gamma_2$ if $z=\gamma_2$; on the other hand, if $\eta>1$, then $z_n$ converges to $\gamma_2$ if $z\neq \gamma_1$ and stays at $\gamma_1$ if $z=\gamma_1$. Therefore, the range of $\eta$ in $(0,1)$ or $(1,\infty)$ is simply a matter of convention, for we can always take $\eta\to\frac{1}{\eta}$ and swap $\gamma_1\leftrightarrow \gamma_2$ to get to the other case. In our convention, we choose $0<\eta<1$, making  $\gamma_1$ the $\emph{stable}$ fixed point and $\gamma_2$ the $\emph{unstable}$ fixed point. 
    \item $\Delta=0$: $\eta=1$. This case corresponds to the $\emph{phase\ transition}$. 
\end{itemize}
\subsection{Time-reversal symmetric case}
We can start one cycle at the middle of a $H_0$-driving region as in \cite{Lapierre:2019rwj}, see Figure \ref{fig:FloquetFig} in the main text. The Hamiltonian in this setup can be written as 
\begin{equation}
\label{FloquetHamiltonianTR}
H(t)=
\left\{
\begin{aligned}
&H_0,\ \ &&0<t\ {\rm mod}(T_0+T_1)<\frac{T_0}{2}\\
&H_1,\ \  &&\frac{T_0}{2}<t\ {\rm mod}(T_0+T_1)<\frac{T_0}{2}+T_1\\
&H_0,\ \ &&\frac{T_0}{2}+T_1<t\ {\rm mod}(T_0+T_1)<T_1+T_0\\
\end{aligned}
\right.
\end{equation}
With the Hamiltonian \eqref{FloquetHamiltonianTR}, the system is invariant  under time-reversal, $t\to -t$. 
\subsubsection{First cycle}
Repeating the procedure in Appendix \ref{Floquet1st} of for the Hamiltonian \eqref{FloquetHamiltonianTR} in the first cycle, we have \cite{Lapierre:2019rwj}
\begin{align}
    z_1=f(z)=\frac{az+b}{cz+d} 
\end{align}
where
\begin{align}
    a=\Big(1+i\frac{T_1}{2}\Big)e^{\frac{iT_0}{2}} &&
b= -i\frac{T_1}{2} &&
c=i\frac{T_1}{2} &&
d=\Big(1-i\frac{T_1}{2}\Big)e^{-i\frac{T_0}{2}}\label{abcdTR}
\end{align}
\subsubsection{$n$-cycle}
The $n$-cycle case also parallels that in Appendix \ref{Floquetnth}, with the coefficients $a,b,c,d$ given by \eqref{abcdTR}. By plugging \eqref{abcdTR} into the general formula \eqref{gammai}, it is straightforward to verify that~\cite{Lapierre:2019rwj}
\begin{align}
    \gamma_1\gamma_2=1 && \Rightarrow && \gamma_2=\gamma_1^*\ ({\rm heating\ phase})
\end{align}
where we have used the fact that in the heating phase $|\gamma_1|=|\gamma_2|=1$. $\gamma_2$ is thus the $\emph{complex\ conjugate}$ of $\gamma_1$ in the heating phase. This then simplifies \eqref{AnBnCnDn} to a particularly nice form \cite{Lapierre:2019rwj}, 
\begin{align}
    A_n=\gamma_1-\eta^n\gamma_2 &&
B_n= -(1-\eta^n) &&
C_n=(1-\eta^n) &&
D_n=-(\gamma_2-\eta^n\gamma_1)\label{AnBnCnDnTR}
\end{align}
which then yields 
\begin{align}
    z_n&=\frac{A_nz+B_n}{C_nz+D_n}=\frac{(\gamma_1-\eta^n\gamma_2)z-(1-\eta^n)}{(1-\eta^n)z-(\gamma_2-\eta^n\gamma_1)}\label{zn}\\
    \overline{z}_n&=\frac{A_n\overline{z}+B_n}{C_n\overline{z}+D_n}=\frac{(\gamma_1-\eta^n\gamma_2)\overline{z}-(1-\eta^n)}{(1-\eta^n)\overline{z}-(\gamma_2-\eta^n\gamma_1)}\label{znbar}
\end{align}
where we also listed the expressions of $\overline{z}_n$ for the anti-holomorphic coordinate $\overline{z}$. 

In the phase transition case, we have, from \eqref{gammai} and \eqref{abcdTR}, 
\begin{align}
    \Delta \propto (a-d)^2+4bc=0 && \gamma=\frac{a-d}{2c}=1\label{gammaPhaseTransition}
\end{align}
i.e.~the two fixed points that are conjugate to each other merge to $z=1$ on the real axis symmetrically. 

\subsection{General cases}
In Section \ref{Floquetsetup}, we have introduced the fundamentals of Floquet CFT built upon SSD CFT with Hamiltonian $H_1$. We remark, however, that Floquet setups for CFT$_2$ could be more general. Floquet CFT based on $H_0$ and $H_{\theta}$ were studied in \cite{Wen:2020wee}. Floquet CFT with $q\ge 2$ Hamiltonian were systematically investigated in \cite{Han:2020kwp}, where it was shown that heating phases always exist, whereas non-heating phases might or might not be present subject to certain criteria \cite{Han:2020kwp}. More generally, one can construct Floquet CFT with deformed Hamiltonian involving the $\emph{full}$ Virasoro algebra (not just the $SL(2)$ sub-algebra) \cite{Fan:2020orx,Lapierre:2020ftq}. Both works indicate that much of the conclusions for Floquet CFT with $H_0$ and $H_1$ generalise to these cases at least qualitatively: the presence/absence of the fixed points (or higher-periodic points) in the stroboscopic evolution of CFT operators characterises the heating/non-heating phases; in the heating phases, energy accumulates at the unstable fixed points rapidly and forms up peaks, while entanglement entropy grows linearly in time if the subregion $A$ includes at least one (but not all) of unstable fixed points; in the non-heating phases, energy and entanglement show quasi-periodic oscillatory behavior.

\section{Details on Floquet CFT at general time}
\subsection{Proof of $z_{{{\rm new}}}$ and $\overline{z}_{{{\rm new}}}$ at general $t$ }\label{znewGeneralTimeProof}
In this appendix, we prove that the formula $\eqref{znewGeneralTime}$ correctly returns to $\eqref{zn}$ at stroboscopic time $t=n(T_0+T_1)$ for general $n\in\mathbb{Z}^+$ using mathematical induction. 

For $n=1$, a direct computation shows that 
\begin{align}
    \begin{pmatrix}
    A_1 & B_1 \\
    C_1 & D_1 
    \end{pmatrix}
=\begin{pmatrix}
    \gamma_1-\eta\gamma_2 & -(1-\eta) \\
    1-\eta & -(\gamma_2-\eta\gamma_1) 
    \end{pmatrix}
=\frac{2}{c}\frac{\sqrt{-4\Delta}}{\sqrt{-4\Delta}-(a+d)}
\begin{pmatrix}
    a & b \\
    c & d 
    \end{pmatrix}
\end{align}
where $A_1, B_1, C_1, D_1$ are given by \eqref{AnBnCnDnTR} with $n=1$, and $a,b,c,d$ are given by \eqref{abcdTR}. 
Therefore $\eqref{znewGeneralTime}$ holds after the completion of the first cycle. In other words, the expression $z_1$ \eqref{abcdTR} is consistent with $z_n$ \eqref{zn} when $n=1$,    
\begin{align}
    z_1=\frac{az+b}{cz+d}=\frac{A_1 z+B_1}{C_1 z+D_1}
\end{align}

Assume that $\eqref{znewGeneralTime}$ reproduces $\eqref{zn}$ at stroboscopic time for $n$ ($n\ge 1$), and consider the $(n+1)$-th cycle. We can then take $z_{{\rm new}}$ at the beginning of the $(n+1)$-th cycle to be $z_n$. The time evolution follows the same logic as that in Appendix \ref{Floquet1st}. After the completion of the $H_1$-driving period, i.e.~at time $t-n(T_0+T_1)=\frac{T_0}{2}+T_1$, $\eqref{znewGeneralTime}$ becomes 
\begin{align}
    z_{{\rm new}}&=\frac{\Big(1+\frac{iT_1}{2}\Big)(e^{\frac{i T_0}{2}}z_n)-\frac{iT_1}{2}}{\frac{iT_1}{2}(e^{\frac{i T_0}{2}}z_n)+\Big(1-\frac{iT_1}{2}\Big)}=e^{-\frac{i T_0}{2}}\frac{\Big(1+\frac{iT_1}{2}\Big)e^{\frac{i T_0}{2}}\cdot z_n-\frac{iT_1}{2}}{\frac{iT_1}{2}\cdot z_n+\Big(1-\frac{i T_1}{2}\Big)e^{-\frac{i T_0}{2}}}=e^{-\frac{i  T_0}{2}}\frac{az_n+b}{cz_n+d}\label{EffOfH1}
\end{align}
where $a,b,c,d$ are given by $\eqref{abcdTR}$. In other words, $H_1$ sends $e^{\frac{i T_0}{2}}z_n$ to $e^{-\frac{i T_0}{2}}z_{n+1}$ after driving for $T_1$. The remaining $H_0$-driving period that lasts for $\frac{T_0}{2}$ then eliminates the $e^{-\frac{i T_0}{2}}$ prefactor in \eqref{EffOfH1}. So, after the completion of the $(n+1)$-th cycle, $\eqref{znewGeneralTime}$ is 
\begin{align}
    z_{n+1}=\frac{az_n+b}{cz_n+d}=\frac{(aA_n+bC_n)z+aB_n+bD_n}{(cA_n+dC_n)z+cB_n+dD_n}
\end{align}
where $A_n, B_n, C_n, D_n$ are given by \eqref{AnBnCnDnTR}. A direct calculation shows that 
\begin{align}
    \begin{pmatrix}
    aA_n+bC_n & aB_n+bD_n \\
    cA_n+dC_n & cB_n+dD_n 
    \end{pmatrix}
=\frac{1}{2}(a+d-\sqrt{-4\Delta})
\begin{pmatrix}
    A_{n+1} & B_{n+1} \\
    C_{n+1} & D_{n+1} 
    \end{pmatrix}
\end{align}
with $A_{n+1}, B_{n+1}, C_{n+1}, D_{n+1}$ in \eqref{AnBnCnDnTR} with $n\to n+1$. So $z_{{\rm new}}$ after the completion of the $(n+1)$-th cycle, i.e.~at $t=(n+1)(T_0+T_1)$ correctly returns to the form \eqref{zn}, 
\begin{align}
    z_{n+1}=\frac{A_{n+1}z+B_{n+1}}{C_{n+1}z+D_{n+1}}
\end{align}

For $\overline{z}_{{\rm new}}$ \eqref{znewbarGeneralTime}, the proof parallels the above with $z\to\overline{z}$ and therefore $z_n\to\overline{z}_n$. 

\subsection{Plots of $\frac{1}{i}w_{{\rm new}}$ and $\frac{1}{i}\overline{w}_{{\rm new}}$}\label{Plotwnewwnewbar}

\begin{figure}[htbp]
\centering
\includegraphics[width=.55\textwidth]{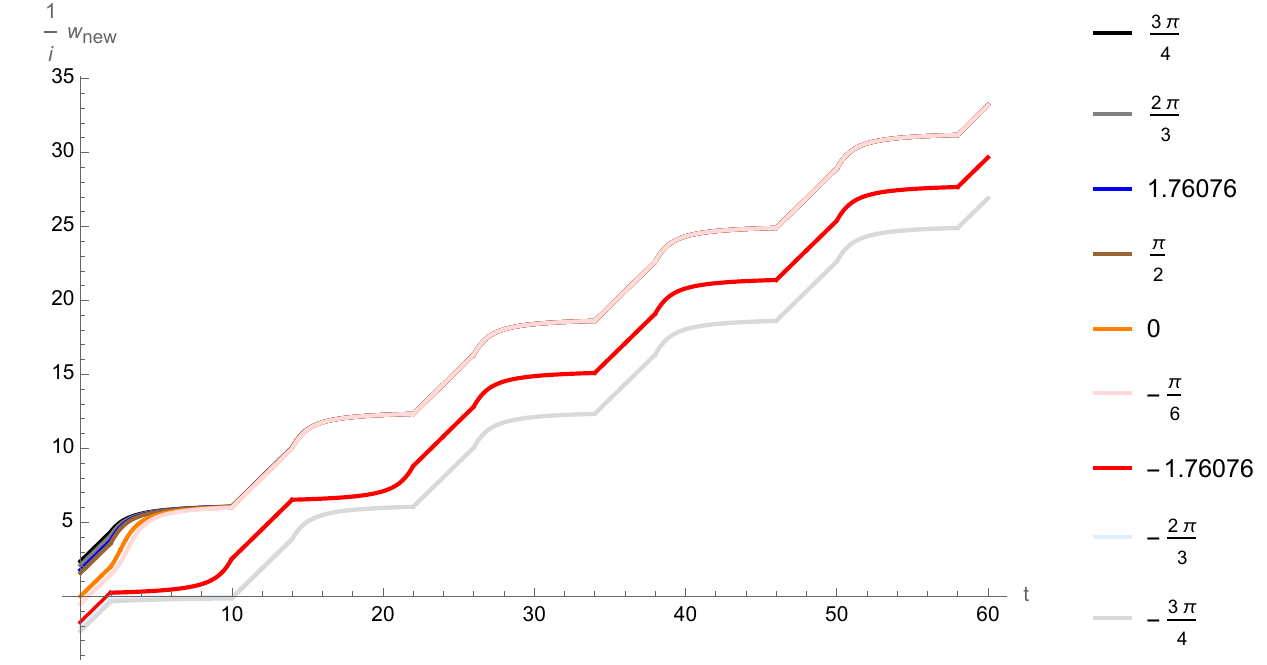}\\[4pt]
\includegraphics[width=.55\textwidth]{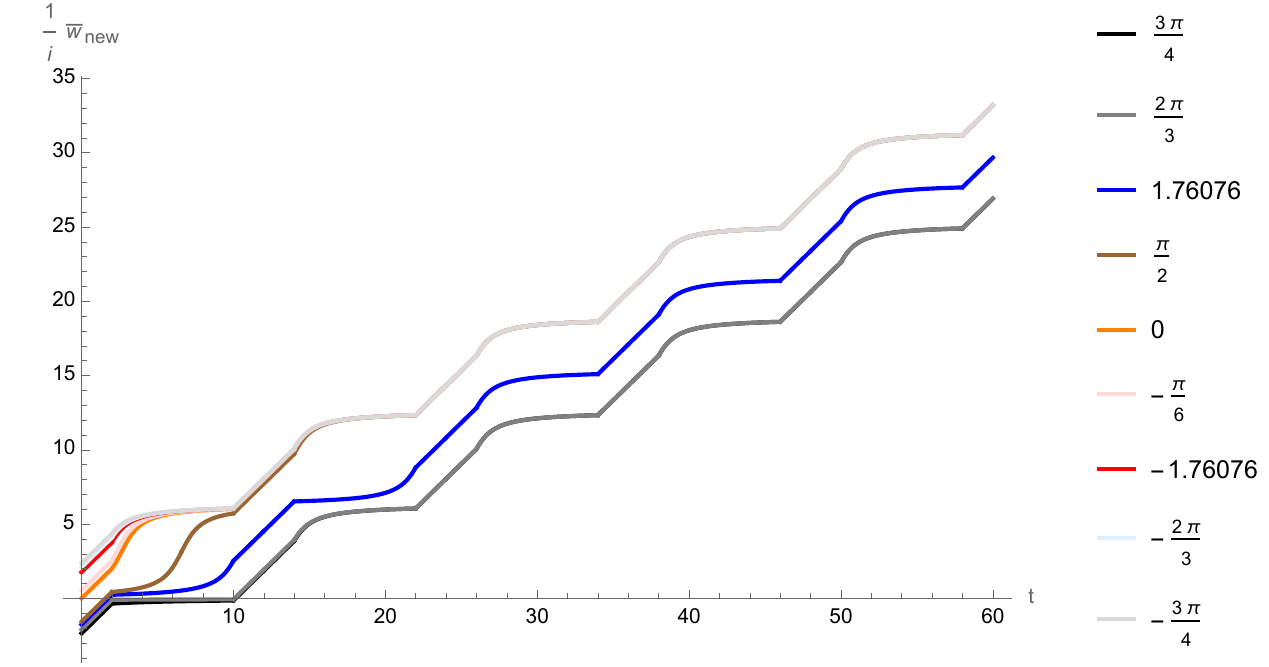}
%\qquad
\caption{Plot of $\frac{1}{i}w_{{{\rm new}}}$ and $\frac{1}{i}\overline{w}_{{{\rm new}}}$ at general $t$ in the heating phase. Here we set $T_0=4$ and $T_1=8$, and plotted the first 5 cycles for $x=-\frac{3\pi}{4},-\frac{2\pi}{3}, \frac{1}{ i}\log \gamma_2\approx -1.76076, -\frac{\pi}{6}, 0, \frac{\pi}{2}, \frac{1}{ i}\log \gamma_1\approx 1.76076, \frac{2\pi}{3}, \frac{3\pi}{4}$. \label{fig:wnewwnewbar}}
\end{figure}

In Figure \ref{fig:wnewwnewbar}, we plot $\frac{1}{i}w_{{{\rm new}}}=\frac{1}{i}\log z_{{{\rm new}}}=
\frac{1}{2}(t_{{{\rm new}}}+x_{{{\rm new}}})$ and $\frac{1}{i}\overline{w}_{{{\rm new}}}=\frac{1}{i}\log \overline{z}_{{{\rm new}}}=\frac{1}{2}(t_{{{\rm new}}}-x_{{{\rm new}}})$. To avoid confusions, we list the analogue of \eqref{znewGeneralTime} for the anti-holomorphic coordinate $\overline{z}=e^{\overline{w}}=e^{-i x}$,
\begin{equation}
\label{znewbarGeneralTime}
\overline{z}_{{{\rm new}}}=
\left\{
\begin{aligned}
& e^{i\mathfrak{t}_{n+1}}\overline{z}_n,\ \ && 0\leq \mathfrak{t}_{n+1}\leq \frac{T_0}{2}\\
& \frac{\Big(1+\frac{i(\mathfrak{t}_{n+1}-T_0/2)}{2}\Big)(e^{\frac{i T_0}{2}}\overline{z}_n)-\frac{i(\mathfrak{t}_{n+1}-T_0/2)}{2}}{\frac{i(\mathfrak{t}_{n+1}-T_0/2)}{2}(e^{\frac{i T_0}{2}}\overline{z}_n)+\Big(1-\frac{i(\mathfrak{t}_{n+1}-T_0/2)}{2}\Big)},\ \ &&\frac{T_0}{2}< \mathfrak{t}_{n+1}\leq \frac{T_0}{2}+T_1\\
& e^{i(\mathfrak{t}_{n+1}-(T_0+T_1))}\overline{z}_{n+1},\ \ &&\frac{T_0}{2}+T_1\leq \mathfrak{t}_{n+1}\leq T_0+T_1\\
\end{aligned}
\right.
\end{equation}
In the first cycle, $\overline{z}_n$ is $\overline{z}=e^{- i x}$.

To understand Figure \ref{fig:wnewwnewbar}, we first notice that $\frac{1}{i}w_{{{\rm new}}}=\frac{1}{i}\log z_{{{\rm new}}}$ is monotonically $\emph{increasing}$ with respect to $x$ when the driving Hamiltonian is $H_1$. In other words, $\frac{1}{i}w_{{{\rm new}}}$ initially with smaller $x\in[-\pi,\pi]$ cannot catch up with those with larger $x$, and $\frac{1}{i}w_{{{\rm new}}}$ curves with different $x$'s therefore do not intersect at general $t>0$. This can be proved by simply taking the $x$ derivative of \eqref{wnew}. A similar analysis on \eqref{wnewbar} shows that $\frac{1}{i}\overline{w}_{{{\rm new}}}=\frac{1}{i}\log \overline{z}_{{{\rm new}}}$ is monotonically $\emph{decreasing}$ with respect to $x$. 

It is observed that both $\frac{1}{i}w_{{{\rm new}}}$ and $\frac{1}{i}\overline{w}_{{{\rm new}}}$ in Figure \ref{fig:wnewwnewbar} asymptote three curves. To understand this phenomenon, first note that at stroboscopic time, $z_n\to\gamma_1$ for $z\neq\gamma_2$ and $z_n=\gamma_2$ for $z=\gamma_2$, see, \eqref{zngamma1}, \eqref{znewLateTime}. Thus, for initial $x\neq\frac{1}{i}\log \gamma_2$, the $\frac{1}{i}w_{{{\rm new}}}$ curve converges to that of $x=\frac{1}{i}\log \gamma_1$, whose stroboscopic values are $w_n=\frac{1}{i}\log \gamma_1$; while for $x=\frac{1}{ i}\log \gamma_2$, $\frac{1}{i}w_{{{\rm new}}}$ follows a different curve that is slightly lower (coloured red in Figure \ref{fig:wnewwnewbar}), with stroboscopic values $w_n=\frac{1}{i}\log \gamma_2$. There is, however, a subtlety, as the stroboscopic $\frac{1}{i}w_n=\frac{1}{ i}\log z_n$ only captures the $\emph{principal}$ value of $w_{{{\rm new}}}$. Due to the monotonicity of $\frac{1}{i}w_{{{\rm new}}}$ with respect to $x$,  $\frac{1}{i}w_{{{\rm new}}}$ curve with $-\pi<x<\frac{1}{i}\log \gamma_2$ should be lower than that with $x=\frac{1}{i}\log \gamma_2$ (i.e.~the red curve in Figure \ref{fig:wnewwnewbar}). These curves thus converge to the $\frac{1}{ i}\log \gamma_1$ curve that is on another Riemann sheet, which results in the curve being $2\pi$ lower than those with $x>\frac{1}{i}\log \gamma_2$. The same reasoning works for $\frac{1}{i}\overline{w}_{{{\rm new}}}$. At stroboscopic time, $\overline{z}_n\to\gamma_1$ for $\overline{z}\neq\gamma_2$ and $\overline{z}_n=\gamma_2$ for $\overline{z}=\gamma_2$. As  $\overline{z}=e^{-ix}$ initially, the unstable fixed point $\overline{z}=\gamma_2$ is instead at $x=-\frac{1}{i}\log \gamma_2=\frac{1}{i}\log \gamma_1$. For initial $x\neq\frac{1}{i}\log \gamma_1$, the $\frac{1}{i}\overline{w}_{{{\rm new}}}$ curve converges to $x=\frac{1}{ i}\log \gamma_2$ (i.e.~$\overline{z}=\gamma_1$), whose stroboscopic values are $\overline{w}_n=\frac{1}{ i}\log \gamma_1$; while for $x=\frac{1}{ i}\log \gamma_1$ (i.e.~$\overline{z}=\gamma_2$), $\frac{1}{i}\overline{w}_{{{\rm new}}}$ follows a different curve that is slightly lower (coloured blue in Figure \ref{fig:wnewwnewbar}), with stroboscopic values $\overline{w}_n=\frac{1}{ i}\log \gamma_2$. Since $\frac{1}{i}\overline{w}_{{{\rm new}}}$ monotonically decreases with $x$, curves with $\frac{1}{ i}\log \gamma_1<x<\pi$ are $2\pi$ lower. 

Averaging $\frac{1}{i}w_{{{\rm new}}}$ and $\frac{1}{i}\overline{w}_{{{\rm new}}}$, the $t_{{{\rm new}}}$ curves are therefore $\pi$ lower for $\frac{1}{i}\log \gamma_1<|x|<\pi$ than for $0<|x|<\frac{1}{ i}\log \gamma_1$ as in Figure \ref{fig:i}; subtracting $\frac{1}{i}w_{{{\rm new}}}$ and $\frac{1}{i}\overline{w}_{{{\rm new}}}$, the $x_{{{\rm new}}}$ curves  thus converge to 0 for  $0<x<\frac{1}{ i}\log \gamma_1$ and $\pm\pi$ for $\frac{1}{i}\log \gamma_1<|x|<\pi$ as in Figure \ref{fig:i}. Note that the reason for $t_{{{\rm new}}}$ to be $\pi$ lower and $x_{{{\rm new}}}$ to converge at $\pi$ (instead of 0) is due to either $\frac{1}{i}w_{{{\rm new}}}$ or $\frac{1}{i}\overline{w}_{{{\rm new}}}$ being $2\pi$ lower. $\frac{1}{i}w_{{{\rm new}}}$ and $\frac{1}{i}\overline{w}_{{{\rm new}}}$ cannot both be $2\pi$ lower, as the $x$ regions for them to be $2\pi$ lower are $-\pi<x<\frac{1}{ i}\log \gamma_2$ and $\frac{1}{ i}\log \gamma_1<x<\pi$, respectively, which do not intersect. 

\section{Two-point functions of twist operators}\label{Calc2pt}
In this appendix, we first evaluate the two-point function $\langle \mathcal{T}_{n}(w_{1,{{\rm new}}},\overline{w}_{1,{{\rm new}}})\mathcal{T}_{-n}(w_{2,{{\rm new}}},\overline{w}_{2,{{\rm new}}})\rangle$ for a CFT$_2$ on a compact space of length $2\pi$, and then evaluate the thermal two-point function $\langle \mathcal{T}_{n}(w_{1,{{\rm new}}},\overline{w}_{1,{{\rm new}}})\mathcal{T}_{-n}(w_{2,{{\rm new}}},\overline{w}_{2,{{\rm new}}})\rangle_{\beta}$ for a CFT$_2$ at temperature $T=\beta^{-1}$ on a non-compact space. Throughout this appendix, we will be working exclusively in $z_{{\rm new}}$ and $w_{{\rm new}}$ coordinate, so we drop the "new" subscript for notation simplicity. Our method is largely based on \cite{Calabrese:2004eu}. Note, however, that we are working in the coordinates 
\begin{align}
    w=\tau+i x && \overline{w}=\tau-i x\label{holocoord}
\end{align}
on the Euclidean cylinder as is mentioned in the main text. This results in a different behavior of the holomorphic and anti-holomorphic parts of the two-point function, in contrast to \cite{Calabrese:2004eu}. We can use the conformal transformation \eqref{expMap} that maps the Euclidean cylinder to the complex plane, where the two-point function of twist operator is simply
\begin{align}
    \langle \mathcal{T}_{n}(z_1,\overline{z}_1)\mathcal{T}_{-n}(z_2,\overline{z}_2)\rangle=\frac{1}{(z_1-z_2)^{2h}}\cdot \frac{1}{(\overline{z}_1-\overline{z}_2)^{2h}}
\end{align}
For the holomorphic part, under a conformal transformation $z\to w$, 
\begin{align}
    \langle \mathcal{T}_{n}(z_1)\mathcal{T}_{-n}(z_2)\rangle=\frac{1}{(z_1-z_2)^{2h}}=\Big(\frac{\partial w_1}{\partial z_1}\Big)^h\Big(\frac{\partial w_2}{\partial z_2}\Big)^h\langle \mathcal{T}_{n}(w_1)\mathcal{T}_{-n}(w_2)\rangle
\end{align}
Substituting \eqref{expMap}, $\langle \mathcal{T}_{n}(w_1)\mathcal{T}_{-n}(w_2)\rangle$ on the Euclidean cylinder is then 
\begin{align}
    \langle \mathcal{T}_{n}(w_1)\mathcal{T}_{-n}(w_2)\rangle=\Big[2\sinh \Big(\frac{1}{2}(w_1-w_2)\Big)\Big]^{-2h}=\Big[2\sinh \Big(\frac{1}{2}((\tau_1-\tau_2)+i(x_1-x_2))\Big)\Big]^{-2h}\label{holo}
\end{align}
The anti-holomorphic part is simply given by $w_i\to\overline{w}_i$, 
\begin{align}
    \langle \mathcal{T}_{n}(\overline{w}_1)\mathcal{T}_{-n}(\overline{w}_2)\rangle=\Big[2\sinh \Big(\frac{1}{2}(\overline{w}_1-\overline{w}_2)\Big)\Big]^{-2h}=\Big[2\sinh \Big(\frac{1}{2}((\tau_1-\tau_2)-i(x_1-x_2))\Big)\Big]^{-2h}\label{antiholo}
\end{align}
Note that \eqref{holo} and \eqref{antiholo} are not equal due to \eqref{holocoord} in our setup. The two-point function $\langle \mathcal{T}_{n}(w_{1,{{\rm new}}},\overline{w}_{1,{{\rm new}}})\mathcal{T}_{-n}(w_{2,{{\rm new}}},\overline{w}_{2,{{\rm new}}})\rangle$ is then given by the product of \eqref{holo} and \eqref{antiholo} and restoring the "new" subscript, which is \eqref{2ptFuncNew} in the main text. 

To calculate the thermal two-point function, $\langle \mathcal{T}_{n}(w_{1,{{\rm new}}},\overline{w}_{1,{{\rm new}}})\mathcal{T}_{-n}(w_{2,{{\rm new}}},\overline{w}_{2,{{\rm new}}})\rangle_{\beta}$ on non-compact space, we simply perform a Wick rotation of the exponential map, and rescale the circumference of the cylinder from $2\pi$ to $\beta$ \cite{Calabrese:2004eu}, 
\begin{align}
    z=e^{\frac{2\pi}{\beta}iw} && \overline{z}=e^{\frac{2\pi}{\beta}i\overline{w}}
\end{align}
this amounts to $w\to iw$ and $2\pi\to\beta$ in \eqref{holo} and \eqref{antiholo}, 
\begin{align}
    \langle \mathcal{T}_{n}(w_1)\mathcal{T}_{-n}(w_2)\rangle_{\beta}&=\Big[\frac{\beta}{\pi}\sin \Big(\frac{\pi}{\beta}(w_1-w_2)\Big)\Big]^{-2h}=\Big[\frac{\beta}{\pi}\sinh \Big(\frac{\pi}{\beta}((\tau_1-\tau_2)+i(x_1-x_2))\Big)\Big]^{-2h}\\
    \langle \mathcal{T}_{n}(\overline{w}_1)\mathcal{T}_{-n}(\overline{w}_2)\rangle_{\beta}&=\Big[\frac{_{\beta}}{\pi}\sin \Big(\frac{\pi}{\beta}(\overline{w}_1-\overline{w}_2)\Big)\Big]^{-2h}=\Big[\frac{_{\beta}}{\pi}\sinh \Big(\frac{\pi}{\beta}((\tau_1-\tau_2)-i(x_1-x_2))\Big)\Big]^{-2h}
\end{align}
The thermal two-point function $\langle \mathcal{T}_{n}(w_{1,{{\rm new}}},\overline{w}_{1,{{\rm new}}})\mathcal{T}_{-n}(w_{2,{{\rm new}}},\overline{w}_{2,{{\rm new}}})\rangle_{\beta}$ is then given by the product of the two, and restoring the "new" subscript.

	\bibliographystyle{JHEP.bst}
	\bibliography{biblio.bib}

\end{document}